\documentclass[12pt]{article}
\pdfoutput=1
\usepackage{amsmath}
\usepackage{graphicx}
\usepackage{enumerate}
\usepackage{natbib}
\usepackage{url} % not crucial - just used below for the URL 

\usepackage[linktoc=all, hidelinks, citecolor = black, urlcolor = blue]{hyperref}  % to hide links (removing color and border)
\hypersetup{
     colorlinks   = false,
     citecolor    = black
}
\usepackage[american]{babel}

\usepackage{mathrsfs}

\usepackage{pdfsync}
\usepackage{color,multirow}
\usepackage{rotating}
\usepackage{threeparttable}
\usepackage{gensymb}
\usepackage{textcomp}
\usepackage{bm}
\usepackage{multirow}
\usepackage{booktabs}
\usepackage{mwe}
\usepackage[]{graphics}
\usepackage{caption} %
\usepackage{adjustbox}
\usepackage{pdflscape}
\usepackage{epstopdf}
\usepackage{enumitem}
\usepackage{amssymb}
\usepackage{amsfonts}

\usepackage{amsthm}

\usepackage{float}
\usepackage[labelformat=simple]{subcaption}

\usepackage{mathtools}
\mathtoolsset{showonlyrefs=true}

\usepackage{setspace}

\newcommand{\diag}{\text{diag}}
\newcommand{\E}{E}
\newcommand{\Var}{\text{Var}}
\newcommand{\Cov}{\text{Cov}}
\newcommand{\Cor}{\text{Cor}}

\newcommand{\R}{\mathbb{R}}
\newcommand{\Nystrom}{Nystr\"{o}m }
\newcommand{\Matern}{Mat\'{e}rn }

\newcommand{\ve}[1]{\bm{{#1}}}
\newcommand{\vesub}[2]{\bm{{#1}}_{#2}}

\newcommand{\vess}[3]{\bm{{#1}}_{#2}^{#3}}

\allowdisplaybreaks
\let \top \relax
\newcommand{\top}{{\mkern-1.5mu\mathsf{T}}}
\usepackage{xr}
\usepackage{titlesec}
\titleformat{\section}{\Large\normalfont\bfseries\setstretch{1}}{\arabic{section}}{1em}{}
\titleformat{\subsection}{\fontsize{12}{15}\bfseries\itshape}
{\thesubsection}{1em}{}
\titleformat{\subsubsection}{\fontsize{12}{15}\itshape}
{\thesubsubsection}{1em}{}
\newcommand{\blind}{1}

\date{}
\begin{document}

\def\spacingset#1{\renewcommand{\baselinestretch}%
{#1}\small\normalsize} \spacingset{1}

%%%%%%%%%%%%%%%%%%%%%%%%%%%%%%%%%%%%%%%%%%%%%%%%%%%%%%%%%%%%%%%%%%%%%%%%%%%%%%

\if1\blind
{
  \title{\bf Modeling Function-Valued Processes with Nonseparable and/or Nonstationary Covariance Structure}
  \author{Evandro Konzen\\
  	School of Mathematics, Statistics \& Physics, Newcastle University, UK\\
  	Jian Qing Shi\thanks{Corresponding author, email: j.q.shi@ncl.ac.uk}\hspace{.2cm}\\
  	School of Mathematics, Statistics \& Physics, Newcastle University, UK\\
  	and \\
  	Zhanfeng Wang\\
    Department of Statistics and Finance, Management School,\\ 
    University of Science and Technology of China, Hefei, China}
  \maketitle
} \fi

\if0\blind
{
  \bigskip
  \bigskip
  \bigskip
  \begin{center}
    {\LARGE\bf Modeling Function-Valued Processes with Nonseparable and/or Nonstationary Covariance Structure}
\end{center}
  \medskip
} \fi

\bigskip
\begin{abstract}
We discuss a general Bayesian framework on modeling multidimensional function-valued processes by using a Gaussian process or a heavy-tailed process as a prior, enabling us to handle nonseparable and/or nonstationary covariance structure. The nonstationarity is introduced by a convolution-based approach through a varying anisotropy matrix, whose parameters vary along the input space and are estimated via a local empirical Bayesian method. For the varying matrix, we propose to use a spherical parametrization, leading to unconstrained and interpretable parameters. The unconstrained nature allows the parameters to be modeled as a nonparametric function of time, spatial location or other covariates. The interpretation of the parameters is based on closed-form expressions, providing valuable insights into nonseparable covariance structures. Furthermore, to extract important information in data with complex covariance structure, the Bayesian framework can decompose the function-valued processes using the eigenvalues and eigensurfaces calculated from the estimated covariance structure. The results are demonstrated by simulation studies and by an application to wind intensity data. Supplementary materials for this article are available online.
\end{abstract}

\noindent%
{\it Keywords:}  Covariance separability; Gaussian process; Spherical parametrization, Varying anisotropy
\vfill

\newpage
\spacingset{1.5} % DON'T change the spacing!
\section{Introduction}\label{sec:Introd}

In multidimensional (or multiway) functional data analysis, we assume that the observed data are realizations of an underlying random process $X(\ve t), \ \ve t \in \mathcal{T} \subset \R^Q$, which has mean function $\mu(\ve t)$ and covariance function $k(\ve t, \ve t') = \Cov\left[X(\ve t),X(\ve t')\right]$.

The accurate estimate of the covariance function, which is one of the key steps in functional principal components analysis (FPCA) and other inference methods for functional data analysis \citep{ramsay2005functional}, is a challenging task. When the dimension of the input space is $Q=2$, the covariance function depends on four arguments and, in the case of sparse designs, nonparametric estimation may suffer from the curse of dimensionality and slow computing. These difficulties are rapidly aggravated as $Q$ becomes larger.

In order to address these issues, many models for two-way functional data (e.g., \cite{chen2012modeling, allen2014generalized, chen2017modelling}) and spatiotemporal data (\cite{banerjee2015hierarchical} and references therein) assume that the covariance function $k(\ve t, \ve t')$ is separable. In other words, they assume that the covariance function can be factorized into the product between $Q$ covariance functions, each one corresponding to one direction.

Besides reducing computational costs and offering simple interpretation, the separability assumption is also useful because it makes it easier to guarantee positive definiteness of the covariance function. However, it does not allow any interaction between the inputs in the covariance structure, and this has motivated recent interest in developing hypothesis tests for separability \citep{aston2017tests, constantinou2017testing,cappello2018testing}.

Although most tend to agree that, ideally, nonseparable models should be used, the description of nonseparability is usually too vague, often limited to its definition. From a practical viewpoint, one would like a clearer understanding on how the nonseparability concept (in other words, interaction between coordinate directions) can be interpreted and how evidence of nonseparability can be shown from a given real dataset. For these reasons, separable covariance models are commonly used not only due to their simplicity but also because it is not clear what is missing in these models.

Several classes of nonseparable covariance functions proposed about two decades ago \citep{cressie1999classes, gneiting2002nonseparable, iaco2002nonseparable, stein2005space} are restricted to the scope of stationarity and many are restricted to limited special cases. Some flexible models based on spectral densities have been proposed (e.g., \cite{stein2005space}), but explicit expressions for the corresponding covariance functions are usually not available.

There are few approaches which consider both nonseparability and nonstationarity in the covariance structure. For example, \cite{jun2007spacetimespheres, jun2008nonstatglobal} apply differential operators with respect to time and spatial coordinates to a stationary process. Although these models can create flexible space-time interaction, the flexibility comes with the cost of difficult understanding on how the coefficients of differential operators affect the resulting covariance structure. In addition, richer models (aiming for flexibility) may be difficult to implement because (i) carefulness is needed to avoid identification problems and (ii) exact likelihood calculation is often not possible. \cite{bruno2009nonsepnonstatozone} also allow for nonseparability and nonstationarity, dealing with the latter through \textit{deformation} \citep{sampson1992deformation}, which consists in transforming the geographical space into another space where stationarity holds. The choice of a suitable transformation is a challenging task. Moreover, \textit{deformation} requires independent replications of the spatial process, which is rare in practice as the observations are usually recorded over time, and therefore some adjustment (e.g., differencing or discarding data) is often needed. We would prefer to include time as a covariate through a modeling approach.

In this paper, we discuss a general Bayesian framework on modeling function-valued processes by using a Gaussian process (GP) or other heavy-tailed processes as a prior, allowing nonseparable and/or nonstationary covariance structure. The nonstationarity is defined by a convolution-based approach \citep{higdon1999non} via a varying kernel. In the case of Gaussian kernel, the nonstationary covariance structure can be simply defined by a varying anisotropy matrix. A local empirical Bayesian approached is used to estimate the hyperparameters involved in the models, including both fixed and varying coefficients.   

We propose to use a spherical parametrization of the varying anisotropy matrix, providing a meaningful interpretation of nonseparability, especially for spatiotemporal data, based a closed-form expression. By using spherical parametrization, the time lag at which two spatial locations have the largest correlation depends on their spatial distance, decay parameters and a degree of nonseparability. The elements of the spherical parametrization can be estimated without any constraint, so that they can be modeled as a nonparametric function of time and/or spatial location, making the model very flexible.

The Bayesian framework provides an efficient approach for obtaining predictive distribution for the unknown underlying regression functions of the processes; in the meantime, it can also decompose the function-valued process using the eigenvalues and eigensurfaces calculated from the estimated covariance structure. A finite number of the eigensurfaces can be used to extract some most important and interpretable information involved in different types of data with complex structure in the spirit of functional principal component analysis.  Nonstationarity and interaction between the coordinate directions can be captured via this flexible approach.   

In Section~\ref{sec:MultiDimProcesses}, we will give a brief introduction on how to define a Bayesian process model for function-valued processes, followed by defining nonstationary covariance structure by a varying kernel or a varying anisotropy matrix in the case of Gaussian kernel via a convolution-based approach. A parametrization method will be discussed and used to model the varying anisotropy matrix, and a local Empirical Bayesian approach will be used to estimate all the hyperparameters included in the covariance structure.  The predictive distribution and decomposition of the random processes are discussed in Section~\ref{sec:GPdecomp}. Some asymptotic theory will also be provided in the section. Simulation studies are presented in Section~\ref{sec:SimulStudies} and an application to wind intensity data in Section~\ref{sec:Application}. Finally, we will give a brief discussion in Section~\ref{sec:Discussion}. Proofs, additional results supporting the simulation studies, an additional application to relative humidity data, and code used to perform the numerical studies are available as supplementary material.

\section{Function-valued Processes with Nonseparable and/or Nonstationary Covariance Structure}\label{sec:MultiDimProcesses}

\addtolength{\textheight}{.3in}%
\subsection{Bayesian Process Models}

Let us consider the following nonlinear functional regression model or a process regression model:
\begin{equation}\label{NLFunReg}
X(\ve t) = f(\ve t) + \varepsilon(\ve t), \quad \varepsilon(\ve t) \sim N(0, \sigma_\varepsilon^2),
\end{equation}
where $\ve t \in  \mathcal{T} \subset \R^Q $ and the unknown nonlinear regression function $f$ is a mapping ${f(\cdot) : \R^Q \rightarrow \R}$. The additive noise $\varepsilon(\ve t)$ is assumed to have normal distribution, but it could have a different distribution (e.g., generalized Gaussian process regression models in \cite{wang2014generalized}).

A variety of models has been proposed to estimate the unknown function $f$. Popular models are based on the approximation $f(\ve t) = \sum_{j=1}^{J} \alpha_j \phi_j(\ve t)$, where $\phi_j$ are basis functions (e.g., smoothing splines \citep{wahba1990spline}). One of the major difficulties  of these frequentist approaches is the \textit{curse of dimensionality} problem in the estimation process when $\ve t$ is multidimensional.

From the Bayesian perspective, the function $f$ is treated as an unknown process (an unknown random function defined in a functional space analogue of a random unknown parameter defined in a conventional Bayesian approach). Therefore, we need to specify a prior distribution over the (random) function $f$ to make probabilistic inference about $f$. One way to do this is by using a Gaussian process (GP) prior.

\setlength{\abovedisplayskip}{6pt}
\setlength{\abovedisplayshortskip}{3pt}
\setlength{\belowdisplayskip}{6pt}
\setlength{\belowdisplayshortskip}{3pt}

The Gaussian process (see e.g., \cite{ohagan1978curvefitting, rasmussen2006gaussian,shi2011gaussian}) is defined as a stochastic process parametrized by its mean function and its covariance function given, respectively, by
\begin{equation}
\mu(\cdot):  \mathcal{T} \rightarrow \R, \ \mu(\ve t) = \E \big[ f(\ve t) \big], \qquad \text{ and } \qquad
k(\cdot, \cdot):  \mathcal{T}^2 \rightarrow \R, \ k(\ve t,\ve t') = \Cov \big[ f(\ve t),f(\ve t') \big]. 
\end{equation}
%\begin{align}
%\mu(\cdot)&:  \mathcal{T} \rightarrow \R, \ \mu(\ve t) = \E \big[ f(\ve t) \big], \label{meanFunctionGP}\\
%k(\cdot, \cdot)&:  \mathcal{T}^2 \rightarrow \R, \ k(\ve t,\ve t') = \Cov \big[ f(\ve t),f(\ve t') \big]. \label{covFunctionGP}
%\end{align}
Henceforth, we will write the GP as 
\begin{equation}\label{GPeq}
f(\cdot) \sim GP \big(\mu(\cdot), k(\cdot,\cdot) \big).
\end{equation}

GP can be seen as a generalization of the multivariate Gaussian distribution to the infinite-dimensional case. When we use a GP prior \eqref{GPeq} for the random function $f$, \eqref{NLFunReg} is referred to as Gaussian process regression (GPR) model. In this case, for any finite $n$ and $\ve t_1, \dots, \ve t_n \in \mathcal{T}$, the joint distribution of $\ve x = \big( x(\ve t_1), \dots,  x(\ve t_n)\big)^\top$ in  \eqref{NLFunReg} is an $n$-variate Gaussian distribution with mean vector $\ve \mu = \big(\mu(\ve t_1), \dots, \mu(\ve t_n)\big)^\top$ and covariance matrix $\ve \Psi_n$ whose $(i,j)$-th entry is given by $\big[\ve \Psi_n\big]_{ij} = k(\ve t_i,\ve t_j) + \delta_{ij} \sigma_\varepsilon^2, \ i,j=1,\dots,n$, where $\delta_{ij}=1$ if $i=j$ and 0 otherwise.

As we will focus on the covariance structure, we will use the mean function estimated via local linear smoother as it is commonly made in FDA (e.g., \cite{yao2005functional}). Other mean models can also be used.

GPR models have become popular for a number of reasons. Firstly, a wide class of nonlinear functions $f$ can be modeled by choosing a suitable prior specification for $k(\cdot, \cdot)$. Other prior distributions can be used for robust heavy-tailed processes \citep{shah14student,wang2017extended,cao2018robust}. This enables us to estimate the covariance structure directly based on the data. In addition, the applicability of GPR models can be readily extended to random process defined on dimensions higher than two. Finally, these models allow to easily quantify the variability of predictions. 

Many recent developments have been made in GPR analysis, including variational GP \citep{tran2015variational}, distributed GP \citep{deisenroth15distribGP}, manifold GP \citep{calandra2016manifold}, linearly constrained GP \citep{jidling2017constrainedGP}, convolutional GP \citep{vanderwilk2017convolutional}, and deep GP \citep{dunlop2018deep}. Some studies investigate connections between GPs with frequentist kernel methods based on reproducing kernel Hilbert spaces \citep{kanagawa2018gaussian}. Finally, many extensions and adaptations have been suggested to apply GPR models to different types of data, such as big data \citep{liu2018GPforBigData}, binary times series \citep{sung2017generalized}, large spatial data \citep{zhang2019GPlargeSpatialdata}, and mixed functional and scalar data in nonparametric functional regression \citep{wang2019GPRforMixed}.

The covariance function $ k(\ve t,\ve t')$ plays a key role in Bayesian process models \eqref{NLFunReg} and \eqref{GPeq}. When the input $\ve t$ is one- or two-dimensional, we can use either a nonparametric covariance  \citep[see e.g.,][]{Hall08} or a parametric one.   
A typical parametric stationary covariance function for the random process $X(\ve t) = f(\ve t) + \varepsilon(\ve t)$ is of the form
\begin{equation}\label{typCovFun}
\Cov \left[X(\ve t), X(\ve t+\ve h) \right] = \sigma^2 g \Big( \sqrt{\ve h^\top \ve A \ve h} \Big) + \sigma_\varepsilon^2 \delta_{\ve h}, 
\end{equation}
where $g$ is a valid correlation function, $\ve A$ is the anisotropy matrix, and $\delta_{\ve h }=1$ if $\ve h =\ve 0$ and $\delta_{\ve h}=0$ otherwise. Suppose that $\ve A=\diag (a_1, \dots, a_Q)$. The hyperparameters $\sigma^2$, $\sigma_\varepsilon^2$ and $1/a_q$ are called the signal variance, the noise variance and the length-scale parameters, respectively. In spatial statistics, these hyperparameters are called the \textit{partial sill}, the \textit{nugget effect} and the \textit{range} parameters, respectively \citep{banerjee2015hierarchical}. The value of $\sigma^2$ controls the vertical scale of variation of $f$.

The diagonal elements of $\ve A$, usually called decay parameters, control how quickly the function $f$ varies on each coordinate direction. The larger the value, the quicker is the variation of $f$ towards the related direction. The off-diagonal elements of $\ve A$ may be non-zero. If $a_{pq} \neq 0$, we say that there exists interaction between the coordinate directions $t_p$ and $t_q$ and covariance functions of the form \eqref{typCovFun} become nonseparable. Large values of $\sigma_\varepsilon^2$ and of $a_q$ both result in more fluctuation of $X$ over ${\ve t}$. The estimated values of these two hyperparameters, however, indicate whether the fluctuation of $X$ over ${\ve t}$ is explained by the signal $f$ or by the noise $\varepsilon$.

The specification of the covariance function is important because it fixes the properties (e.g., stationarity, separability) of the underlying function $f$ that we want to infer. Several families of stationary covariance functions can be chosen, such as the powered exponential, rational quadratic, and \Matern families \citep{shi2011gaussian}. Each family has adjustable parameters which allow separate effects for each coordinate in $\ve t$ and can be inferred from the data. Selection of covariance functions is discussed in \cite{rasmussen2006gaussian} and \cite{shi2011gaussian}.

When ${\ve t}$ is multi-dimensional, a general nonparametric covariance cannot usually be used due to the curse of dimensionality. One way to address the problem is to assume a separable covariance function
\begin{equation}\label{separab}
k(\ve t,\ve t') = k_1(t_1,t_1') \cdots k_Q(t_Q,t_Q').
\end{equation}
That is, if it can be factorized into the product between marginal covariance functions, each one corresponding to one dimension, then it can be modeled nonparametrically \citep[see e.g.,][]{chen2017modelling, rougier2017representation}.

In this paper, we propose a semiparametric approach for the estimation of a flexible covariance function in such way we can relax the assumptions of stationarity and separability. The nonstationarity over ${\ve t}$ is defined by a convolution-based approach via a varying kernel, whose parameters are modeled nonparametrically. In particular, we propose to use a suitable parametrization for the varying anisotropy matrix, allowing unconstrained estimation.

\subsection{Nonstationary Covariance Functions}\label{Sec:nonstatcov}

The linear covariance function $k(\ve t,\ve t') = \sum_{q=1}^{Q} a_q \ve t_q \ve t_q'$ \citep{shi2011gaussian} is an example of nonstationary covariance function.  Its simplicity, though, is of limited use for modeling complex covariance structures and it is often used together with other covariance functions (e.g., \cite{wang2014generalized}).

\cite{higdon1999non} propose a constructive, convolution-based approach to account for nonstationarity in the covariance function. A spatial process $f(\cdot)$ is represented as the convolution of a Gaussian white noise process $z(\cdot)$ with a kernel $k_{\ve t}$, that is,
\begin{equation}\label{Higdon}
f(\ve t) = \int_{\R^Q} k_{\ve t} (\ve u ) z(\ve u ) d\ve u ,
\end{equation}
where the nonstationarity is achieved by considering a spatially-varying kernel $k_{\ve t}$. The covariance function of \eqref{Higdon} takes the form
\begin{equation}\label{HigdonCovFun}
\Cov \big[ f(\ve t),f(\ve t') \big] = \int_{\R^Q} k_{\ve t} (\ve u ) k_{\ve t'} (\ve u ) d\ve u
\end{equation}
and is positive definite provided that $\sup \int_{\R^Q} k_{\ve t} (\ve u )^2 d\ve u   < \infty $. 

The convolution-based approach has become popular mainly because specifying a kernel which satisfies the above condition is much easier than specifying a covariance function directly. \cite{higdon2002space} suggests different process convolution specifications to build flexible space and space-time models.

\cite{paciorek2006} show that the covariance function \eqref{HigdonCovFun} is valid in every Euclidean space $\R^Q, \ Q=1,2,\dots$. They also note that if we assume a Gaussian kernel $k_{\ve t} (\ve u)  = (2 \pi)^{-Q/2} |\ve A (\ve t)|^{1/2} \exp \big\{-(1/2)(\ve t-\ve u)^\top \ve A (\ve t) (\ve t-\ve u)\big\}$, the covariance function of $f(\cdot)$ will be of the form
\begin{equation}\label{CovFunHigdon99}
\Cov \big[ f(\ve t),f(\ve t') \big] = \sigma^2 | \ve  A(\ve t) | ^{-1/4} | \ve  A(\ve t') | ^{-1/4} \bigg| \frac{\ve  A^{-1}(\ve t) + \ve  A^{-1}(\ve t')}{2} \bigg| ^{-1/2} \exp \{ -Q_{\ve t\ve t'}\},
\end{equation}
where
\begin{equation}\label{Qdist}
Q_{\ve t\ve t'} = (\ve t - \ve t')^\top \bigg( \frac{\ve  A^{-1}(\ve t) + \ve  A^{-1}(\ve t')}{2} \bigg) ^{-1} (\ve t - \ve t').
\end{equation}
A more general class for nonstationary covariance functions given by
\begin{equation}\label{NScovfun}
\Cov \big[ f(\ve t),f(\ve t') \big] = \sigma(\ve t) \sigma(\ve t') | \ve A(\ve t) | ^{-1/4} | \ve A(\ve t') | ^{-1/4} \bigg| \frac{\ve A^{-1}(\ve t) + \ve  A^{-1}(\ve t')}{2} \bigg| ^{-1/2} g\Big( \sqrt{Q_{\ve t\ve t'}}\Big),
\end{equation}
where $g(\cdot)$ is a valid isotropic correlation function. The Gaussian process regression model with nonseparable and nonstationary covariance function \eqref{NScovfun} will be referred to as NSGP.

Even if the anisotropy matrix is assumed to be constant ($\ve  A(\ve t) = \ve  A$), the covariance function \eqref{NScovfun} is also nonstationary. In this special case, the nonstationarity is introduced through scaling of a stationary process \citep[Section~3.2]{banerjee2015hierarchical}. In other words, if a stationary process $V({\ve t})$ has mean $0$, variance $1$ and correlation function $\rho$, then $Z({\ve t}) = \sigma({\ve t})V({\ve t})$ is a nonstationary process with covariance function $\Cov \left[ Z ({\ve t}),Z({\ve t}') \right] =  \sigma({\ve t}) \sigma({\ve t}') \rho({\ve t} - {\ve t}')$. The composite Gaussian process model \citep{ba2012composite} also uses this idea to allow varying volatility.

The varying anisotropy matrix $\ve  A(\ve t)$ measures how quickly varying is the fluctuation of the random processes over $\ve t$ and one may want to allow $\ve  A(\ve t)$ to vary with respect to $\ve t$. Both $\sigma(\cdot)$ and $\ve A(\cdot)$ can also vary over $\ve \tau \in {\cal T^*} \subset \R^{Q^*},$ where $Q^* \leq Q$. This $\ve \tau$ can represent, for example, time or spatial coordinates, accounting for time-varying or spatially-varying parameters, or both. This provides a flexible way to model nonstationary and nonseparable covariance structure. We will use the observed data to estimate the covariance structure nonparametrically. The details will be discussed in the Subsection~\ref{subsubsec:VAMparametrisation}.

If $g$ is, for example, a (squared) exponential function, it is easy to see that if and only if we can factorize $\sigma(\ve t) = \sigma(t_1)\cdots\sigma(t_Q)$ and have zero off-diagonal elements in $\ve  A(\ve t)$, then a separable covariance function \eqref{separab} is obtained.

\subsubsection{Local Interpretation of the Varying Anisotropy Matrix}\label{subsubsec:VAMinterp}

In order to visualize the presence of nonseparability, we should not look directly at the covariance function, but rather to the corresponding correlation function. Let $\ve s$ be the spatial location and $\tau$ the time. Under separability, for a certain location $\ve s$ the temporal covariance function $\Cov \big[ f(\ve s,\tau), f(\ve s,\tau') \big] = k_{\ve s} \big(\ve s, \ve s\big) k_\tau \big(\tau,\tau'\big)$ can change with respect to $\ve s$ if a nonstationary model for $k_{\ve s}$ is used. By contrast, the temporal correlation function $\text{Cor} \big[ f(\ve s,\tau), f(\ve s,\tau') \big] = k_\tau (\tau,\tau') / \sqrt{k_\tau (\tau,\tau) k_\tau (\tau',\tau')}$ does not change with $\ve s$ under the separability assumption. Therefore, we should look at correlation functions to analyze nonseparability.

\begin{figure*}[t!]
	\centering
	\begin{subfigure}[t]{0.5\textwidth}
		\centering
	\includegraphics[width=1\linewidth]{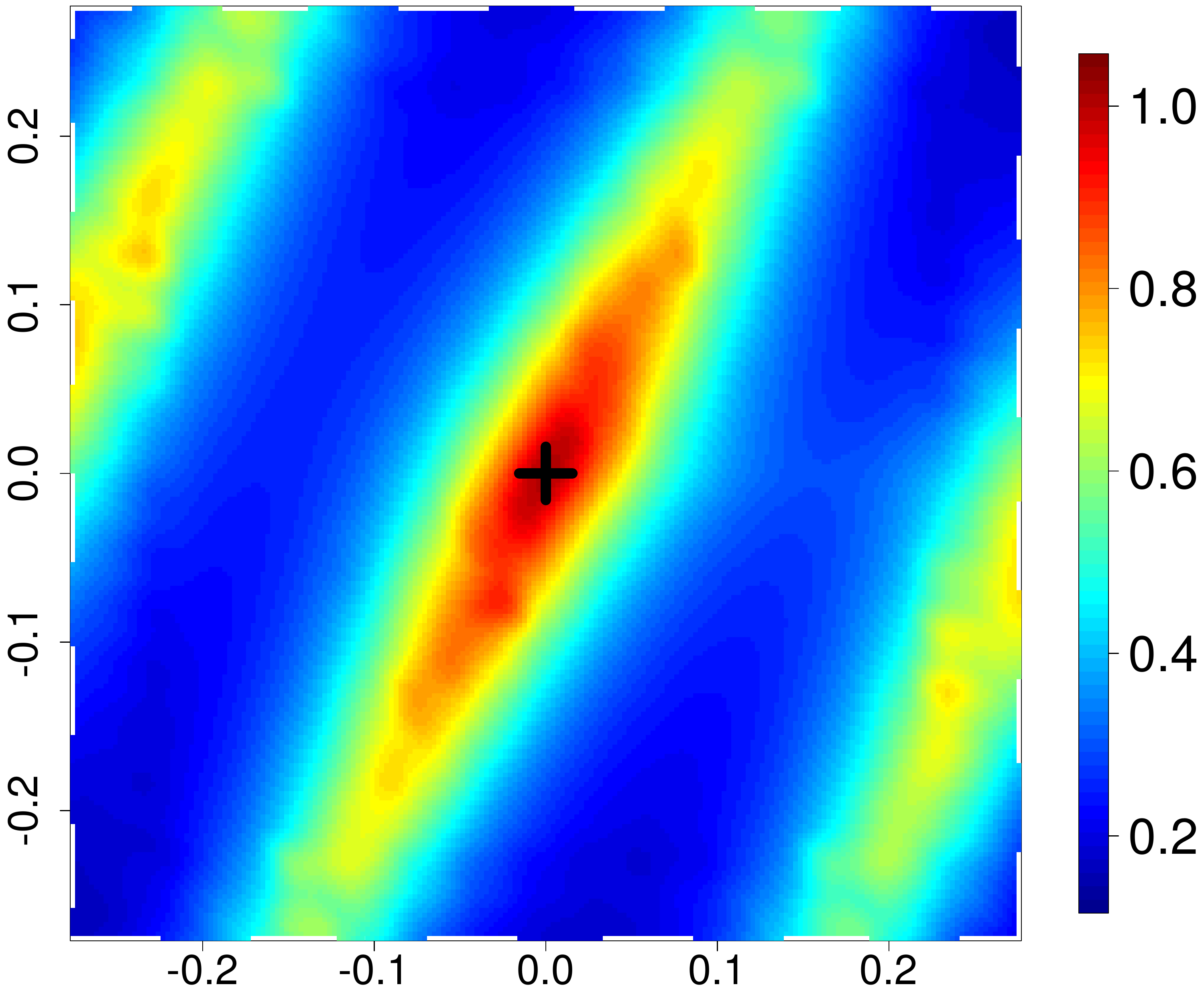}
		\caption{}
		\label{Fig:simConvsubfig-a}
	\end{subfigure}%
	~ 
	\begin{subfigure}[t]{0.5\textwidth}
		\centering
	\includegraphics[width=0.83\linewidth]{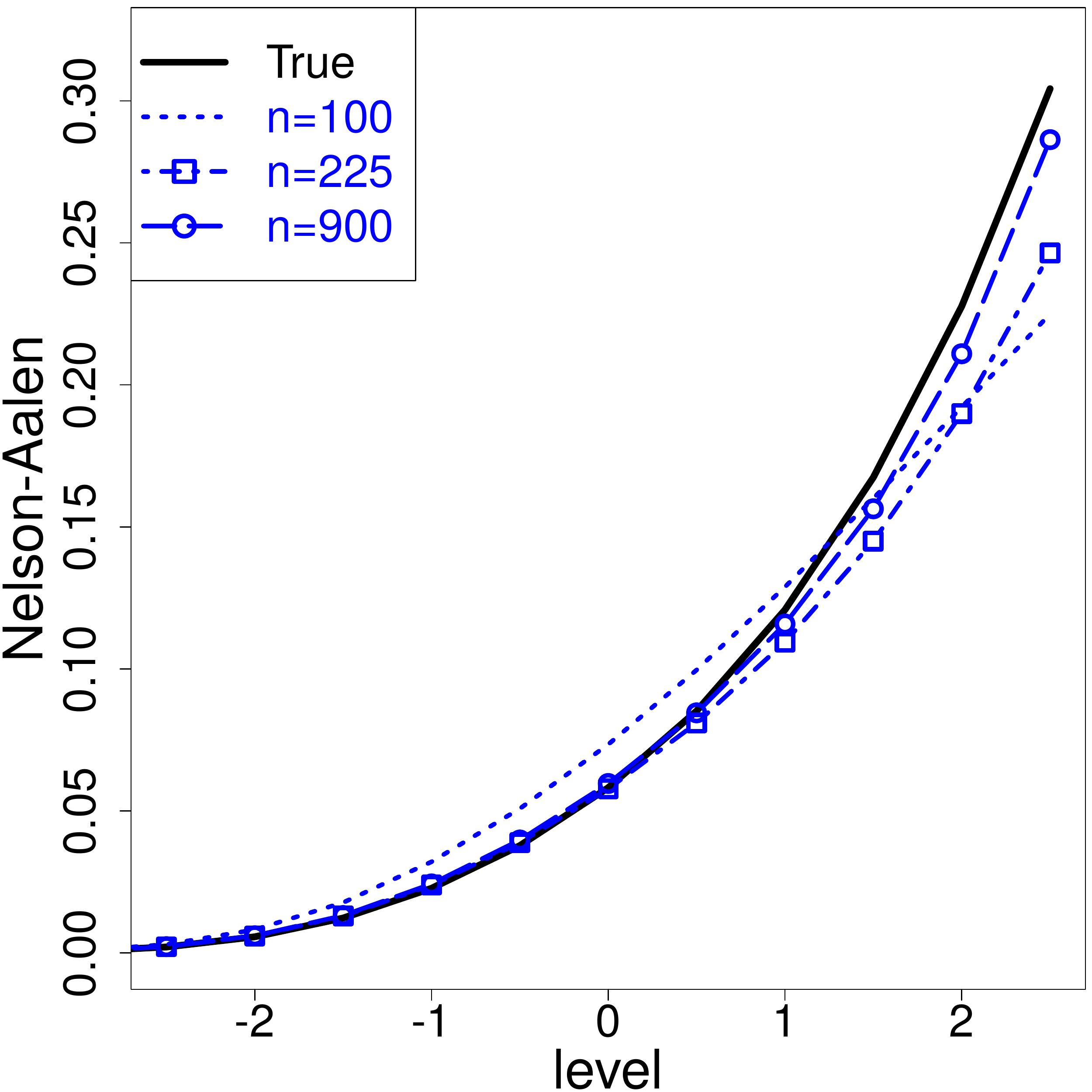}
		\caption{}
		\label{Fig:simConvsubfig-b}
	\end{subfigure}
	\caption{(a) Heatmap of the nonseparable, nonstationary correlation function used in Subsection~\ref{secSim_converg}: $\Cor \big[ X(s_1,s_2), X(s_1',s_2') \big]$ plotted along $s_1-s_1'$ (x-axis) and $s_2-s_2'$ (y-axis) at $s_1'=s_2'=0.31$. (b) Nelson-Aalen plots for simulation study in Subsection~\ref{secSim_converg}. The continuous line is based on the true covariance function, while the other lines are based on the NSGP covariance function estimated by using different sample sizes $n$.}
\end{figure*}

Figure~\ref{Fig:simConvsubfig-a} illustrates the heatmap of a nonseparable, nonstationary correlation function \eqref{NScovfun} of a two-dimensional function-valued process $X(s_1,s_2)$. This correlation function is used in the simulation study of Subsection~\ref{secSim_converg}. The nonseparability aspect is clear in Figure~\ref{Fig:simConvsubfig-a} because of the diagonally oriented ellipses. Note also that, in contrast to commonly used stationary correlation kernels, \eqref{NScovfun} with a spatially-varying anisotropy matrix does not make the correlation function decay monotonically with respect to $\ve s - \ve s'$. Relaxing this assumption is especially useful, for example, if one coordinate is time and we want a flexible model which accommodates complex seasonality patterns in the correlation structure.

Suppose the parameters in \eqref{NScovfun} vary smoothly along a subset $\ve \tau \in {\cal T^*} \subset \R^{Q^*}, \ Q^* \leq Q$. Thus, we can say that model \eqref{NScovfun} is locally stationary, i.e. with locally constant parameters, and so \eqref{NScovfun} becomes
\begin{equation}\label{NScovfunLocal}
\Cov \big[ f(\ve t),f(\ve t') \big] = \sigma^2 g\Big( \sqrt{Q_{\ve t\ve t'}}\Big), \qquad Q_{\ve t\ve t'} = (\ve t - \ve t')^\top \ve A (\ve t - \ve t').
\end{equation}
If the correlation function $g$ is powered exponential,
\begin{equation}\label{sqExpCorr}
g(\ve t, \ve t') = \Cor \big[f(\ve t), f(\ve t')\big] = \exp \Big\{- \big[(\ve t - \ve t')^\top \ve A (\ve t - \ve t')\big]^p \Big\}, \qquad 0 < p \leq 2.
\end{equation}
Let $\ve t = (s, \tau)^\top$, where $s$ is the spatial location and $\tau$ the time. At the same time point, two locations $s$ and $s'$ might be not highly correlated; but they can be highly correlated with some time lag $\tau-\tau'$. We are interested in the time lag at which these locations have the largest correlation. That is, given locations $s$ and $s'$ and a time point $\tau'$, we want to find
\begin{equation}\label{taumax}
\tau^*(s,s', \tau') = \underset{\tau}{\mathrm{argmax}} \ \Cor \big[f(s,\tau), f(s',\tau')\big].
\end{equation}
If we use the correlation function \eqref{sqExpCorr}, then \eqref{taumax}  is solved by
\begin{align}\label{taumax2}
\tau^*(s,s',\tau') &= \underset{\tau}{\mathrm{argmin}} \ \ve A_{11} (s-s')^2 + \ve A_{22} (\tau-\tau')^2 + 2 \ve A_{12} (s-s') (\tau-\tau') \\
&= \tau' + \frac{\ve A_{12}}{\ve A_{22}} (s'-s).
\end{align}
In separable models, $\ve A_{12}=0$ and thus the maximum correlation between locations $s$ and $s'$ always occurs with no time lag, i.e. $\tau^*(s,s',\tau') - \tau' = 0$. Expressions similar to \eqref{taumax2} can be obtained for $Q>2$; see Section~\ref{sec:SupplVAM} of the Supplementary Material.

\subsubsection{Parametrization of the Varying Anisotropy Matrix}\label{subsubsec:VAMparametrisation}

We must ensure positive definiteness of the anisotropy matrix $\ve A(\ve t)$ in \eqref{NScovfun}. This can be done by using different parametrizations. For example, \cite{higdon1998atlantic,higdon1999non,risser2017locallik} use geometrically-based parametrizations which capture local anisotropy by rotating and stretching coordinate directions. \cite{paciorek2006} suggest using a spectral decomposition. However, these methods are either designed for some special cases or are difficult to provide a clear interpretation of its elements.

\cite{pinheiro1996unconstrained} present other five parametrizations for a covariance matrix, one of which is the spherical parametrization, a particularly interesting strategy because it provides direct interpretation of parameters in terms of variances and correlations. We propose using the spherical parametrization for $\ve  A(\ve t)$ and interpret the parameters in terms of decay parameters and directions of dependence between the inputs.

As discussed above, the off-diagonal elements of $\ve A(\ve t)$ have to be zero to produce a separable covariance function. Therefore, a value which is distant from zero indicates nonseparable covariance structure due to the interaction between the coordinate directions of $\ve t$ in the way the process fluctuates over $\ve t$.

We will consider the Cholesky decomposition
\begin{equation}%\label{Chol}
\ve A (\ve t) = \ve A (\ve \tau) = \ve B(\ve \tau )^\top \ve B(\ve \tau ),
\end{equation}
where $\ve B$ is an $Q \times Q$ upper triangular matrix (including the main diagonal). Positiveness of the main diagonal entries of $\ve B$ ensures that $\ve A$ is positive definite.

We will follow closely the exposition of \cite{pinheiro1996unconstrained} to explain the spherical parametrization. Let $\ve B_q$ denote the $q$-th column of $\ve B$ and $\beta_q$ denote the spherical coordinates of the first $q$ elements of $\ve B_q$. Therefore, we have
\begin{align*}
\left[B_q \right]_1 &=  \left[\beta_q \right]_1 \cos(\left[\beta_q \right]_2), \\
\left[B_q \right]_2 &=  \left[\beta_q \right]_1 \sin(\left[\beta_q \right]_2) \cos(\left[\beta_q \right]_3), \\
& \dots , \\
\left[B_q \right]_{q-1} &=  \left[\beta_q \right]_1 \sin(\left[\beta_q \right]_2) \cdots \cos(\left[\beta_q \right]_q), \\
\left[B_q \right]_q &=  \left[\beta_q \right]_1 \sin(\left[\beta_q \right]_2)  \cdots \sin(\left[\beta_q \right]_q).	
\end{align*}

We can show that $\ve A_{qq} = \left[\beta_q \right]_1^2 $ and $\rho_{1q} = \cos(\left[\beta_q \right]_2), \ q=2,\dots,Q$, with $-1 < \rho_{1q} < 1$. This means that we can interpret the values of $\ve B$ in terms of the decay parameters and directions of dependence (hereafter called degree of nonseparability) in $\ve A$.

Now, in the two-dimensional case $Q=2$, \eqref{taumax2} becomes
\begin{equation}\label{taumax2new}
\tau^*(s,s',\tau') = \tau' + \frac{\left[\beta_1 \right]_1}{\left[\beta_2 \right]_1} \cos(\left[\beta_2 \right]_2) (s'-s).
\end{equation}
In other words, the time lag at which the maximum correlation between locations $s$ and $s'$ occurs depends on (i) the spatial distance $s'-s$, (ii) the (square root of) decay parameters related to time and location, and (iii) the degree of nonseparability ${\rho_{12} = \cos(\left[\beta_2 \right]_2) \in (-1,1)}$. 

The cosine of the angle between two random variables can be seen as the Pearson's correlation coefficient from a geometric perspective. In the spherical parametrization of the anisotropy matrix, the cosine of the spherical coordinates $\left[\beta_q \right]_2, \ q=2,\dots, Q$, measures the interaction between directions $\ve t_1$ and $\ve t_q$. If $\cos(\left[\beta_q \right]_2) \neq 0$ for some $q$, the covariance structure is nonseparable.

In \eqref{taumax2new}, in the separable case,  $\cos(\left[\beta_2 \right]_2) =0$ and therefore $\tau^*(s,s',\tau') - \tau' =0$. In the nonseparable case, i.e. $\cos(\left[\beta_2 \right]_2) \neq 0$, a separable model tend to underestimate (overestimate) the linear dependence between the locations when in fact they are strongly (weakly) correlated with some time lag.

The spherical parametrization is unique if $\left[\beta_q \right]_1  > 0,  \ q=1,\dots,Q,$ and $\left[\beta_q \right]_p \in (0,\pi)$, ${q=2,\dots,Q,} \ \ p=2,\dots,q$. We can then easily proceed with an unconstrained estimation by defining a new vector of parameters $\ve \alpha$ including $\log (\left[\beta_q \right]_1),$ $q=1,\dots,Q,$ and $\log \big(\left[\beta_q \right]_p/(\pi - \left[\beta_q \right]_p) \big), \ q=2,\dots,Q, \ \ p=2,\dots,q$. The upper triangular matrix $\ve B$ can be reparametrized by $\ve \alpha$. Each element ${\ve \alpha_j = \ve \alpha_j (\ve \tau)}$, for $j=1,\dots,Q(Q+1)/2$, depends on $\ve \tau$ if the covariance structure is nonstationary.

The unconstrained estimation of each element of $\ve \alpha$ allows $\ve \alpha$ to be modeled as a nonparametric function of ${\ve \tau}$.  In addition, the spherical parametrization has some other advantages over other parametrizations in that: (i) it is uniquely defined and can be readily extended for any $Q>2$, which is difficult when implementing geometrically-based parametrizations; (ii) it has about the same computational efficiency as the Cholesky parametrization applied directly; and (iii) we can make interpretations on the spherical coordinates $\left[\beta_q \right]_p$.

A geometrical interpretation of the spherical parametrization can be seen in \cite{rapisarda2007parameterizing}. Other parametrizations based on Cholesky decomposition has been widely discussed. \cite{zhang2015joint} mention that unconstrained nature of the parametrization of the Cholesky factor allows to represent angles of the spherical parametrization via regression as
functions of some covariates, an idea also used by \cite{pourahmadi1999joint} and \cite{leng2010semiparametric} when parametrizing covariance matrices using a modified Cholesky decomposition.

If the covariance structure depends along one coordinate direction $\tau \subset \R$ (i.e. $Q^* = 1$, e.g., time-varying parameters), many nonparametric methods can be used, e.g.,
\begin{equation}\label{bsplines}
\ve \alpha_j(\tau) = \sum_{l=1}^{L} \ve \theta_{j l} \ve \gamma_{j l}(\tau),
\end{equation}
where $\ve \gamma_l$ form B-spline basis functions \citep{deBoor2001book}. This representation ensures that the resulting function is smooth and still very flexible as we can change the degree of the piecewise polynomials and the number and location of knots. The locations of the knots are usually the quantiles of $\tau$, but they can be chosen differently; we can also allow discontinuities in derivatives by repeating knots at the same location. The gain of adding more knots comes with the cost of increasing the number of coefficients to be estimated. Typically, the number of knots is chosen by cross-validation. Some other methods can be used, e.g.,  \cite{ba2012composite} use a Gaussian kernel regression model.

For multidimensional $\ve \tau \subset \R^{Q^*}$ (e.g., spatially-varying parameters), we can construct multivariate B-splines basis function by taking the product of the $Q^*$ univariate basis.

An alternative method is to use a Gaussian process to model each $\ve \alpha_j (\ve \tau)$ using a parametric covariance function. Let $\ve \alpha_{ji} = \ve \alpha_{j}(\ve \tau_i), \ i=1,\dots,n$. Then we define 
\begin{equation}\label{thetaGP}
(\ve \alpha_{j1}, \dots, \ve \alpha_{jn}) \sim N\big(\ve 0, \ve K_j(\ve \theta_j)\big),
\end{equation}
where $\ve K_j$ is an $n \times n$ covariance matrix where its $(i,i')$-th element is calculated by the covariance function $k_j(\ve \tau_i, \ve \tau_{i'}; \ve \theta_j)$, depending on unknown parameter $\ve \theta_j$. In practice, we may use the same covariance function for $j=1, \dots, Q$ and for $j=Q+1, \dots, Q(Q+1)/2$. This method can cope with the large dimensional cases, i.e. $Q^*>1$. 

\subsubsection{Model Learning}

We now denote the covariance function constructed by \eqref{NScovfun} and the above parametrization methods by $k(\ve t, \ve t'; \ve \theta)$ for any $\ve t, \ve t' \in \R^Q$, where $\ve \theta$ includes all the unknown parameters in \eqref{bsplines} if B-splines are used or in \eqref{thetaGP} if GPRs are used; in addition, $\ve \theta$ includes $\log(\sigma^2)$ in \eqref{CovFunHigdon99}. We will use an empirical Bayesian approach to estimate the unknown parameters and thus the nonstationary covariance structure.

For a given set of observed data ${\cal D} = \{ \ve x, \ve t \} = \{ (x_i, t_{i1},\dots, t_{iQ}), \ 1,\dots,n \}$, a GPR model for \eqref{NLFunReg} can be written as
\begin{align}\label{GPR2}
x_i| f_i  & \overset{\text{{\tiny i.i.d.}}}{\sim}  N (f_i, \sigma_\varepsilon^2), \\
(f_1, \dots, f_n)  & \overset{\hphantom{\text{i.i.d.}}}{\sim}  GP \big(\ve 0, k(\cdot, \cdot; \ve \theta) \big),
\end{align}
where the covariance function $k(\cdot, \cdot; \ve \theta)$ may be nonstationary, constructed using the methods discussed above. Thus, the marginal distribution of $\ve x$ given $\ve \theta$ is 
\begin{equation}\label{margDist}
p(\ve x| \ve \theta) = \int p(\ve x| \ve f) p(\ve f| \ve \theta) d\ve f,
\end{equation}
where $p(\ve x| \ve f) = \prod_{i=1}^{n} \zeta(f_i)$, with $\zeta(f_i)$ denoting the normal probability density function with mean $f_i$ and variance $\sigma_\varepsilon^2$, and $\ve f = \big(f(t_1),\dots,f(t_n)\big)^\top \sim N(\ve 0, \ve K_n)$, where $\big[\ve K_n\big]_{ij} = k(\ve t_i,\ve t_j)$, $i,j=1,\dots,n$. For convenience, $\ve \theta$ includes the parameter $\sigma_\varepsilon^2$ as well. For Gaussian data defined in \eqref{GPR2}, the marginal distribution of $\ve x$ is $N(\ve 0, \ve \Psi_n)$, where $\ve \Psi_n = \ve K_n + \sigma_\varepsilon^2 \ve I_n$, the marginal log-likelihood of $\ve \theta$ is given by
\begin{equation}\label{LogLikTheta}
\ell(\ve \theta|\mathcal{D}) = -\frac{1}{2} \log |\ve \Psi_n(\ve \theta)|  -\frac{1}{2} \ve x'\ve \Psi_n(\ve \theta)^{-1} \ve x  -\frac{n}{2} \log 2 \pi.
\end{equation}
The estimates of $\ve \theta$ in \eqref{LogLikTheta} are called empirical Bayes estimates as they are obtained by using observed data \citep{carlin2008bayesian}.

To reduce the computational costs when calculating the determinant and the inverse of $\ve \Psi_n$ in \eqref{LogLikTheta}, we can instead use local likelihood estimation (LLE) \citep{tibshirani1987locallik}. In the LLE, instead of maximizing \eqref{LogLikTheta} directly, we maximize 
\begin{equation}\label{LocalLogLikTheta}
\ell_k(\ve \alpha_k|\mathcal{D}_k) = -\frac{1}{2} \log |\ve \Psi(\ve \alpha_k)|  -\frac{1}{2} \ve x_k'\ve \Psi(\ve \alpha_k)^{-1} \ve x_k  -\frac{n_k}{2} \log 2 \pi
\end{equation}
locally, where $k$ is the index of location $\ve t_k$. Estimates of $\ve \alpha_k$ are obtained by considering only the data in the neighborhood of $\ve t_k$, that is, ${\cal D}_k= \big\{ (\ve x_i, \ve t_i): \{  || \ve t_i - \ve t_k  || < r \} \big\}$, where $r$ is a predefined radius. Using the available observations in the neighborhood of $\ve t_k$ is important as the behavior of the covariance function near the origin determines properties of the process \citep{Stein1999book}. 
\cite{risser2017locallik} suggest a mixture component approach in which they estimate the spatially varying parameters $\ve \alpha_k, \  k=1, \dots, k_{max}$, locally and then, for any arbitrary location $\ve t$, $\ve \alpha(\ve t)$ is obtained by averaging, respectively, $\ve \alpha_{k}, \ k=1, \dots, k_{max}$, with a weight function depending on the distance between $\ve t_k$ and $\ve t$.

A special case is when the nonstationarity depends on one coordinate direction as discussed around equation \eqref{bsplines}. We can use B-spline basis functions and then estimate the corresponding coefficients $\ve \theta_{j l}$ by maximizing \eqref{LogLikTheta}, yielding unconstrained $\tau$-varying estimates of the continuous functions $\ve \alpha_j (\tau)$. In practice, we may simply estimate the unconstrained $\ve \alpha_{jk}$ locally for some locations via \eqref{LocalLogLikTheta} (i.e. assuming $\ve \alpha_{jk}$ is constant within a neighborhood) and then regress these estimates to obtain smooth functions $\ve \alpha_j(\tau)$ over $\tau$, using a nonparametric approach, e.g., B-splines.

\section{Prediction and Decomposition of Function-valued Processes}\label{sec:GPdecomp}

\subsection{Bayesian Prediction and Decomposition}

Let us consider the GPR model \eqref{GPR2}. The posterior distribution $p(\ve f | {\cal D}, \sigma_\varepsilon^2)$ is a multivariate Gaussian distribution with $\E \left[\ve f | {\cal D}, \sigma_\varepsilon^2\right] = \ve K_n(\ve K_n+\sigma_\varepsilon^2 \ve I_n)^{-1} \ve x$ and $\Var \left[ \ve f | {\cal D}, \sigma_\varepsilon^2 \right] = \sigma_\varepsilon^2 \ve K_n(\ve K_n+\sigma_\varepsilon^2 \ve I_n)^{-1}$.

The marginal distribution of $\ve x$ is $N(\ve 0, \ve \Psi_n)$, where $\ve \Psi_n = \ve K_n + \sigma_\varepsilon^2 \ve I_n$. Therefore, we can easily make predictions of test data at locations $\ve t$ given the observed data ${\cal D}$. The posterior distribution $p(f(\ve t) | {\cal D})$ also has multivariate normal distribution, with 
\begin{align}
E \left[f(\ve t) | {\cal D}\right] &= \vess{k}{n}{\top}(\ve t)(\vesub{K}{n}+\sigma_\varepsilon^2 \vesub{I}{n})^{-1}\ve x, \label{poster_f} \\
\Var \left[ f(\ve t) | {\cal D} \right] &= k(\ve t,\ve t) - \vess{k}{n}{\top}(\ve t)(\vesub{K}{n}+\sigma_\varepsilon^2 \vesub{I}{n})^{-1} \vess{k}{n}{}(\ve t),
\end{align}
where 
$\ve x=\big(x(\ve t_1), \dots, x(\ve t_n) \big)^\top$, 
$\ve K_n=\big(k(\ve t_i,\ve t_j)\big)_{n\times n}$, and $\ve k_n(\ve t)= \big(k(\ve t_1,\ve t), \dots,k(\ve t_n,\ve t) \big)^\top$.

However, the predictive distribution becomes much more complicated for non-Gaussian data (see, e.g., \cite{wang2014generalized}). We may therefore consider using the decomposition methods detailed below.

Once the covariance function $k(\cdot, \cdot)$ is estimated, we can obtain its eigenfunctions $\phi(\cdot)$ via \Nystrom approximation method. Thus a finite GPR approximation can be obtained as in \eqref{KLdec} by using only the first $J$ eigenfunctions. This allows us to make predictions at any arbitrary location $\ve t$ given observed data $\ve x$ and a finite number of components $\phi(\cdot)$ similarly as in FPCA:
\begin{equation}\label{KLdec}
X(\ve t) = \mu(\ve t) + \sum_{j=1}^{\infty} \xi_j \phi_j(\ve t),
\end{equation}
where $\mu$ is the mean function, $\xi_j$ are uncorrelated random variables and $\phi_j$ are eigenfunctions of the covariance operator of $X$ (Karhunen-Lo\'{e}ve expansion), i.e. $\phi_j$ are solutions to the equation 
\[
\int k(\ve t, \ve t')\phi(\ve t') d \ve t' = \lambda \phi(\ve t).
\]

The decomposition \eqref{KLdec} is especially useful to identify the main modes of variation in the data. In addition, the covariance function $k(\cdot, \cdot)$ is $(2 \times Q)$-dimensional, which makes its visualization rather difficult; therefore, it might be important to look at its eigenfunctions to identify some of features of the covariance function. 

The eigenvalue $\lambda_j$ is the variance of $X$ in the principal direction $\phi_j$ and the cumulative fraction of variance explained by the first $J$ directions is given by
\begin{equation}\label{CFVEeq}
\text{CFVE}_J = \frac{\sum_{j=1}^J \lambda_j}{\sum_{j=1}^M \lambda_j}, \quad \text{ where }M \text{ is large}.
\end{equation}

Note that the decomposition \eqref{KLdec} is based on the covariance function $k(\ve t, \ve t'), \ \ve t, \ve t' \in \R^Q$, constructed and learned by the methods discussed in the previous section. It can model the covariance structure even if it is nonstationary or nonseparable. By contrast, most of the existing methods are based on the separable assumption for the multidimensional case, (i.e. $Q>1$). For example, \cite{chen2017modelling} suggest using tensor product representations, namely Product FPCA and Marginal FPCA, in which the eigensurfaces $\phi_j(\cdot)$ in \eqref{KLdec} are assumed to be the product of eigenfunctions estimated separately in each coordinate direction. In the model of Product FPCA, the two-dimensional function-valued process $X$ is represented as
$\label{ProdFPCA}
X(s,\tau) = \mu(s,\tau) + \sum_{j=1}^{\infty}\sum_{l=1}^{\infty} \chi_{jl} \phi_{j}(s) \psi_l(\tau),
$
where $\phi$ and $\psi$ are the eigenfunctions of the marginal covariance functions for the $Q=2$-dimensional case. This is a special case of \eqref{KLdec}; we use it only when we are sure that the data has a separable covariance structure \citep{aston2017tests}. In general, we should use \eqref{KLdec}.

\subsection{Asymptotic Theory}

In this subsection, we provide asymptotic theory for the decomposition and Bayesian prediction based on a Gaussian process prior with a general covariance structure discussed above. The proofs are given in Section~\ref{sec:ProofsSuppl} of the Supplementary Material. 

In equation \eqref{KLdec}, $\xi_j$ are independent normal random variables and $\phi_j(\cdot)$ are the eigenfunctions of the kernel function $k(\cdot, \cdot)$. Therefore, the eigenfunctions are orthonormal satisfying 
\begin{equation}\label{EigEq}
\int k(\ve t,\ve t') \phi_j (\ve t') d \ve t' = \lambda_j \phi_j(\ve t), \ \ \ \int \phi_i(\ve t) \phi_j(\ve t) d \ve t = \delta_{ij},
\end{equation}
where $\lambda_1 \geq \lambda_2 \geq \dots \geq 0$ are the eigenvalues of $k(\cdot, \cdot)$ and $\delta_{ij}$ is the Kronecker delta. 

Let $X^c(\ve t) = X(\ve t) - \mu(\ve t)$ and $\Phi (f) = \int_{\mathcal{T}} k(\ve t,\cdot) f(\ve t) d \ve t$ be an operator for $f \in L^2 (\mathcal{T})$. In fact,
\begin{equation}\label{xiScores}
\xi_j = \langle X^c(\cdot), \phi_j(\cdot) \rangle = \int X^c(\ve t) \phi_j(\ve t) d \ve t
\end{equation}
has mean $0$ and variance $\lambda_j$.

\vskip 10pt
\noindent{\bf Theorem 1.}
\textit{	For $J \geq 1$, for which  $\lambda_J>0$, the functions $\left\{\phi_j, \ j=1,\dots,J \right\}$ provide the best finite dimensional approximation to $X^c(\ve t)$ with respect to minimizing criterion
	\begin{equation}\label{BestFiniteDimApprEq}
	\underset{g_1,\dots, g_J \in L^2 (\mathcal{T})}{\mathrm{argmin}} E \Big[ || X^c(\ve t) - \sum_{j=1}^{J} g_j(\ve t) \xi_j^* ||^2  \Big],
	\end{equation}
	where $g_1,\dots, g_J \in L^2 (\mathcal{T})$ are orthonormal and $\xi_j^* = \langle X^c(\cdot), g_j(\cdot) \rangle = \int X^c(\ve t) g_j(\ve t) d \ve t $. The minimising value is $\sum_{j=J+1}^{\infty} \lambda_j$. }

This theorem is similar to Theorem 1 in \cite{chen2017modelling}; but the latter provides the best finite approximation under the separability assumption. The above theorem is true for a very general covariance structure even if it is nonstationary or nonseparable.

The following theorem provides the convergence rates also under a general covariance structure.

\vskip 10pt
\noindent{\bf Theorem 2.}
\textit{Suppose conditions C1 - C3 (see Supplementary Material) hold and $\hat \mu(\ve t)$ satisfies $\sup_{\ve t}|\hat \mu(\ve t)-\mu(\ve t)|=O_p\big(\{\log(n)/n\}^{1/2}\big)$. Therefore, for $1\leq j\leq J$, 
	\begin{align}
	&||k_{\hat\theta}(\cdot,\cdot)-k_\theta(\cdot,\cdot)||=O_p(\{\log(n)/n\}^{1/2}),\label{th-2-1}\\
	&||\hat \lambda_j-\lambda_j||=O_p(\{\log(n)/n\}^{1/2}),\label{th-2-2}\\
	&||\hat\phi_j(\cdot)-\phi_j(\cdot)||=O_p(\{\log(n)/n\}^{1/2}),\label{th-2-3}\\
	&||\hat\xi_j-\xi_j||=O_p(\{\log(n)/n\}^{1/2}).\label{th-2-4}
	\end{align}
}

We now look at the relationship between the Bayesian prediction and the decomposition based on Karhunen-Lo\'{e}ve expansion. Using models \eqref{NLFunReg} and \eqref{GPeq}, where $f \sim GP(0,{k})$ with ${k}={k}_\theta$ and $\varepsilon(\ve t)$ is a Gaussian error process  $GP(0,k_\varepsilon)$ with ${k_\varepsilon(\ve t, \ve t')=\sigma_\varepsilon^2I(\ve t=\ve t')}$.
Hence, $X \sim GP(0,\tilde{k}_\theta)$ where $\tilde{k}_\theta={k}_\theta+k_\varepsilon$. Given $\ve f = \big(f(\ve t_1), \dots,f(\ve t_n) \big)^\top$, we use ${E \left[f(\ve t) | \ve f \right] = \vess{k}{n}{\top}(\ve t)\vesub{K}{n}^{-1}\ve f}$ to estimate $f(\ve t)$. Given the observed data ${\cal D}$, we use \eqref{poster_f} to estimate $f(\ve t)$.

In addition, from Karhunen-Lo\'{e}ve expansion we have
\begin{align}\label{KLdecompffstar}
f(\ve t)=\sum_{j=1}^\infty \phi_{j}(\ve t) \xi_{j}, \qquad X(\ve t)=\sum_{j=1}^\infty \tilde{\phi}_{j}(\ve t) \tilde{\xi}_{j},
\end{align}
where $\phi_{j}(\cdot)$ and $\tilde{\phi}_{j}(\cdot)$ are the eigenfunctions of $k_\theta$ and $\tilde{k}_\theta$, respectively, and $\lambda_{1}\geq\lambda_{2}\geq\cdots\geq 0$ and $\tilde{\lambda}_{1}\geq\tilde{\lambda}_{2}\geq\cdots\geq 0$ their corresponding eigenvalues.  The truncated sum of \eqref{KLdecompffstar} will be
\begin{align}\label{fRandomFunction}
f_{n}(\ve t)=\sum_{j=1}^n \phi_{j}(\ve t) \xi_{j}, \qquad X_{n}(\ve t)=\sum_{j=1}^n \tilde{\phi}_{j}(\ve t) \tilde{\xi}_{j}. 
\end{align}
\vskip 10pt
\noindent{\bf Theorem 3}
\textit{Under the conditions in Theorem 2, $\E \left[f(\ve t)| \ve f \right] = f_{n}(\ve t)+o_p(1)$. Moreover, under model \eqref{NLFunReg}, $\E \left[f(\ve t)|{\cal D}\right] = X_{n}(\ve t)+o_p(1)$. }
\vskip 10pt

This theorem indicates that the Bayesian prediction and Karhunen-Lo\'{e}ve expansion provide similar results. This gives flexibility in functional data analysis. If we are mainly interested in a predictive model for Gaussian data, we may just use the Bayesian prediction. The implementation is fairly efficient if the sample size is not very large. However, if we are also interested in how the covariance is structured, we may study the leading eigenfunctions and corresponding eigenvalues; more discussion will be given in the next sections. The finite dimensional approximation also provides a way to develop efficient approximation for big data (e.g., \Nystrom method \citep[p.42]{shi2011gaussian}) and for non-Gaussian data.

\section{Simulation Studies}\label{sec:SimulStudies}

In this section, we provide two examples of simulated data with nonseparable, nonstationary covariance structure. In order to assess the covariance function estimated by various methods, we primarily use the Nelson-Aalen cumulative hazard estimator \citep{garside2020topol}, which is briefly explained in Section~\ref{sec:NelsonAalenSuppl} of the Supplementary Material. All analyses were conducted in a mixture of R and C++ code, using the \texttt{Rccp} \citep{RccpPackage} and \texttt{RcppArmadillo} \citep{RccpArmadilloPackage} packages for integration of compiled C++ code with R. An additional simulation study for a three-dimensional function-valued process ($Q=3$) is provided in Section~\ref{sec:SupplSim3d} of the Supplementary Material.

\subsection{Simulation Study 1}\label{secSim_2D_cheb}

Let $X(s_1, s_2) = f(s_1, s_2) + \varepsilon(s_1, s_2), \ s_q \in {\cal T}_q \subset \left[-1,1\right]$, $q=1,2$, be a function-valued process, where $f$ has zero mean function and covariance function given by 
\[
\Cov \left[f(\ve s), f(\ve s')\right] = \sum_{j=1}^{20} b_j \phi_j(s_1+s_2) \phi_j(s_1'+s_2'),
\]
where $ \phi_j(\cdot)$ are Chebyshev polynomials on $\in \left[-1,1\right]$, $b_j = 10 j^{-3/2}$, and the noise variance is $\sigma_\varepsilon^2(s_1, s_2) = 0.01$. The basis functions of the form $\phi_j(s_1+s_2)$ are clearly nonseparable and produce a nonseparable covariance structure. We simulate $100$ surfaces on a grid with $n_1 \times n_2 = 20 \times 20 = 400$ equally spaced points. 

Then we estimate the covariance structure of the simulated data by NSGP model. The unconstrained parameters $\ve \alpha_{pq}$, related to the elements $\ve A_{pq}$, are modeled using B-spline basis functions, similarly to \eqref{bsplines}, but now considering that $\ve \alpha_{pq}$ change along two coordinate directions:
\begin{equation}\label{bsplines2d}
\ve \alpha_{pq}(s_1,s_2) = \sum_{l=1}^{L} \sum_{m=1}^{M}\ve \theta_{j l m} \ve \gamma_{l}^{(1)}(s_1) \ve \gamma_{m}^{(2)}(s_2),  \qquad \qquad p,q=1,2,
\end{equation}
where $\ve \gamma_{l}^{(1)}$ stands for the $\ell$-th basis function defined on $\mathcal{T}_1$,  and $\ve \gamma_{m}^{(2)}$ stands for the $m$-th basis function defined on $\mathcal{T}_2$. In this simulation study, we use $L=M=5$ B-spline basis functions with evenly spaced knots.

We also implement two separable models: Marginal and Product FPCA \citep{chen2017modelling}. These models are based on the product of the first few eigenfunctions (we use the first six) of marginal covariance functions. The marginal covariance functions are estimated non-parametrically, following \cite{yao2005functional} and by using the \texttt{fdapace} \citep{fdapacePackage} package. R code for obtaining marginal eigenfunctions is available at \url{https://www.stat.pitt.edu/khchen/pub.html}.

\begin{figure*}[t!]
	\centering
	\begin{subfigure}[b]{0.5\textwidth}
		\centering
	\includegraphics[width=0.92\linewidth]{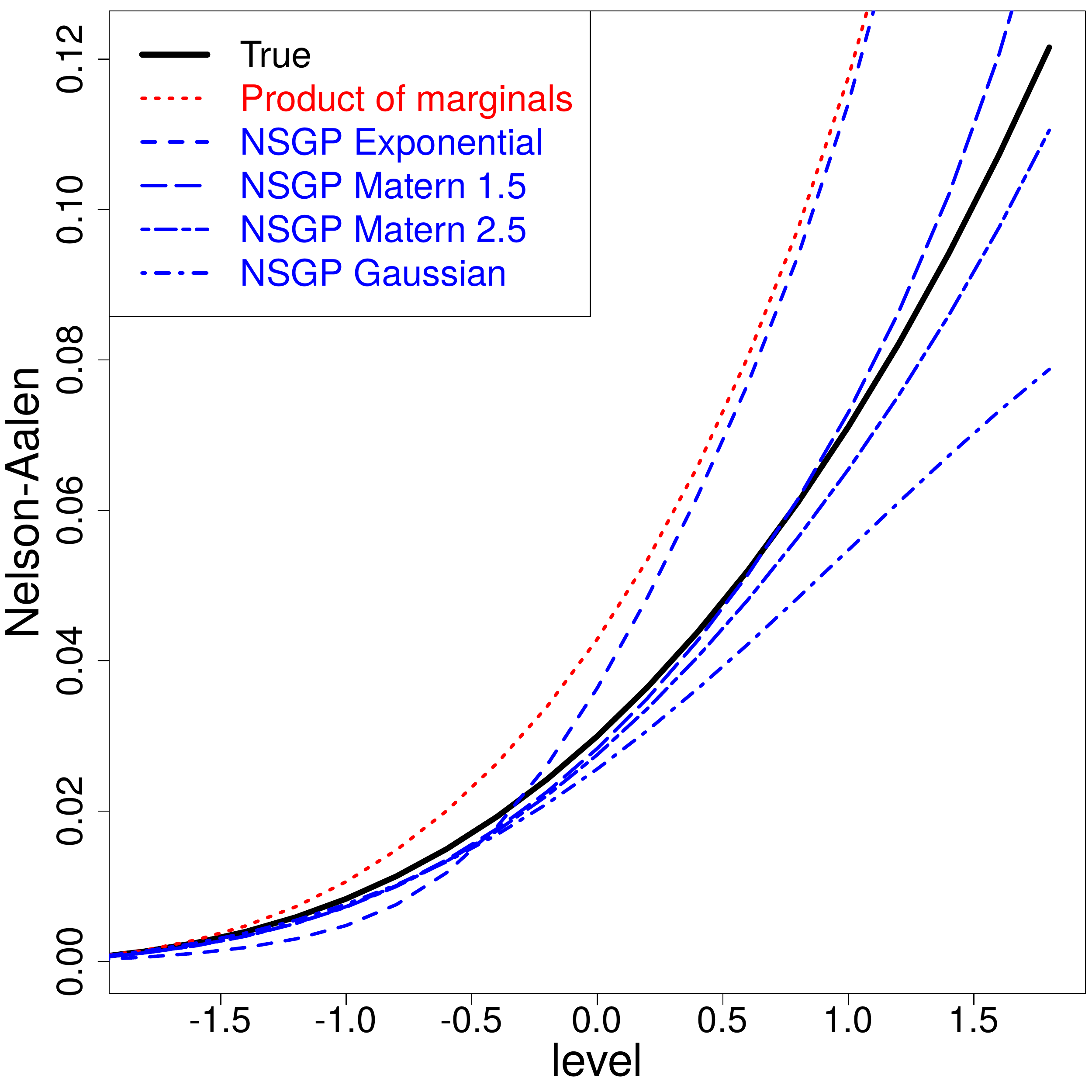}
		\caption{}
		\label{Fig:simChebsubfig-a}
	\end{subfigure}%
	~ 
	\begin{subfigure}[b]{0.5\textwidth}
		\centering
	\includegraphics[width=0.92\linewidth]{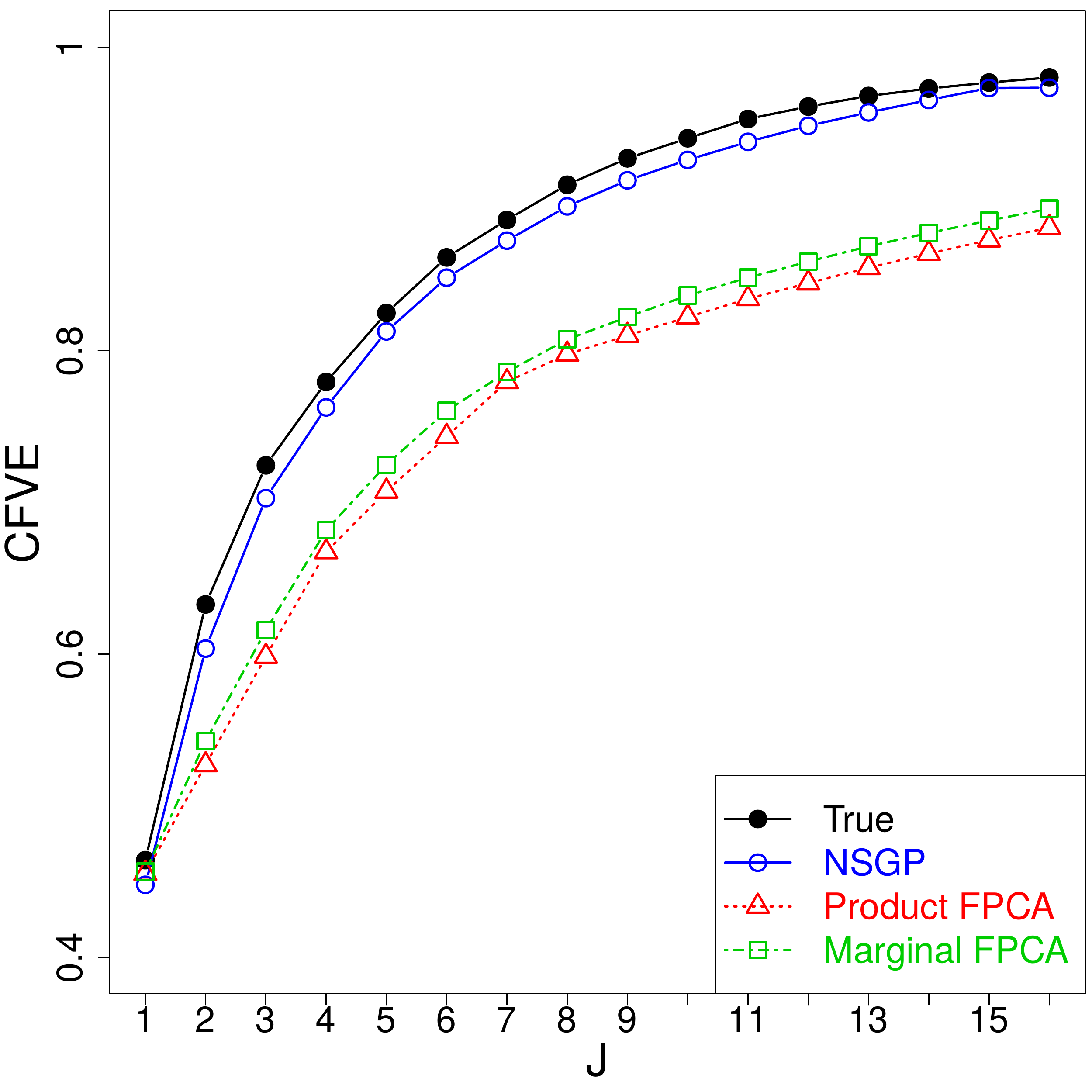} 
		\caption{}
		\label{Fig:simChebsubfig-b}
	\end{subfigure}
	\caption{Nelson-Aalen and CFVE results for data with covariance structure based on Chebyshev polynomials. (a) Nelson-Aalen plots: the continuous line is based on the true covariance function; the dotted line on the separable model; and the dashed lines on NSGP model with different specifications for $g$. (b) CFVE obtained by the true eigensurfaces, and by eigensurfaces obtained by NSGP model (with \Matern $\nu=1.5$) and by product representation models.}
\end{figure*}

Figure~\ref{Fig:simChebsubfig-a} shows Nelson-Aalen plots obtained by different methods, including NSGP models and a separable model. Provided that a suitable correlation function $g$ is chosen, NSGP model obtains Nelson-Aalen estimates close to the true one, while the separable model does not. The covariance function of the separable model used for producing the results in 

Figure~\ref{Fig:simChebsubfig-a} is the product between the marginal covariance functions estimated non-parametrically, and not simply the covariance structure resulting from the product of the first few eigenfunctions. The contribution of the leading eigensurfaces of the methods in terms of CFVE \eqref{CFVEeq} can be seen in Figure~\ref{Fig:simChebsubfig-b}. The figure shows that, when using more than one component, NSGP model based on \Matern correlation function ($\nu=1.5$) is preferable to separable models.

\begin{figure}[H]
	\centering 
	\includegraphics[width=0.85\linewidth]{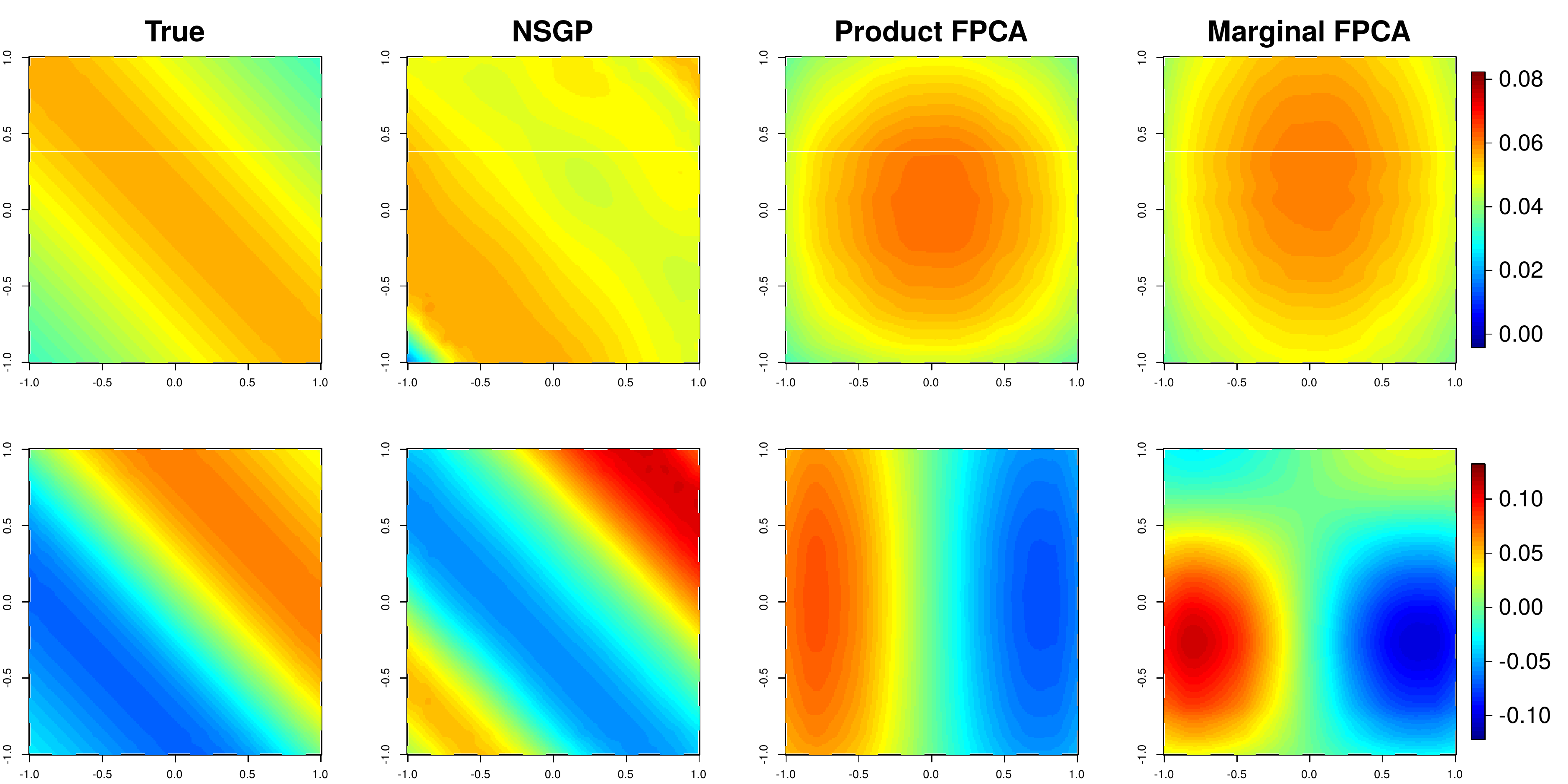} 
	\caption{First two leading eigensurfaces $\phi_j(s_1, s_2)$ of the true (based on Chebyshev polynomials) model (first column) and the corresponding eigensurfaces $\hat{\phi}_j(s_1, s_2)$ estimated by NSGP model (with \Matern $\nu=1.5$) (second column), Product FPCA (third) and Marginal FPCA (fourth).}
	\label{eigensurf_cheb}
\end{figure}

Figure~\ref{eigensurf_cheb} indicates that this NSGP model obtains more accurate estimates of the leading eigensurfaces than separable models do. The diagonal shapes of the true leading eigensurfaces suggest strong nonseparability aspects in the covariance function, aspects which are not captured by the models which assume covariance separability. In this example, the eigensurfaces have diagonally oriented shapes because they are polynomials of the sum $s_1+s_2$, and later eigensurfaces change faster along the input domain as they are polynomials of higher order. The third and fourth leading eigensurfaces can be seen in Section~\ref{sec:SupplsecSim_2D_cheb} of the Supplementary Material. 

\subsection{Simulation Study 2}\label{secSim_converg}

The purpose of this simulation study is to assess the estimation accuracy of a given spatially-varying anisotropy matrix for different sample sizes. We simulate ten realizations from a function-valued process $X(s_1, s_2)$,  where $s_q \in {\cal T}_q \subset \left[0,1\right], \ q=1,2$, from \eqref{NLFunReg}, where $f$ is zero-mean Gaussian process with covariance function \eqref{NScovfun}, exponential correlation kernel $g(\cdot)$ and $\sigma(s_1,s_2)=1$. The elements of the true varying anisotropy matrix $\ve A(s_1,s_2)$ are:
\begin{align}
& \ve A_{11}(s_1,s_2) = \exp\big(6 \cos(10 s_1 - 5 s_2)\big), \qquad \ve A_{22}(s_1,s_2) = \exp\big(\sin(6 s_1^3) + \cos(6 s_2^4)\big),\\
& \ve A_{12}(s_1,s_2) = \{\ve A_{11}(s_1,s_2)\ve A_{22}(s_1,s_2)\}^{1/2}\rho_{12}(s_1,s_2), \quad \rho_{12}(s_1,s_2) = \tanh\big((s_1^2+s_2^2)/2\big).
\end{align}

The unconstrained parameters associated to the varying anisotropy matrix are modeled via \eqref{bsplines2d}, using $L=M=6$ B-spline basis functions with evenly spaced knots. To assess the estimation accuracy of the elements of $\ve A(s_1,s_2)$, we employ the integrated squared error
\begin{equation}\label{eq:ISE}
\text{ISE}\Big(\log \hat{\ve A}_{pq} (s_1,s_2)\Big) = \int_{\mathcal{T}_1 } \int_{\mathcal{T}_2} \left[ \log \ve  A_{pq}(s_1,s_2) - \log \hat{\ve A}_{pq}(s_1,s_2) \right]^2 ds_2 ds_1.
\end{equation} 
ISE results in Table~\ref{Table:ISESimConv} show that the estimates of $\ve A(s_1,s_2)$ are very good for fairly small sample sizes ($20 \times 20$), and can be obtained without much computational time. The implementation was conducted on a 16GB, 2.20GHz Linux machine. 

Nelson-Aalen plots (Figure~\ref{Fig:simConvsubfig-b}) show that the estimated model becomes more consistent with the true one as the sample size increases. Estimated eigensurfaces and corresponding CFVE can be seen in Section~\ref{sec:SupplSimConv} of the Supplementary Material.

\begin{table}[H]
	\centering
	\caption{ISE for the elements of the varying anisotropy matrix which was estimated by using different sample sizes $n$. The last column displays the estimation time.}
	\begin{tabular}{crrrr}
		\toprule
		$n$ & $\log \hat{\ve A}_{11}(s_1,s_2)$ & $\log \hat{\ve A}_{22}(s_1,s_2)$ & $\hat{\rho}_{12}(s_1,s_2)$ & time (hours) \\ 
		\midrule
		100 & 68.9580 & 32.6482 & 0.9002 & 0.17 \\ 
		225 & 10.6267 & 12.4810 & 0.2883 & 0.42 \\ 
		400 & 1.2665 & 0.3339 & 0.2365 & 0.89 \\ 
		900 & 1.7725 & 0.1105 & 0.3265 & 2.54 \\ 
		\bottomrule
	\end{tabular}
	\label{Table:ISESimConv}
\end{table}

\section{Application to Wind Intensity Data}\label{sec:Application}

In this application, we aim to illustrate the benefits of using NSGP model and some comparison with separable, stationary models. In order to examine nonseparability aspects, as discussed previously, we look at data standardized by the standard deviation across realizations and then analyze the correlation structure. We interpret the empirical correlation function as the \textit{true} one. We highlight, though, that the empirical correlation function is only available because there exist multiple realizations and that these are on the same grid.

\begin{figure*}[t!]
	\centering
	\begin{subfigure}[b]{0.4\textwidth}
		\centering
		\includegraphics[width=1\linewidth]{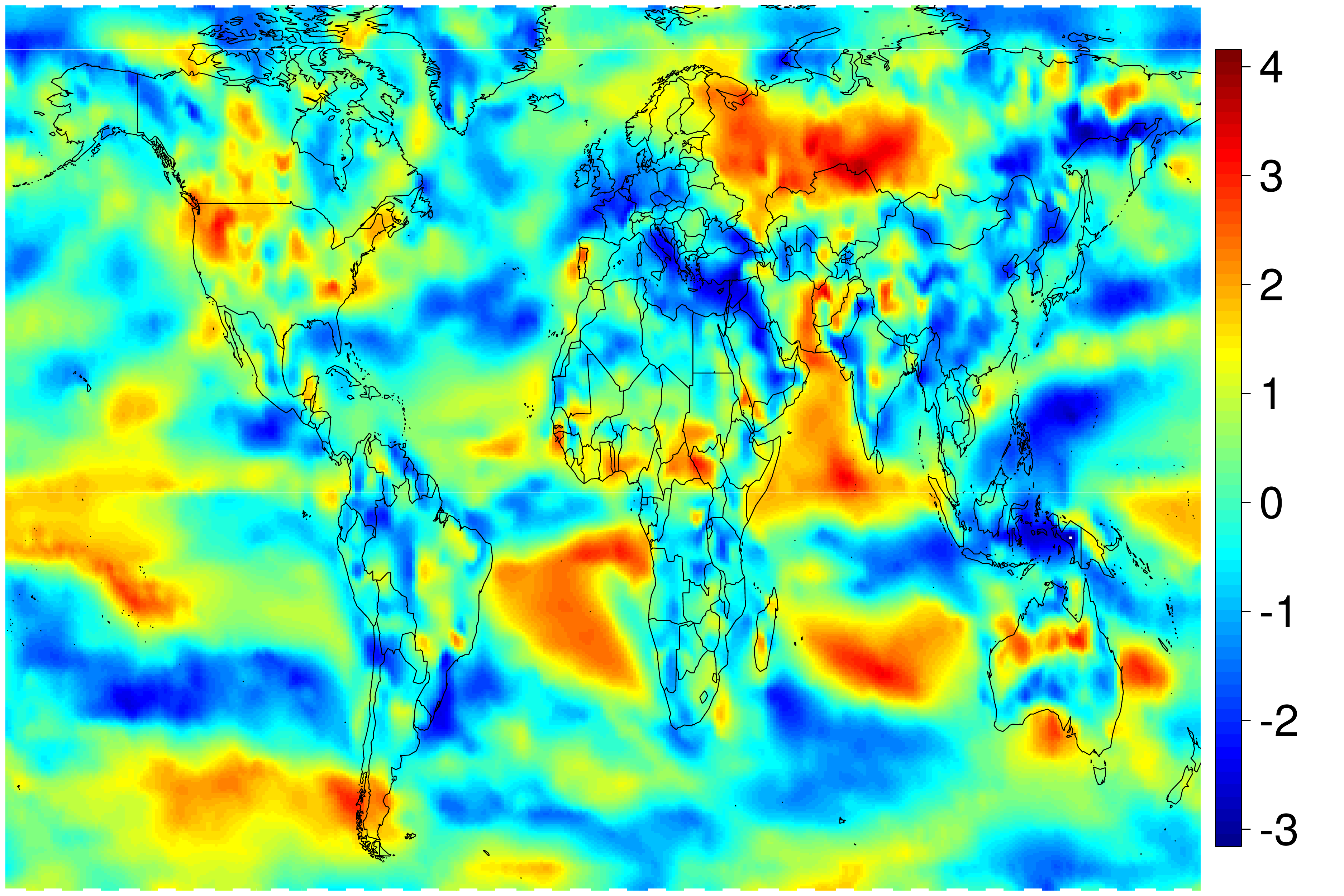}
		\caption{}
		\label{Fig:windsubfig-a} \vspace{0.2cm}
	\end{subfigure}%
	~ 
	\begin{subfigure}[b]{0.4\textwidth}
		\centering
		\includegraphics[width=1\linewidth]{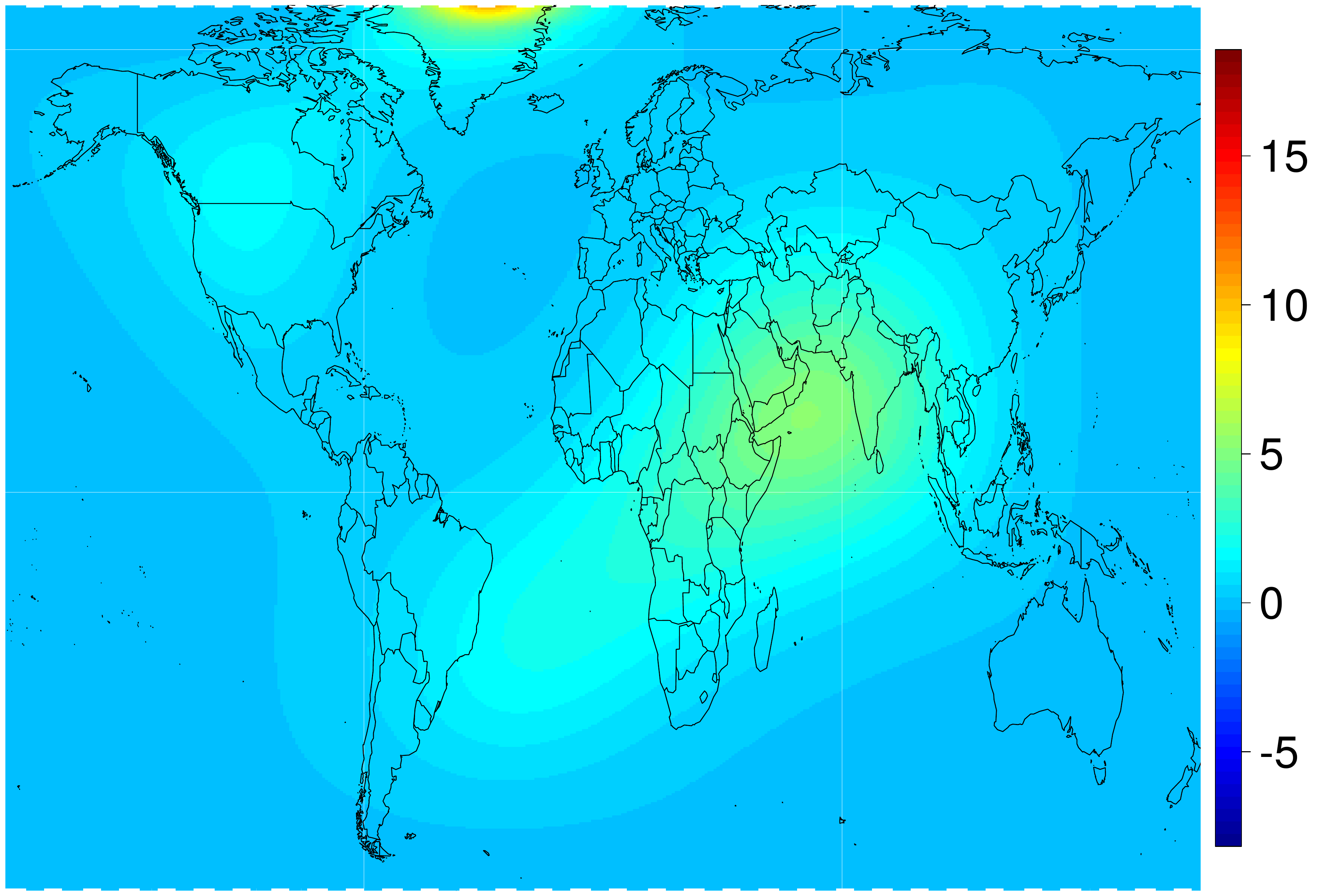}
		\caption{}
		\label{Fig:windsubfig-b} \vspace{0.2cm}
	\end{subfigure}
	~ 
	\begin{subfigure}[b]{0.4\textwidth}
		\centering
		\includegraphics[width=1\linewidth]{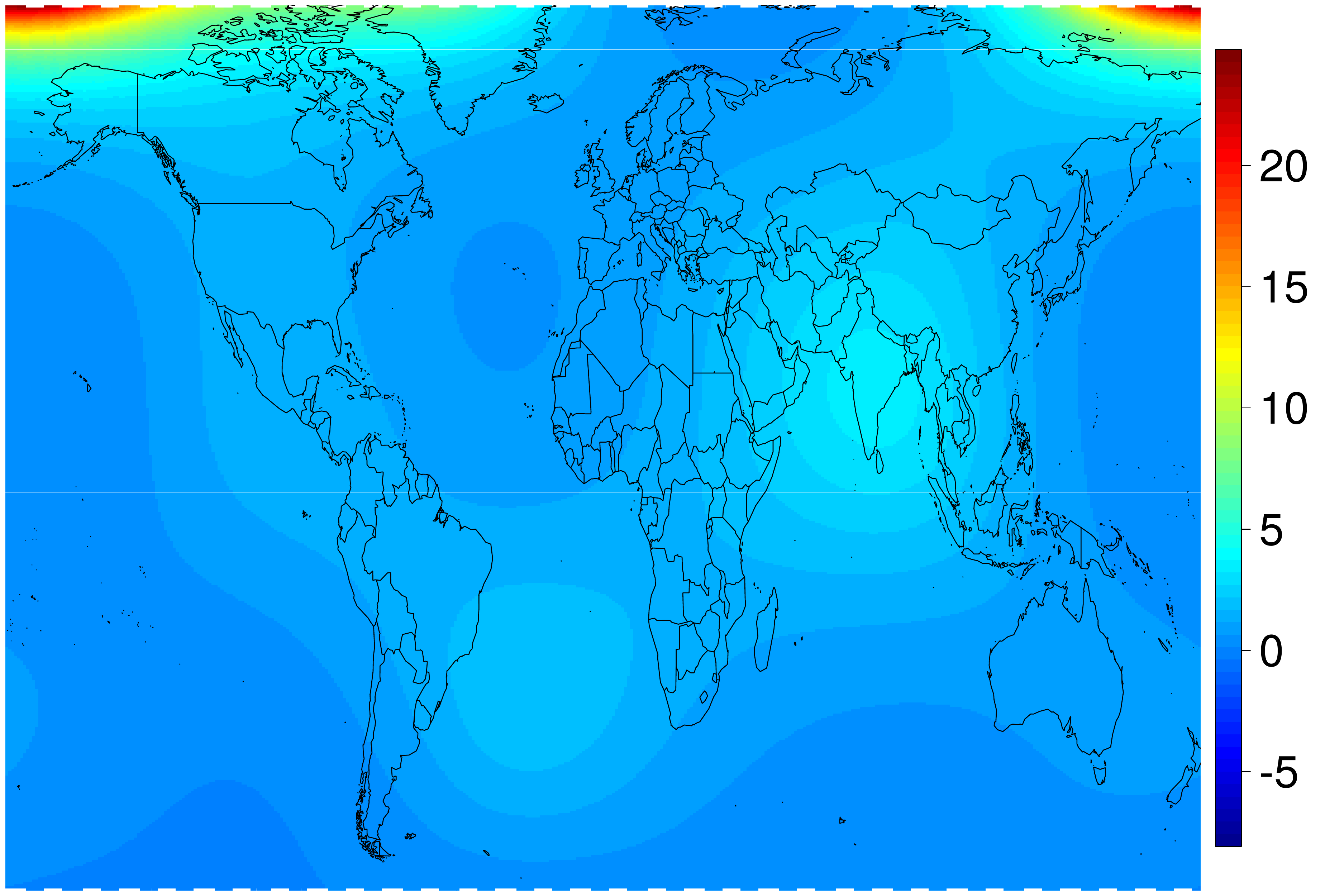}
		\caption{}
		\label{Fig:windsubfig-c}
	\end{subfigure}
	~ 
	\begin{subfigure}[b]{0.4\textwidth}
		\centering
		\includegraphics[width=1\linewidth]{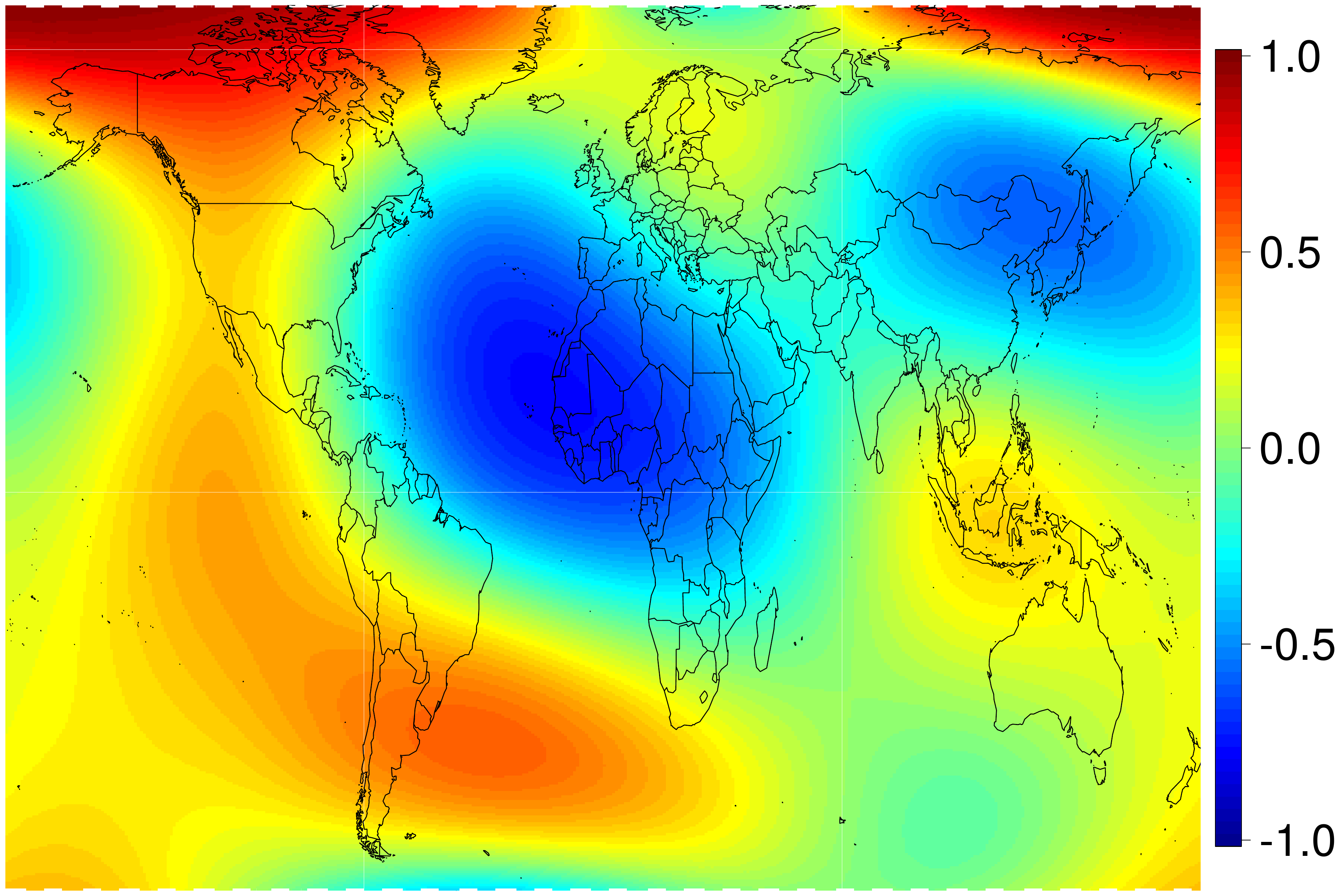}
		\caption{}
		\label{Fig:windsubfig-d}
	\end{subfigure}
	\caption{(a) One of the 30 realizations from the annual wind intensity data. (b), (c) and (d) show the elements of the varying anisotropy matrix obtained by NSGP model with \Matern ($\nu=1.5$) correlation function $g$: (b) $\ve A_{11}(s_1,s_2)$; (c) $\ve A_{22}(s_1,s_2)$; and (d) degree of nonseparability $\rho_{12}(s_1,s_2)$.}
	\label{Fig:windmaps}
\end{figure*}

The application is to annual wind intensities at $10$m in 2006. The dataset is used by \cite{garside2020topol} and consists of $30$ realizations observed on a $51 \times 96$ (latitude $\times$ longitude) grid. At each point in the grid, the data are standardized by the mean and standard deviation of the $30$ values at that location, so that we assume zero mean function and unit marginal variance. A realization is shown in Figure~\ref{Fig:windsubfig-a}.

%Community Earth System Model (CESM) Large Ensemble project \citep{kay2015community}.

\begin{figure}[H]
	\centering
	\includegraphics[width=0.45\linewidth]{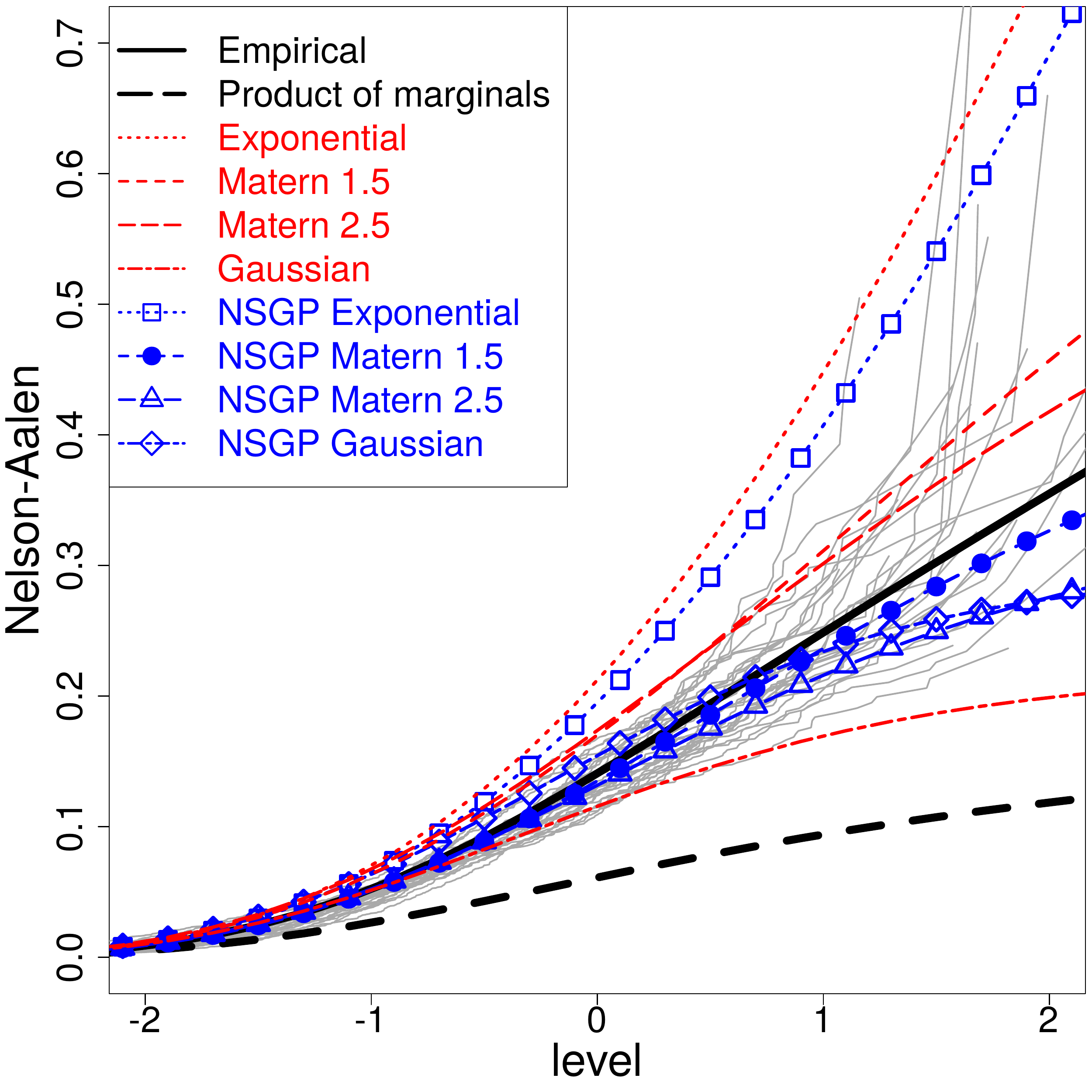}
	\caption{Nelson-Aalen plots for wind intensity data. Grey lines represent the results for the $30$ realizations. The continuous, thick line is based on the empirical correlation function. The dashed, thick line is on the product representation model. The other lines are results based on separable, stationary Gaussian process models and NSGP models with different specifications for $g$. }
	\label{Fig:ClimData_NelsonAalen}
\end{figure}

Figure~\ref{Fig:ClimData_NelsonAalen} displays Nelson-Aalen results for the $30$ realizations and for a few competing models. The empirical correlation function seems to represent well the overall behavior of the $30$ realizations, and is interpreted as the Nelson-Aalen for the true correlation function. 

We fit a Gaussian process regression model with four parametric separable, stationary correlation functions -- Exponential, \Matern ($\nu=1.5$), \Matern ($\nu=2.5$) and Gaussian. We also fit NSGP model \eqref{NScovfun} with $\sigma(s_1,s_2) = 1$ and with the same four specifications for the correlation function $g$. The unconstrained elements of the spatially-varying anisotropy matrix are modeled via \eqref{bsplines2d}, using $L=M=5$ B-spline basis functions with evenly spaced knots. Cyclic B-spline basis functions are used for the longitudinal direction.

Figure~\ref{Fig:ClimData_NelsonAalen} suggests that the model based on the product of marginal covariance functions is not consistent with the data. With regards to Gaussian process-based models, for each correlation kernel the introduction of nonseparability and nonstationarity improves the results in terms of Nelson-Aalen estimates. For example, when  \Matern ($\nu=1.5$) is used, the separable, stationary model overestimates the cumulative hazard function, while NSGP model based on the same correlation kernel seems to provide a much better fit.

Figure~\ref{Fig:windmaps} displays the elements of the varying anisotropy matrix estimated by NSGP model with \Matern ($\nu=1.5$) correlation function $g$. The estimates of both spatially varying decay parameters $\ve A_{11}(s_1,s_2)$ and $\ve A_{22}(s_1,s_2)$ are in general lower in oceanic regions, suggesting that the spatial process is smoother in these regions in both north-south and west-east directions. The estimate of the degree of nonseparability $\rho_{12}(s_1,s_2)$ is between $-0.765$ and $0.968$, indicating strong interaction between longitude and latitude in the covariance structure in some regions. Large positive values indicate that the spatial process changes more quickly towards north-eastward and south-westward directions. In the Southern Cone -- the southernmost areas of South America -- that is expected because of a typical wind (locally known as \textit{Minuano}) that brings low temperature towards north-east. Similarly, large negative values (in the center of the figure) can be associated to the well-known winds coming from the west coast of Africa to the Caribbean Sea.

\section{Discussion}\label{sec:Discussion}

Whereas nonparametric models for the covariance function are flexible but difficult to estimate in multidimensional domains due to the curse of dimensionality problem, parametric models can easily be estimated but their flexibility is limited by the choice of parametric covariance function, which is usually either stationary or separable. We proposed to use a flexible, convolution-based approach which allows for nonstationarity and nonseparability, crucial properties to achieve good fit of the covariance function, extract the most important modes of variation in the data and obtain better estimates of uncertainty in predictions. This approach is readily applied to multidimensional domains.

The unconstrained estimation of the parameters in the varying anisotropy matrix enables us to model them as a function of time or spatial location easily. They can be further modeled as a function of time (or spatially) dependent covariates or other covariates. In any of these cases, the function can be represented by a variety of basis functions, among which we have found B-splines basis very suitable for ensuring smoothness and being flexible. In particular, our proposed spherical parametrization for $\ve A_{Q \times Q}(\ve t)$, which allows us to easily deal with input dimensions higher than two, is specified by a decomposition whose elements have statistical interpretation. 

When estimating $\tau$-varying parameters using B-spline basis in \eqref{bsplines}, using all the data (rather than the local likelihood estimation (LLE)) may require a potentially high computational cost. However, when computational costs are not prohibitive, this approach should be preferable to the LLE approach, whose performance heavily depends on the neighborhood size $r$. In the LLE, if a small neighborhood is used (e.g., in order to model very local features), one might obtain unstable local estimates. On the other hand, if a large neighborhood is used (something necessary when data are sparse), then the local stationarity assumption may be no longer appropriate and local estimates might be very biased.

Instead of using empirical Bayes estimates, we could have defined a hyperprior distribution for $\ve \theta$. In this case, our knowledge (posterior distribution) about $\ve \theta$ is updated as more data are observed. Finding the mode of the posterior density is a way to find what we call the \textit{maximum a posteriori} (MAP) estimate of $\ve \theta$. When we use a non-informative or a uniform prior distribution, the MAP estimates are precisely the same as the empirical Bayes estimates \citep{shi2011gaussian}.

The decomposition of GPs may be important for developing efficient approximation for big data, non-Gaussian data \citep{wang2014generalized} and heavy-tailed data \citep{shah14student,wang2017extended,cao2018robust}. For non-Gaussian and heavy-tailed data, the decomposition might be used instead of their predictive distributions which are usually complicated. It can also be important for further analysis of scalar-on-functions or function-on-functions regression models, where we try to reduce the dimension of data by using a small number of components. Our proposed approach can be especially important when these components (eigensurfaces) are nonseparable, as we have seen in the simulation studies.

Convolution-based GP methods can also be used to deal with multivariate GP outputs. For example, dependence between spatial processes is discussed in \cite{ver1998constructing} and used by \cite{boyle2004dependent} to work with multiple outputs. Considering nonstationarity in the multivariate setting is a topic for future research.

\bigskip
\begin{center}
	{\large\bf SUPPLEMENTARY MATERIAL}
\end{center}

\begin{description}
	\item[Supplementary Material:] It contains proofs, technical derivations, additional numerical results, and an additional application to relative humidity data% (supplem.pdf)
%	\item[R code:] R code (with C++ functions within it) used to generate the results reported in the article (zipped file)
\end{description}

\if1\blind
{
\section*{Acknowledgments}
The authors thank Prof. Robin Henderson of Newcastle University, UK, for the wind intensity data analyzed in Section~\ref{sec:Application}.
} \fi

\if0\blind
{

} \fi

\addtolength{\textheight}{0.7in}%
\bibliographystyle{agsm}
\bibliography{references}

@article{Hall08,
	author={Hall, P. and M\"uller, H.-G. and and Yao, F.},
	year={2008},
	title={Modelling Sparse Generalized Longitudinal
	Observations with Latent Gaussian Processes},
	journal={Journal of the Royal Statistical Society. Series B
	(Methodological)},
	volume={70},
	number={}, 
	page={703-723}}

@book{shi2011gaussian,
  title={Gaussian process regression analysis for functional data},
  author={Shi, Jian Qing and Choi, Taeryon},
  year={2011},
  publisher={{CRC Press}}
}

@book{Basawa1980,
author={Basawa, I. V. and Prakasa Rao, B. L. S.},
year={1980}, 
title={Statistical Inference for Stochastic Processes},
publisher={Academic Press}
}

@book{Bosq2000,
author={Bosq, D.},
year={2000},
title={Linear Processes in Function Spaces: Theory and Applications},
publisher={Lectures notes in statistics, Springer Verlag}}

@article{wang2014generalized,
  title={{Generalized Gaussian process regression model for non-Gaussian functional data}},
  author={Wang, Bo and Shi, Jian Qing},
  journal={Journal of the American Statistical Association},
  volume={109},
  number={507},
  pages={1123--1133},
  year={2014},
  publisher={Taylor \& Francis}
}

@book{ramsay2005functional,
  title={Functional Data Analysis},
   author =  {Ramsay, J. and Silverman, B. W.},
  year={2005},
  publisher={Springer},
  edition = {2}
}

@inproceedings{boyle2004dependent,
  title={{Dependent Gaussian processes}},
  author={Boyle, Phillip and Frean, Marcus},
  booktitle={Advances in neural information processing systems},
  pages={217--224},
  year={2004}
}

@book{rasmussen2006gaussian,
  title={{Gaussian Processes for Machine Learning}},
  author={Rasmussen, C.E. and Williams, C.K.I.},
  year={2006},
  publisher={University Press Group Limited}
}

@article{chen2017modelling,
  title={Modelling function-valued stochastic processes, with applications to fertility dynamics},
  author={Chen, Kehui and Delicado, Pedro and M{\"u}ller, Hans-Georg},
  journal={Journal of the Royal Statistical Society: Series B (Statistical Methodology)},
  volume={79},
  number={1},
  pages={177--196},
  year={2017},
  publisher={Wiley Online Library}
}

@article{yao2005functional,
  title={Functional data analysis for sparse longitudinal data},
  author={Yao, Fang and M{\"u}ller, Hans-Georg and Wang, Jane-Ling},
  journal={Journal of the American Statistical Association},
  volume={100},
  number={470},
  pages={577--590},
  year={2005},
  publisher={Taylor \& Francis}
}

@book{wahba1990spline,
  title={Spline models for observational data},
  author={Wahba, Grace},
  volume={59},
  year={1990},
  publisher={Siam}
}

@article{ba2012composite,
author = "Ba, Shan and Joseph, V. Roshan",
doi = "10.1214/12-AOAS570",
fjournal = "The Annals of Applied Statistics",
journal = "Ann. Appl. Stat.",
month = "12",
number = "4",
pages = "1838--1860",
publisher = "The Institute of Mathematical Statistics",
title = {{Composite Gaussian process models for emulating expensive functions}},
volume = "6",
year = "2012"
}

@book{banerjee2015hierarchical,
  title={Hierarchical modeling and analysis for spatial data},
  author={Banerjee, Sudipto and Carlin, Bradley P and Gelfand, Alan E},
  year={2015},
  publisher={{CRC Press}},
  edition = {2nd}
}

@article{aston2017tests,
  title={Tests for separability in nonparametric covariance operators of random surfaces},
  author={Aston, John AD and Pigoli, Davide and Tavakoli, Shahin and others},
  journal={The Annals of Statistics},
  volume={45},
  number={4},
  pages={1431--1461},
  year={2017},
  publisher={Institute of Mathematical Statistics}
}

@article{rougier2017representation,
  title={A representation theorem for stochastic processes with separable covariance functions, and its implications for emulation},
  author={Rougier, Jonathan},
  journal={Preprint, ArXiv:1702.05599},
  year={2017}
}

@article{constantinou2017testing,
  title={Testing separability of space-time functional processes},
  author={Constantinou, Panayiotis and Kokoszka, Piotr and Reimherr, Matthew},
  journal={Biometrika},
  volume={104},
  number={2},
  pages={425--437},
  year={2017},
  publisher={Oxford University Press}
}

@article{cappello2018testing,
  title={Testing the type of non-separability and some classes of space-time covariance function models},
  author={Cappello, C and De Iaco, S and Posa, D},
  journal={Stochastic Environmental Research and Risk Assessment},
  volume={32},
  number={1},
  pages={17--35},
  year={2018},
  publisher={Springer}
}

@article{gneiting2002nonseparable,
  title={Nonseparable, stationary covariance functions for space--time data},
  author={Gneiting, Tilmann},
  journal={Journal of the American Statistical Association},
  volume={97},
  number={458},
  pages={590--600},
  year={2002},
  publisher={Taylor \& Francis}
}

@article{cressie1999classes,
  title={Classes of nonseparable, spatio-temporal stationary covariance functions},
  author={Cressie, Noel and Huang, Hsin-Cheng},
  journal={Journal of the American Statistical Association},
  volume={94},
  number={448},
  pages={1330--1339},
  year={1999},
  publisher={Taylor \& Francis Group}
}

@article{iaco2002nonseparable,
  title={Nonseparable space-time covariance models: some parametric families},
  author={De Iaco, S and Myers, Donald E and Posa, D},
  journal={Mathematical Geology},
  volume={34},
  number={1},
  pages={23--42},
  year={2002},
  publisher={Springer}
}

@article{allen2014generalized,
  title={A generalized least-square matrix decomposition},
  author={Allen, Genevera I and Grosenick, Logan and Taylor, Jonathan},
  journal={Journal of the American Statistical Association},
  volume={109},
  number={505},
  pages={145--159},
  year={2014},
  publisher={Taylor \& Francis}
}

@article{chen2012modeling,
  title={Modeling repeated functional observations},
  author={Chen, Kehui and M{\"u}ller, Hans-Georg},
  journal={Journal of the American Statistical Association},
  volume={107},
  number={500},
  pages={1599--1609},
  year={2012},
  publisher={Taylor \& Francis}
}

@InProceedings{shah14student,
  title = 	 {{Student-t Processes as Alternatives to Gaussian Processes}},
  author = 	 {Amar Shah and Andrew Wilson and Zoubin Ghahramani},
  pages = 	 {877--885},
  year = 	 {2014},
  volume = 	 {33},
  series = 	 {Proceedings of Machine Learning Research},
  address = 	 {Reykjavik, Iceland},
  month = 	 {22--25 Apr},
  publisher = 	 {PMLR}
}

@article{cao2018robust,
author = {Chunzheng Cao and Jian Qing Shi and Youngjo Lee},
title ={Robust functional regression model for marginal mean and subject-specific inferences},
journal = {Statistical Methods in Medical Research},
volume = {27},
number = {11},
pages = {3236-3254},
year = {2018}
}

@article{wang2017extended,
title = {{Extended $t$-process regression models}},
journal = {Journal of Statistical Planning and Inference},
volume = {189},
pages = {38 - 60},
year = {2017},
issn = {0378-3758},
doi = {https://doi.org/10.1016/j.jspi.2017.05.006},
author = {Zhanfeng Wang and Jian Qing Shi and Youngjo Lee},
}

@book{horvath2012inference,
  title={Inference for functional data with applications},
  author={Horv{\'a}th, Lajos and Kokoszka, Piotr},
  volume={200},
  year={2012},
  publisher={Springer Science \& Business Media}
}

@book{carlin2008bayesian,
  title={Bayesian methods for data analysis},
  author={Carlin, Bradley P and Louis, Thomas A},
  year={2008},
  publisher={CRC Press}
}

@article{paciorek2006,
author = {Paciorek, Christopher J. and Schervish, Mark J.},
title = {Spatial modelling using a new class of nonstationary covariance functions},
journal = {Environmetrics},
year = {2006},
volume = {17},
number = {5},
pages = {483-506},
doi = {10.1002/env.785}
}

@article{risser2017locallik,
   author = {Mark Risser and Catherine Calder},
   title = {{Local Likelihood Estimation for Covariance Functions with Spatially-Varying Parameters: The convoSPAT Package for R}},
   journal = {Journal of Statistical Software, Articles},
   volume = {81},
   number = {14},
   year = {2017},
   issn = {1548-7660},
   pages = {1--32},
   doi = {10.18637/jss.v081.i14}
}

@article{ver1998constructing,
  title={Constructing and fitting models for cokriging and multivariable spatial prediction},
  author={Ver Hoef, Jay M and Barry, Ronald Paul},
  journal={Journal of Statistical Planning and Inference},
  volume={69},
  number={2},
  pages={275--294},
  year={1998},
  publisher={Elsevier}
}

@article{higdon1998atlantic,
author="Higdon, David",
title={{A process-convolution approach to modelling temperatures in the North Atlantic Ocean}},
journal="Environmental and Ecological Statistics",
year="1998",
month="Jun",
day="01",
volume="5",
number="2",
pages="173--190",
issn="1573-3009"
}

@article{higdon1999non,
  title={Non-stationary spatial modeling},
  author={Higdon, Dave and Swall, J and Kern, J},
  journal={Bayesian statistics},
  volume={6},
  number={1},
  pages={761--768},
  year={1999}
}

@incollection{higdon2002space,
  title={Space and space-time modeling using process convolutions},
  author={Higdon, Dave},
  booktitle={Quantitative methods for current environmental issues},
  pages={37--56},
  year={2002},
  publisher={Springer}
}

@article{stein2005space,
  title={Space--time covariance functions},
  author={Stein, Michael L},
  journal={Journal of the American Statistical Association},
  volume={100},
  number={469},
  pages={310--321},
  year={2005},
  publisher={Taylor \& Francis}
}

@article{pinheiro1996unconstrained,
  title={Unconstrained parametrizations for variance-covariance matrices},
  author={Pinheiro, Jos{\'e} C and Bates, Douglas M},
  journal={Statistics and computing},
  volume={6},
  number={3},
  pages={289--296},
  year={1996},
  publisher={Springer}
}

@article{rapisarda2007parameterizing,
  title={Parameterizing correlations: a geometric interpretation},
  author={Rapisarda, Francesco and Brigo, Damiano and Mercurio, Fabio},
  journal={IMA Journal of Management Mathematics},
  volume={18},
  number={1},
  pages={55--73},
  year={2007},
  publisher={Oxford University Press}
}

@article{zhang2015joint,
issn = {1369-7412},
journal = {Journal of the Royal Statistical Society: Series B (Statistical Methodology)},
pages = {219--238},
volume = {77},
number = {1},
year = {2015},
title = {A joint modelling approach for longitudinal studies},
author = {Zhang, Weiping and Leng, Chenlei and Tang, Cheng Yong},
}

@article{pourahmadi1999joint,
  title={Joint mean-covariance models with applications to longitudinal data: Unconstrained parameterisation},
  author={Pourahmadi, Mohsen},
  journal={Biometrika},
  volume={86},
  number={3},
  pages={677--690},
  year={1999},
  publisher={Oxford University Press}
}

@article{leng2010semiparametric,
  title={Semiparametric mean--covariance regression analysis for longitudinal data},
  author={Leng, Chenlei and Zhang, Weiping and Pan, Jianxin},
  journal={Journal of the American Statistical Association},
  volume={105},
  number={489},
  pages={181--193},
  year={2010},
  publisher={Taylor \& Francis}
}

@article{ohagan1978curvefitting,
 ISSN = {00359246},
 author = {A. O'Hagan},
 journal = {Journal of the Royal Statistical Society. Series B (Methodological)},
 number = {1},
 pages = {1--42},
 publisher = {[Royal Statistical Society, Wiley]},
 title = {Curve Fitting and Optimal Design for Prediction},
 volume = {40},
 year = {1978}
}

@article{sampson1992deformation,
 author = {Paul D. Sampson and Peter Guttorp},
 journal = {Journal of the American Statistical Association},
 number = {417},
 pages = {108--119},
 publisher = {[American Statistical Association, Taylor & Francis, Ltd.]},
 title = {Nonparametric Estimation of Nonstationary Spatial Covariance Structure},
 volume = {87},
 year = {1992}
}

@inproceedings{vanderwilk2017convolutional,
  title = {{Convolutional Gaussian Processes}},
  author={Van der Wilk, Mark and Rasmussen, Carl Edward and Hensman, James},
  booktitle={Advances in Neural Information Processing Systems 30},
  pages={2849--2858},
  year={2017}
}

@article{dunlop2018deep,
  title={{How Deep Are Deep Gaussian Processes?}},
  author={Dunlop, Matthew M and Girolami, Mark A and Stuart, Andrew M and Teckentrup, Aretha L},
  journal={Journal of Machine Learning Research},
  volume={19},
  number={54},
  pages={1--46},
  year={2018},
  publisher={Journal of Machine Learning Research}
}

@article{kanagawa2018gaussian,
  title={Gaussian Processes and Kernel Methods: A Review on Connections and Equivalences},
  author={Kanagawa, Motonobu and Hennig, Philipp and Sejdinovic, Dino and Sriperumbudur, Bharath K},
  journal={Preprint, ArXiv:1807.02582},
  year={2018}
}

@inproceedings{jidling2017constrainedGP,
  title={Linearly constrained Gaussian processes},
  author={Jidling, Carl and Wahlstr{\"o}m, Niklas and Wills, Adrian and Sch{\"o}n, Thomas B},
  booktitle={Advances in Neural Information Processing Systems},
  pages={1215--1224},
  year={2017}
}

@InProceedings{deisenroth15distribGP,
  title = 	 {{Distributed Gaussian Processes}},
  author = 	 {Marc Deisenroth and Jun Wei Ng},
  booktitle = 	 {Proceedings of the 32nd International Conference on Machine Learning},
  pages = 	 {1481--1490},
  year = 	 {2015},
  editor = 	 {Francis Bach and David Blei},
  volume = 	 {37},
  series = 	 {Proceedings of Machine Learning Research},
  address = 	 {Lille, France},
  month = 	 {07--09 Jul},
  publisher = 	 {PMLR}
}

@article{tran2015variational,
  title={{The variational Gaussian process}},
  author={Tran, Dustin and Ranganath, Rajesh and Blei, David M},
  journal={Preprint, ArXiv:1511.06499},
  year={2015}
}

@inproceedings{calandra2016manifold,
  title={{Manifold Gaussian processes for regression}},
  author={Calandra, Roberto and Peters, Jan and Rasmussen, Carl Edward and Deisenroth, Marc Peter},
  booktitle={Neural Networks (IJCNN), 2016 International Joint Conference on},
  pages={3338--3345},
  year={2016},
  organization={IEEE}
}

@article{liu2018GPforBigData,
  title={{Understanding and Comparing Scalable Gaussian Process Regression for Big Data}},
  author={Liu, Haitao and Cai, Jianfei and Ong, Yew-Soon and Wang, Yi},
  journal={Preprint, ArXiv:1811.01159},
  year={2018}
}

@article{sung2017generalized,
  title={{A generalized Gaussian process model for computer experiments with binary time series}},
  author={Sung, Chih-Li and Hung, Ying and Rittase, William and Zhu, Cheng and Wu, CF},
  journal={Preprint, ArXiv:1705.02511},
  year={2017}
}

@article{wang2019GPRforMixed,
title = {Gaussian process methods for nonparametric functional regression with mixed predictors},
journal={Computational Statistics \& Data Analysis},
volume = {131},
pages = {80--90},
year = {2019},
doi = {https://doi.org/10.1016/j.csda.2018.07.009},
author = {Wang, Bo and Xu, Aiping}
}

@article{zhang2019GPlargeSpatialdata,
  title={{Smoothed full-scale approximation of Gaussian process models for computation of large spatial datasets}},
  author={Zhang, Bohai and Sang, Huiyan and Huang, Jianhua Z},
  journal={Statistica Sinica},
  volume={29},
  pages={1711--1737},
  year={2019}
}

@article{tibshirani1987locallik,
	author = {Robert Tibshirani and Trevor Hastie},
	journal = {Journal of the American Statistical Association},
	number = {398},
	pages = {559--567},
	publisher = {[American Statistical Association, Taylor & Francis, Ltd.]},
	title = {Local Likelihood Estimation},
	volume = {82},
	year = {1987}
}

@book{Stein1999book,
	title={Interpolation of Spatial Data: Some Theory for Kriging},
	author={Stein, Michael L},
	year={1999},
	publisher={Springer: NY}
}

@book{deBoor2001book,
	title={A Practical Guide to Splines},
	author={de Boor, C},
	year={2001},
	publisher={New York: Springer}
}

@misc{cisl_rda_ds093_0,
 author = "{Saha et al.}",
 title = "{NCEP Climate Forecast System Reanalysis (CFSR) 6-hourly Products, January 1979 to December 2010}",
 publisher  = "Research Data Archive at the National Center for Atmospheric Research, Computational and Information Systems Laboratory",
 address = {Boulder CO},
 year  = "2010",
 url = "https://doi.org/10.5065/D69K487J"
}

@article{jun2007spacetimespheres,
 author = {Mikyoung Jun and Michael L. Stein},
 journal = {Technometrics},
 number = {4},
 pages = {468--479},
 publisher = {[Taylor & Francis, Ltd., American Statistical Association, American Society for Quality]},
 title = {An Approach to Producing Space: Time Covariance Functions on Spheres},
 volume = {49},
 year = {2007}
}

@article{jun2008nonstatglobal,
  title={Nonstationary covariance models for global data},
  author={Jun, Mikyoung and Stein, Michael L},
  journal={The Annals of Applied Statistics},
  volume={2},
  number={4},
  pages={1271--1289},
  year={2008},
  publisher={Institute of Mathematical Statistics}
}

@article{bruno2009nonsepnonstatozone,
  title={A simple non-separable, non-stationary spatiotemporal model for ozone},
  author={Bruno, Francesca and Guttorp, Peter and Sampson, Paul D and Cocchi, Daniela},
  journal={Environmental and ecological statistics},
  doi = {10.1007/s10651-008-0094-8},
  volume={16},
  number={4},
  pages={515--529},
  year={2009},
  publisher={Springer}
}

@TECHREPORT{garside2020topol,
    author = {Garside, K and Henderson, R and Johnson, H and Makarenko, I},
    title = {Topological event history analysis},
    institution = {Newcastle University, UK},
    year = {2020}
}

@article{RccpPackage,
    title = {{Rcpp}: Seamless {R} and {C++} Integration},
    author = {Dirk Eddelbuettel and Romain Fran\c{c}ois},
    journal = {Journal of Statistical Software},
    year = {2011},
    volume = {40},
    number = {8},
    pages = {1--18}
  }

@article{RccpArmadilloPackage,
    title = {{RcppArmadillo: Accelerating R with high-performance C++ linear algebra}},
    author = {Dirk Eddelbuettel and Conrad Sanderson},
    journal = {Computational	Statistics and Data Analysis},
    year = {2014},
    volume = {71},
    month = {March},
    pages = {1054--1063}
  }

@Manual{fdapacePackage,
    title = {fdapace: Functional Data Analysis and Empirical Dynamics},
    author = {Yaqing Chen and Cody Carroll and Xiongtao Dai and Jianing Fan and Pantelis Z. Hadjipantelis and Kyunghee Han and Hao Ji and Hans-Georg Mueller and Jane-Ling Wang},
    year = {2019},
    note = {R package version 0.5.1}
  }

\end{document}

\def\spacingset#1{\renewcommand{\baselinestretch}%
{#1}\small\normalsize} \spacingset{1}

%%%%%%%%%%%%%%%%%%%%%%%%%%%%%%%%%%%%%%%%%%%%%%%%%%%%%%%%%%%%%%%%%%%%%%%%%%%%%%

\if1\blind
{
  \title{Supplementary Material for: \\ \bf Modeling Function-Valued Processes with Nonseparable and/or Nonstationary Covariance Structure}
  \author{Evandro Konzen\\
  	School of Mathematics, Statistics \& Physics, Newcastle University, UK\\
  	Jian Qing Shi\thanks{Corresponding author, email: j.q.shi@ncl.ac.uk}\hspace{.2cm}\\
  	School of Mathematics, Statistics \& Physics, Newcastle University, UK\\
  	and \\
  	Zhanfeng Wang\\
    Department of Statistics and Finance, Management School,\\ 
    University of Science and Technology of China, Hefei, China}
  \maketitle
} \fi

\if0\blind
{
  \bigskip
  \bigskip
  \bigskip
  \begin{center}
    {\LARGE Supplementary Material for: \\ \bf Modeling Function-Valued Processes with Nonseparable and/or Nonstationary Covariance Structure}
\end{center}
  \medskip
} \fi

\bigskip

%\newpage
\spacingset{1.5} % DON'T change the spacing!

\addtolength{\textheight}{0.5in}%
\section{Proofs} \label{sec:ProofsSuppl}

\subsection{Proof of Theorem 1} \label{th1}

We show that 
\begin{align*}
& E \Big[ || X^c(\ve t) - \sum_{j=1}^{J} g_j(\ve t) \xi_j^* ||^2 \Big] \\
& = E \langle  X^c(\cdot) -  \sum_{j=1}^{J} g_j(\cdot) \xi_j^*, X^c(\cdot) -  \sum_{j=1}^{J} g_j(\cdot) \xi_j^*\rangle \\
& = E \Big[ \langle X^c(\cdot), X^c(\cdot) \rangle - \sum_{j=1}^{J} \big( \langle X^c(\cdot), g_j(\cdot) \rangle \big)^2  \Big] \\
& = E ||X^c ||^2 -  \sum_{j=1}^{J} E \Big[  \int_{\mathcal{T}}\int_{\mathcal{T}}  X^c(\ve t) X^c(\ve t')      g_j(\ve t) g_j(\ve t') d \ve t d \ve t'  \Big] \\
& = E ||X^c ||^2 - \sum_{j=1}^{J} \int_{\mathcal{T}}\int_{\mathcal{T}}  k(\ve t,\ve t') g_j(\ve t) g_j(\ve t') d \ve t d \ve t' \\
& = E ||X^c ||^2 - \sum_{j=1}^{J} \langle \Phi(g_j), g_j  \rangle.
\end{align*}
Hence, the minimizing problem \eqref{BestFiniteDimApprEq} becomes to maximize $\sum_{j=1}^{J} \langle \Phi(g_j), g_j  \rangle$ with respect to $g_1,\dots, g_J \in L^2 (\mathcal{T})$. Since the operator $\Phi$ is symmetric, positive definite Hilbert-Schmidt, following Theorem 3.2 in \cite{horvath2012inference} the proof is completed.

\subsection{Proof of Theorem 2} \label{th2} 

Without loss of generality, we consider $Q=2$. Then $\ve t=(s,\tau)^\top$ with $s,\tau \in \R$.
Let $\{\hat\lambda_j,j=1,2, \dots\}$ and $\{\hat\phi_j(\cdot),j=1,2, \dots\}$ be the eigenvalues and eigenfunctions of the covariance function
$\hat k(\ve t,\ve t')=k_{\hat\theta}(\ve t,\ve t')$, where $\ve t=(s,\tau)^\top,~\ve t'=(s',\tau')^\top$, and
$$\hat\xi_j= \langle X^c(\cdot), \hat\phi_j(\cdot)  \rangle =\int X^c(\ve t)\hat\phi_j(\ve t) d \ve t.$$

Let  $p(\ve x_l^c;\ve{\theta})=p(x_{1}^c,\dots,x_l^c;\ve{\theta})$ be the density function of $\ve x_l^c$. Let $\vesub{\theta}{0}$ be the true value of $\ve{\theta}$ and $p_{l}(\ve{\theta})$ be the conditional density of $\ve x_l^c$ for  given  $\ve x_{l-1}^c$. Actually, for every $l\geq 1$,
$$p_{l}(\ve{\theta})=p(\ve x_l^c;\ve{\theta})/p(\ve x_{l-1}^c;\ve{\theta}).$$
It shows that $p_{l}(\ve{\theta})=N(\mu_{l|l-1},\sigma_{l|l-1}^2)$ with
\begin{align}
\mu_{l|l-1}=\vesub{k}{l}\vess{K}{l-1}{-1}\ve x_{l-1}^c,~~\sigma_{l|l-1}^2=k(\ve t_l,\ve t_l)-\vesub{k}{l}\vess{K}{l-1}{-1}\vess{k}{l}{\top},\nonumber
\end{align}
where $\vesub{k}{l}=\big(k(\ve t_1,\ve t_l), \dots,k(\ve t_{l-1},\ve t_l)\big)^\top.$
Assume that $p_{l}(\ve{\theta})$ is twice differentiable with respect to $\ve{\theta}$. Let $\phi_{l}(\ve{\theta})=\log p_{l}(\ve{\theta})$, $\vesub{U}{l}(\ve{\theta})=\dot{\phi}_{l}(\ve{\theta})$ and $\vesub{V}{l}(\ve{\theta})=\ddot{\phi}_{l}(\ve{\theta})$, where $\dot{g}$ and $\ddot{g}$
are the first and second derivatives of function $g(\ve{\theta})$ with respect to $\ve{\theta}$, respectively.
Without loss of generality, we consider the parameter with one dimension. Then $\vesub{U}{l}(\ve{\theta})$ and $\vesub{V}{l}(\ve{\theta})$ are scalars $U_{l}(\theta)$ and $V_{l}(\theta)$, and denoted by $U_{l}=U_{l}(\theta_{0})$ and $V_{l}=V_{l}(\theta_{0})$.
For the proof of Theorem 2, we need  the following conditions: \\
(C1) $\sup_{s,\tau}|\mu(s,\tau)|<\infty$.\\
(C2) The covariance function $k_\theta(\ve t, \ve t')$ has thrice continuous derivative with respect to $\theta$, and is continuous, differentiate and square-integrable on $\ve t, \ve t'$. For eigenvalues and eigenvectors of $k_\theta$, assume $\delta_j>0$ and $\phi_j(s,\tau)$ is square-integrable, where $\delta_j=\min\{\lambda_1-\lambda_2,\lambda_{j-1}-\lambda_j,\lambda_j-\lambda_{j+1}\}$.

Define  $i_{k}(\theta_{0})=\Var[U_{k}|\mathscr{F}_{k-1}]=E[U_{k}^{2}|\mathscr{F}_{k-1}]$, where $\mathscr{F}_{k-1}=\sigma(x_1^c, \dots,x_{l-1}^c) $. Let $I_{n}(\theta_{0})=\sum_{k=1}^{n}i_{k}(\theta_{0})$,  $S_{n}=\sum_{k=1}^{n}U_{k}$ and $S_{n}^{*}=\sum_{k=1}^{n}V_{k}+I_{n}(\theta_{0})$.
It shows that $S_{n}$ and $S_{n}^*$ are zero-mean martingales with respect to $\sigma$-filtration $\mathscr{F}_{n}$.
The third condition is\\
(C3) Assume
\begin{enumerate}
	\item $n^{-1}|\sum_{k=1}^n V_k|\stackrel{p}{\rightarrow} i(\theta_0)$, and $n^{-1/2}S_n\stackrel{L}{\rightarrow} N(0, i({\theta}_{0}))$ for some non-random function $i({\theta}_{0})>0$, 
	\item For all $\varepsilon>0$ and $\eta>0$, there exists $\delta>0$ and $n_0>0$ such that for all $n>n_0$, $P\{n^{-1}
	|\sum_{k=1}^n (V(\theta)-V_k) |>\eta, |\theta-\theta_0|<\delta \}<\varepsilon$, 
	\item $n^{-1}\sum_{k=1}^{n}E|W_{k}(\theta)|<M<\infty$ for all $\theta$ and $n$, where $W_{k}(\theta)$ is the third derivative of $\phi_{k}(\theta)$ with respect to $\theta$.
\end{enumerate}

Under conditions (C2) and (C3), one can easily show that the conditions of Theorem 2.2 in Chapter 7 of \cite{Basawa1980} %Basawa and Rao(1980)
holds. Hence,
$\hat\theta$ is a consistent estimator of $\theta_0$ and has asymptotically normality,
$$n^{-1/2}(\hat\theta-\theta_0)\stackrel{L}{\rightarrow} N(0, i({\theta}_{0})^{-1}),$$
which indicates that
$$||\hat\theta-\theta_0||=O_p(\{\log(n)/n\}^{1/2}).$$
Since the covariance function $k_\theta$ is thrice continuously differentiate on $\theta$, we have
$$||k_{\hat\theta}(\cdot,\cdot)-k_\theta(\cdot,\cdot)||=O_p(\{\log(n)/n\}^{1/2}).$$

From Lemma 4.2 in \cite{Bosq2000}, it follows that for all $j$,
\begin{align}\label{value-ieq}
||\hat \lambda_j-\lambda_j||\leq ||k_{\hat\theta}(\cdot,\cdot)-k_\theta(\cdot,\cdot)||,
\end{align}
and similar to Lemma 4.3 in \cite{Bosq2000}, we have, for fixed $j$,
\begin{align}\label{eigenfun-ieq}
||\hat\phi_j(\cdot)-\phi_j(\cdot)||\leq 2\sqrt{2}\delta_j^{-1} ||k_{\hat\theta}(\cdot,\cdot)-k_\theta(\cdot,\cdot)||,
\end{align}
where $\delta_j=\min\{\lambda_1-\lambda_2,\lambda_{j-1}-\lambda_j,\lambda_j-\lambda_{j+1}\}$.
Then \eqref{value-ieq} and \eqref{eigenfun-ieq} give that
\begin{align}
&||\hat \lambda_j-\lambda_j||=O_p(\{\log(n)/n\}^{1/2}),\nonumber\\
&||\hat\phi_j(\cdot)-\phi_j(\cdot)||=O_p(\{\log(n)/n\}^{1/2}).\nonumber
\end{align}

For $\xi_j$, we show that
\begin{align}
&|\hat{\xi}_j-\xi|=\left|\int(Z(s,\tau)-\hat{\mu}(s,\tau))\hat{\phi}_j(s,\tau)ds d\tau-\int(Z(s,\tau)-\mu(s,\tau))\phi_j(s,\tau)ds d\tau\right|\nonumber\\
\leq&\left|\int(Z(s,\tau)-\mu(s,\tau))(\hat{\phi}_j(s,\tau)-\phi_j(s,\tau))dsd\tau\right|\nonumber\\
&+\left|\int(\hat{\mu}(s,\tau)-\mu(s,\tau))(\hat{\phi}_j(s,\tau)-\phi_j(s,\tau))dsd\tau\right|\nonumber\\
&+\left|\int(\hat{\mu}(s,\tau)-\mu(s,\tau))\phi_j(s,\tau)dsd\tau\right|.\nonumber
\end{align}
Hence, from condition (C2), (\ref{value-ieq}) and
$\sup_{s,\tau}|\hat \mu(s,\tau)-\mu(s,\tau)|=O_p[\{\log(n)/n\}^{1/2}]$, it shows that
\begin{align}
||\hat\xi_j-\xi_j||=O_p(\{\log(n)/n\}^{1/2}).\nonumber
\end{align}

\subsection{Proof of Theorem 3}  \label{th3}

Let $\vesub{K}{n}= \big(k(\ve t_i,\ve t_j) \big)_{n\times n}$ be a Gram matrix, and
$\lambda_1^{(n)}\geq\lambda_2^{(n)}\geq\cdots\geq \lambda_n^{(n)}\geq 0$ be the eigenvalues of $\vesub{K}{n}$,
and $\vesub{V}{j,n},~j=1, \dots,n$ be the eigenvectors of $\vesub{K}{n}$. Then from the \Nystrom approximation method, we show that
\begin{align}
&\sqrt{n}V_{hj,n}=\phi_{j}(\ve t_h)+O_p \Big(\frac{1}{\sqrt{n}} \Big), \qquad \frac{\lambda_j^{(n)}}{n}=\lambda_{j}+O_p \Big(\frac{1}{\sqrt{n}} \Big), \nonumber \\
&\frac{\sqrt{n}}{\lambda_j^{(n)}}\vess{k}{n}{\top}(\ve t)\vesub{V}{j,n} =\phi_{j}(\ve t) +O_p \Big(\frac{1}{\sqrt{n}} \Big), \label{phiapp}
\end{align}
where $V_{hj,n}$ is the $h$-th element of $\vesub{V}{j,n}$, and $\vesub{k}{n}(\ve t)=\big(k(\ve t_1,\ve t), \dots,k(\ve t_n,\ve t) \big)^\top$.
Due to $\E \left[\xi_{j}\right]=0$ and $\Var \left[\xi_{j}\right]=\lambda_{j}\rightarrow 0$ as $j\rightarrow \infty$, it follows from \eqref{phiapp} that
\begin{align}
f_{n}(\ve t)=&\sum_{j=1}^n \phi_{j}(\ve t) \xi_{j}=\sum_{j=1}^n\frac{\sqrt{n}}{\lambda_j^{(n)}}\vess{k}{n}{\top}(\ve t)\vesub{V}{j,n}\xi_{j}+o_p(1).\label{nytrom}
\end{align}

In addition, we show that
\begin{align}\label{xi}
\xi_{j}=&\lambda_{j}  \langle f,\phi_{j}  \rangle \nonumber\\
=& \lambda_{j} \langle f,\frac{\sqrt{n}}{\lambda_j^{(n)}}\vess{k}{n}{\top}(\cdot)\vesub{V}{j,n} \rangle+O_p \Big(\frac{1}{\sqrt{n}}\Big)\nonumber\\
=&\frac{\sqrt{n}\lambda_{j}}{\lambda_j^{(n)}}\vess{V}{j,n}{\top}  \langle f,\vesub{k}{n}(\cdot)  \rangle +O_p \Big(\frac{1}{\sqrt{n}}\Big)\nonumber\\
=&\frac{\sqrt{n}\lambda_{j}}{\lambda_j^{(n)}}\vess{V}{j,n}{\top}\ve f +O_p \Big(\frac{1}{\sqrt{n}}\Big).
\end{align}
Hence, we have
\begin{align}\label{nytromappro}
f_{n}(\ve t)
=& \sum_{j=1}^n\frac{\sqrt{n}}{\lambda_j^{(n)}}\vess{k}{n}{\top}(\ve t)\vesub{V}{j,n}\frac{\sqrt{n}\lambda_{j}}{\lambda_j^{(n)}}\vess{V}{j,n}{\top}\ve f+o_p(1)\nonumber\\
=&\vess{k}{n}{\top}(\ve t)\sum_{j=1}^n\frac{1}{\lambda_j^{(n)}}\vesub{V}{j,n}\vess{V}{j,n}{\top}\ve f +o_p(1)\nonumber\\
=&\vess{k}{n}{\top}(\ve t)\vess{K}{n}{-1}\ve f+o_p(1),
\end{align}
which indicates that $E \left[f(\ve t) | \ve f \right] = \vess{k}{n}{\top}(\ve t)\vesub{K}{n}^{-1}\ve f = f_{n}(\ve t)+o_p(1)$.

The \Nystrom approximation is also applied to $X(\ve t)$ and $\varepsilon(\ve t)$, respectively, and we have
\begin{align}
X_{n}(\ve t) & = \tvess{k}{n}{\top}(\ve t)\tvess{K}{n}{-1}\ve x + o_p(1),\label{yappro}\\
%&f_{n}(\ve t)=\vess{k}{n}{\top}(\ve t)\vess{K}{n}{-1}\vesub{f}{n},\label{fappro}\\
\varepsilon_{n}(\ve t) & = \vess{k}{\varepsilon n}{\top}(\ve t)\vess{K}{\varepsilon n}{-1}\vesub{\varepsilon}{}+o_p(1), \label{fappro}
\end{align}
where $\ve x=\big(x(\ve t_1), \dots, x(\ve t_n) \big)^\top$, $\vesub{\varepsilon}{}=\big(\varepsilon(\ve t_1), \dots,\varepsilon(\ve t_n)\big)^\top$,
$\tilde{k}=k+k_\varepsilon$,\\
$\big[\tilde{\ve K}_n\big]_{ij} = \tilde{k}(\ve t_i,\ve t_j)$,
% $\tvesub{K}{n}=\big(\tilde{k}(\ve t_i,\ve t_j)\big)_{n\times n}$, 
$\tvesub{k}{n}(\ve t)= \big(\tilde{k}(\ve t_1,\ve t), \dots,\tilde{k}(\ve t_n,\ve t) \big)^\top$,  
$\big[\ve K_{\varepsilon n}\big]_{ij} = k_\varepsilon(\ve t_i,\ve t_j)$,
%$\vesub{K}{\varepsilon n}= \big(k_\varepsilon(\ve t_i,\ve t_j) \big)_{n\times n}$,
and \\ $\vesub{k}{\varepsilon n}(\ve t)=\big(k_\varepsilon(\ve t_1,\ve t), \dots,k_\varepsilon(\ve t_n,\ve t) \big)^\top$.
From the definition of $k_\varepsilon$, we know that $\vesub{k}{\varepsilon n}(\ve t)=\ve 0$ and $\tvesub{K}{n}=\vesub{K}{n}+\sigma_\varepsilon^2 \vesub{I}{n}$.
Hence, it follows that
\[
f_{n}(\ve t)=X_{n}(\ve t)-\varepsilon_{n}(\ve t)=\vess{k}{n}{\top}(\ve t)(\vesub{K}{n}+\sigma_\varepsilon^2 \vesub{I}{n})^{-1}\ve x+o_p(1),
\]
which suggests that
\[
\E \left[  f(\ve t)|{\cal D} \right] =\vess{k}{n}{\top}(\ve t)(\vesub{K}{n}+\sigma_\varepsilon^2 \vesub{I}{n})^{-1}\ve x=X_{n}(\ve t)+o_p(1).
\]

\newpage

\setlength{\abovedisplayskip}{6pt}
\setlength{\abovedisplayshortskip}{3pt}
\setlength{\belowdisplayskip}{6pt}
\setlength{\belowdisplayshortskip}{3pt}

\section{Supporting Information for Subsection~\ref{subsubsec:VAMinterp}} \label{sec:SupplVAM}

An expression similar to equation \eqref{taumax2} can be found for the case $Q=3$. If ${\ve t = (s_1, s_2, \tau)^\top \in \R^3}$,
\begin{align}\label{taumax3}
\tau^*(s_1, s_2, s_1', s_2', \tau') &= \underset{\tau}{\mathrm{argmin}} \ 
\ve A_{11} (s_1-s_1')^2 + \ve A_{22} (s_2-s_2')^2 + \ve A_{33} (\tau-\tau')^2 + \\
& \qquad \qquad \qquad 2 \ve A_{12} (s_1-s_1')(s_2-s_2') + \\
& \qquad \qquad \qquad 2 \ve A_{13} (s_1-s_1')(\tau-\tau') + \\
& \qquad \qquad \qquad 2 \ve A_{23} (s_2-s_2')(\tau-\tau')\\
&= \tau' + \frac{\ve A_{13}(s_1'-s_1) + \ve A_{23}(s_2'-s_2)}{\ve A_{33}}.
\end{align}
For cases $Q>3$, the further extension is straightforward.

\section{Supporting Information for Subsection~\ref{secSim_2D_cheb}} \label{sec:SupplsecSim_2D_cheb}

\begin{figure}[H]
	\centering 
	\includegraphics[width=0.85\linewidth]{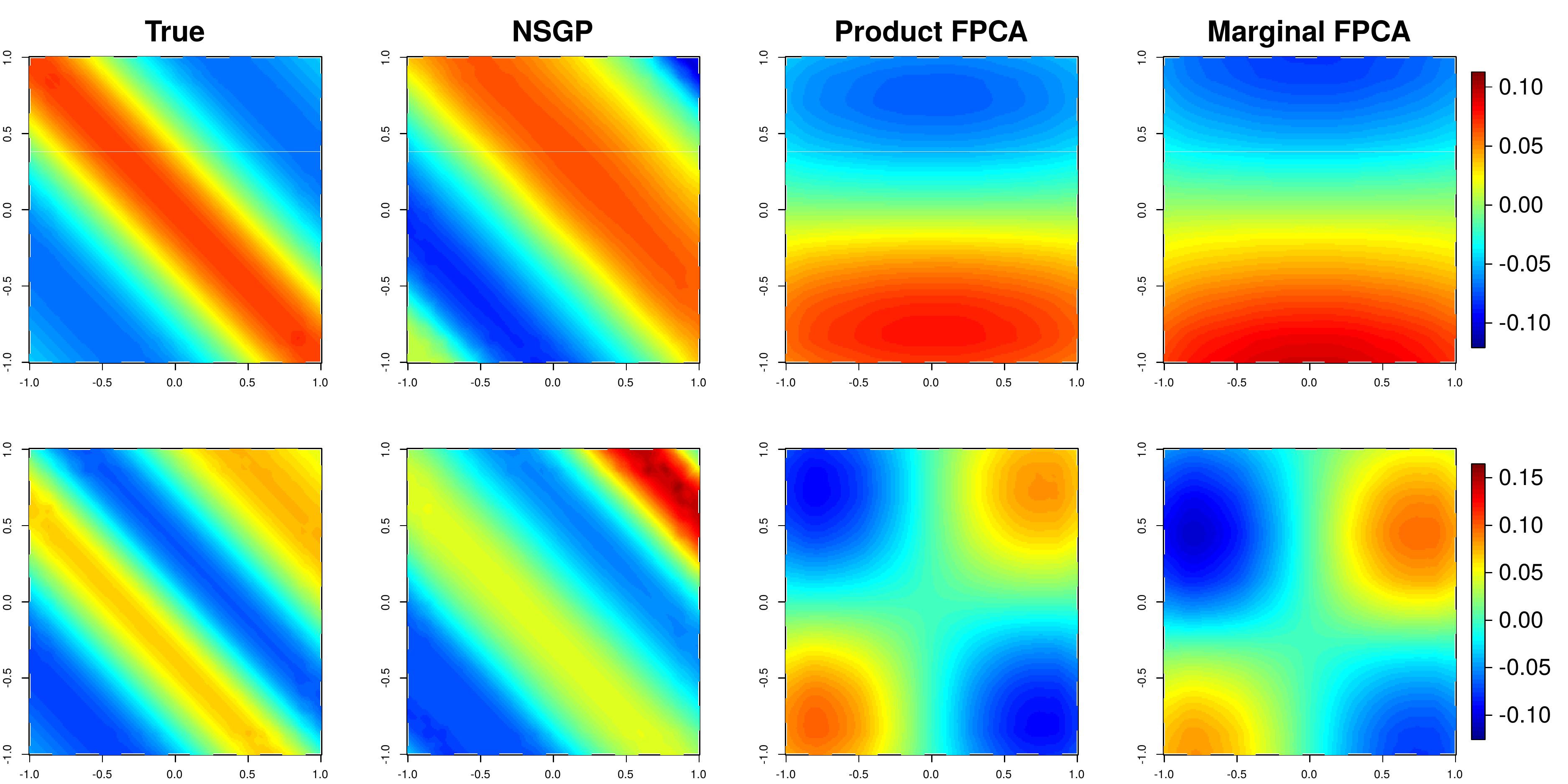} 
	\caption{Third and four leading eigensurfaces $\phi_j(s_1, s_2)$ of the true (based on Chebyshev polynomials) model (first column) and the corresponding eigensurfaces $\hat{\phi}_j(s_1, s_2)$ estimated by NSGP model (with \Matern $\nu=1.5$) (second column), Product FPCA (third) and Marginal FPCA (fourth).}
	\label{eigensurf_cheb_3_4}
\end{figure}

\section{Supporting Information for Subsection~\ref{secSim_converg}} \label{sec:SupplSimConv}

\begin{figure}[H]
	\centering 
	\includegraphics[width=0.6\linewidth]{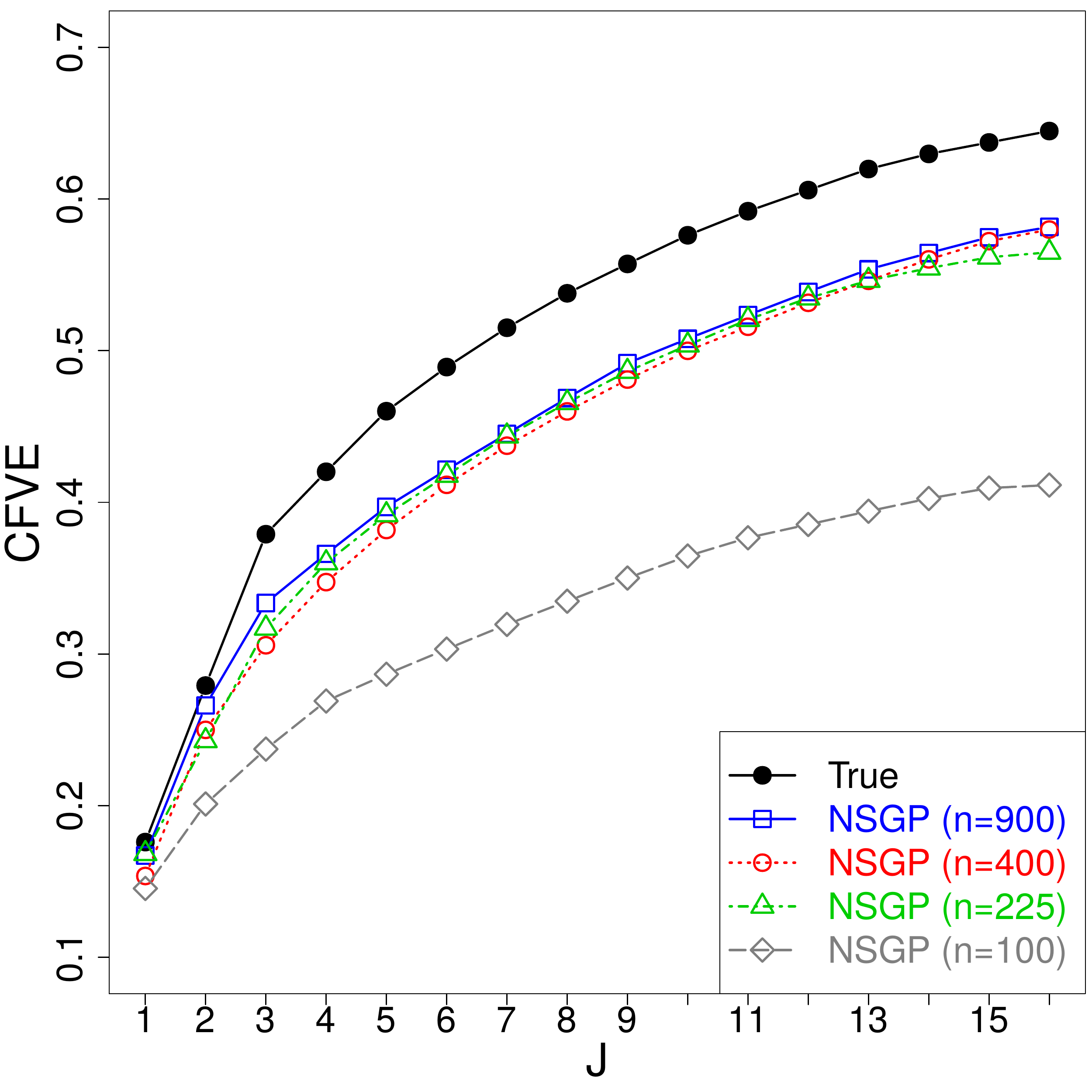} 
	\caption{Cumulative FVEs obtained by using eigensurfaces of the true covariance function and of the covariance functions estimated by NSGP model using data with different sample sizes.}
	\label{CFVE_simConv}
\end{figure}

\begin{figure}[H]
	\centering 
	\includegraphics[width=1\linewidth]{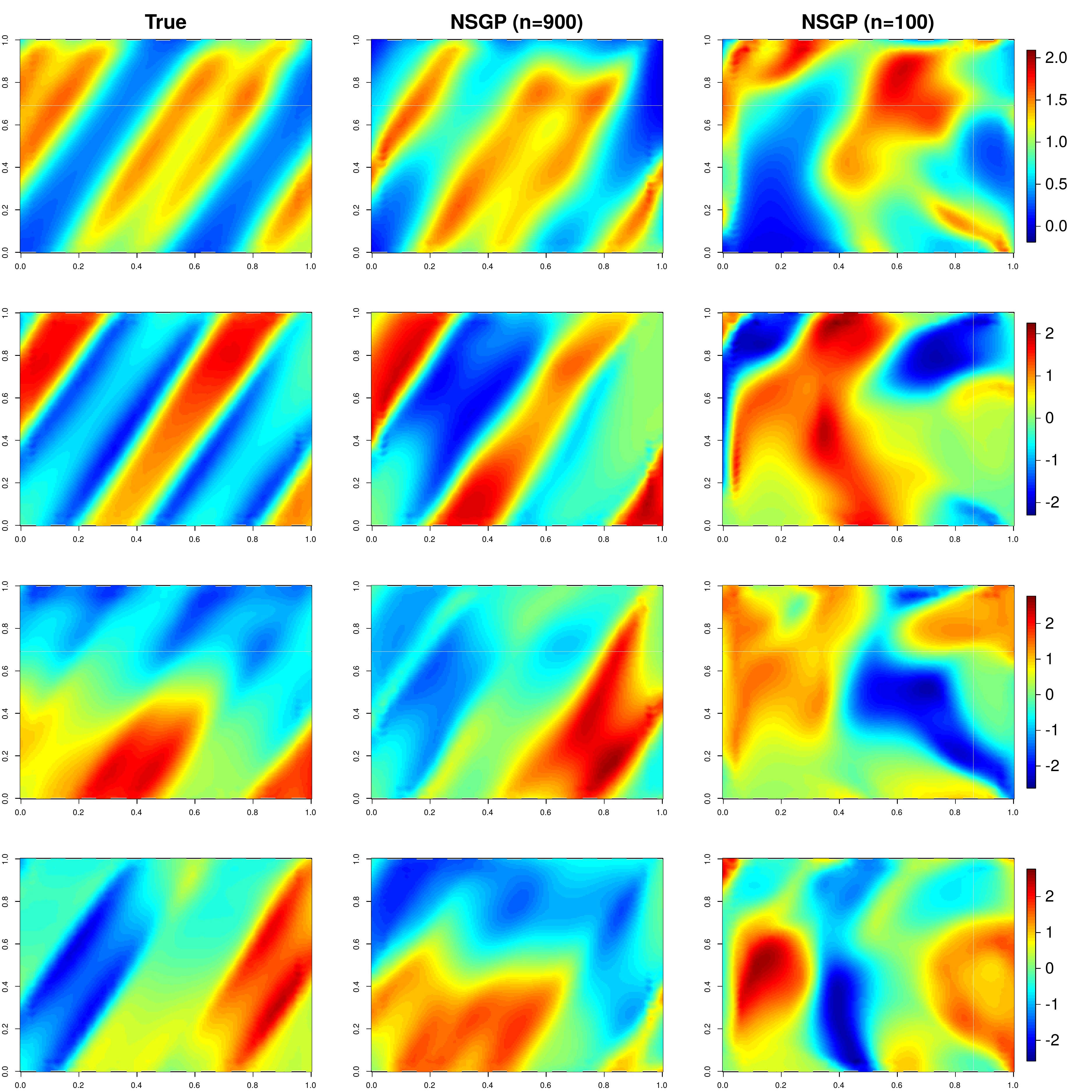} 
	\caption{First four leading eigensurfaces $\phi_j(s_1, s_2)$ of the true covariance structure of data from Simulation Study 2 (first column) and the corresponding eigensurfaces $\hat{\phi}_j(s_1, s_2)$ estimated by NSGP model for data with different sample sizes: $n=100$ and $900$.}
	\label{eigensurf_simConv}
\end{figure}

Figure~\ref{CFVE_simConv} displays the CFVE for the leading eigensurfaces of the simulation study of Subsection~\ref{secSim_converg}. The leading eigensurfaces are shown in Figure~\ref{eigensurf_simConv}. The eigensurfaces from NSGP model ($n=900$) seem to identify very closely the main modes of variation in the data. The third and fourth true eigensurfaces have similar contribution in explaining the variation in the data: $4.12\%$ and $3.99\%$, respectively, and this similarity made NSGP model identify these components in reverse order. 

\addtolength{\textheight}{0.3in}%

\section{Additional Simulation Study: Case $Q=3$} \label{sec:SupplSim3d}

In this example, we simulate a three-dimensional function-valued process $X(\tau, s_1, s_2)$,  where $\tau, s_1, s_2 \in \left[0,1\right]$, from \eqref{NLFunReg}, where $f$ is zero-mean student $t$-process with six degrees of freedom, covariance function \eqref{NScovfun} and squared exponential correlation kernel $g(\cdot)$. The noise variance is $\sigma_\varepsilon^2 = 0.1$. We assume that $\sigma^2(\cdot)$ and $\ve A(\cdot)$ in $\eqref{NScovfun}$ depend only on $\tau$. This example is comparable to spatiotemporal models which have time and spatial coordinates as input variables and time-varying coefficients (e.g. dynamic linear models in \cite{banerjee2015hierarchical}).

In the data generating process, $\sigma^2(\tau) = \exp(\tau)$ and
\begin{equation}
\ve A(\tau) = \begin{bmatrix}
1  & 0               & 0                 \\[0.3em]
0  & 1               & \rho_{23}(\tau)  \\[0.3em]
0  & \rho_{23}(\tau) & 1
\end{bmatrix},
\end{equation}
with degree of nonseparability $\rho_{23}(\tau) = -0.95 \ \tau$. That is, the interaction between coordinate directions $s_1$ and $s_2$ in the covariance function increases with respect to $\tau$. As $\tau$ increases, $\rho_{23}$ becomes more distant from zero, and thus models which assume separable covariance structure might be not suitable.  

In order to apply Product FPCA to the $Q=3$-dimensional setting, we use its natural extension: 
$\label{ProdFPCA3d}
X(\tau, s_1,s_2) = \mu(\tau, s_1,s_2) + \sum_{j=1}^{\infty}\sum_{l=1}^{\infty}\sum_{m=1}^{\infty} \chi_{jlm} \phi_{j}(\tau) \psi_{l}(s_1) \zeta_{m}(s_2)
$. Note that, if we look for a common maximum number of components $J=L=M$ in each direction, we will need more than $J \cdot L \cdot M$ realizations of $X$. An extension of Marginal FPCA to the $Q=3$ case would also be possible, but its implementation would require an even larger sample size or a smaller number of components in each direction. 

We have simulated 100 realizations of $X$ and set $J=L=M=4$ for Product FPCA. For NSGP model, we have used $5$ B-spline basis functions for the seven time-varying parameters -- the log overall variance $\log \sigma(\tau)$ and the six unconstrained elements related to $\ve A(\tau)$.

\begin{figure}[H]
	\centering 
	\includegraphics[width=1\linewidth]{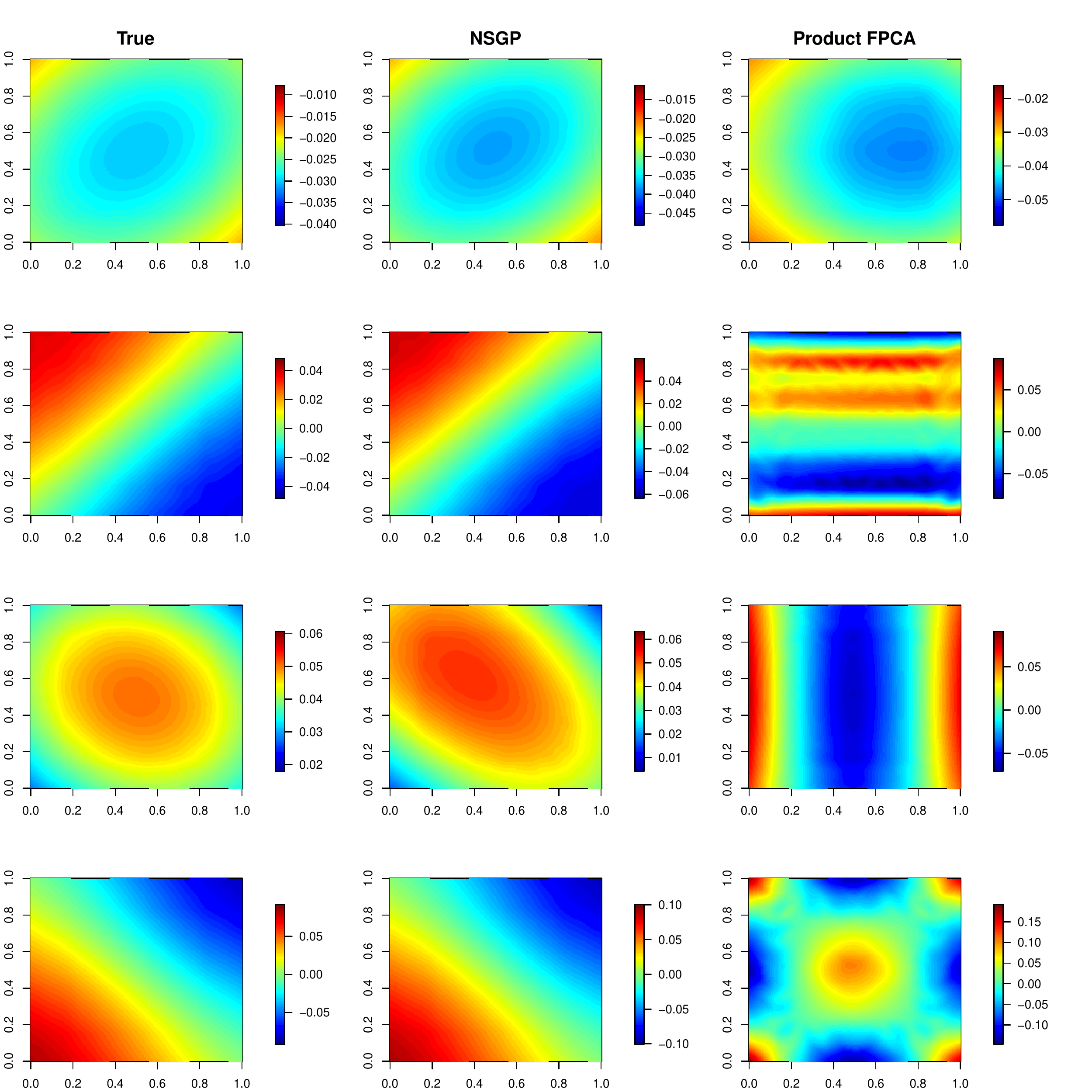} 
	\caption{First four leading eigensurfaces $\phi_j(0.125, s_1, s_2)$ of the true model (left column) and the corresponding estimated eigensurfaces $\hat{\phi}_j(0.125, s_1, s_2)$ of NSGP model (center) and Product FPCA model (right).}
	\label{sim_Q3Res_Eigensurf_tau0125}
\end{figure}

\begin{figure}[H]
	\centering 
	\includegraphics[width=1\linewidth]{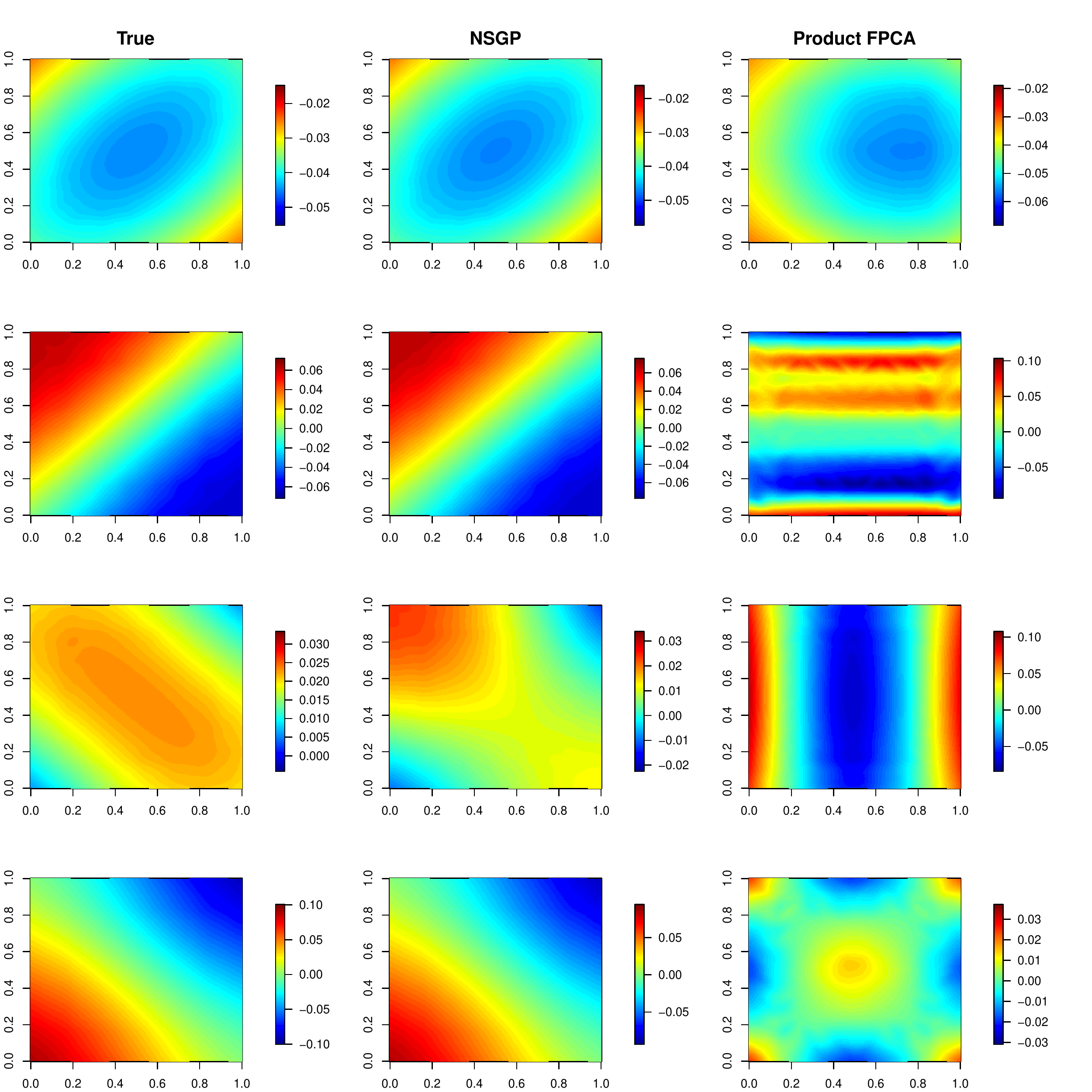} 
	\caption{First four leading eigensurfaces $\phi_j(0.5, s_1, s_2)$ of the true model (left column) and the corresponding estimated eigensurfaces $\hat{\phi}_j(0.5, s_1, s_2)$ of NSGP model (center) and Product FPCA model (right).}
	\label{sim_Q3Res_Eigensurf_tau05}
\end{figure}

\begin{figure}[H]
	\centering 
	\includegraphics[width=1\linewidth]{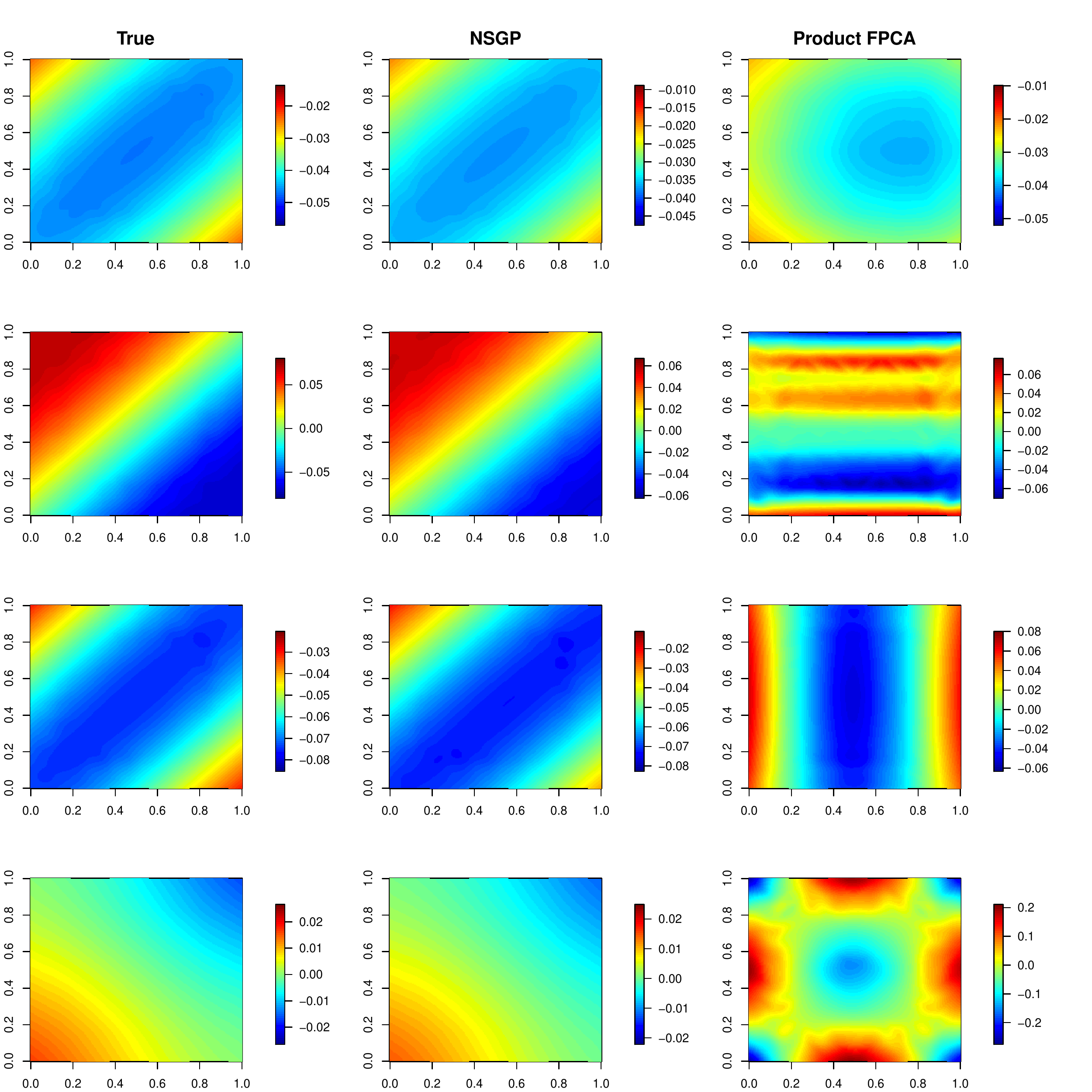} 
	\caption{First four leading eigensurfaces $\phi_j(1, s_1, s_2)$ of the true model (left column) and the corresponding estimated eigensurfaces $\hat{\phi}_j(1, s_1, s_2)$ of NSGP model (center) and Product FPCA model (right).}
	\label{sim_Q3Res_Eigensurf_tau1}
\end{figure}

Figures~\ref{sim_Q3Res_Eigensurf_tau0125}, \ref{sim_Q3Res_Eigensurf_tau05}, and \ref{sim_Q3Res_Eigensurf_tau1}, show, for $\tau=0.125,0.5$, and $1$, respectively, the true leading eigensurfaces $\phi(\tau, s_1, s_2)$ and the corresponding estimates obtained by NSGP and Product FPCA \citep{chen2017modelling} models. Although the data were simulated from a student $t$-process, NSGP model obtains eigensurfaces quite similar to the true ones. The model clearly identifies that the interaction between $s_1$ and $s_2$ becomes stronger as $\tau$ increases -- see how the diagonal orientation of ellipses (especially in the first and third components) is more evident as $\tau$ increases from Figure~\ref{sim_Q3Res_Eigensurf_tau0125} to Figure~\ref{sim_Q3Res_Eigensurf_tau1}. This aspect cannot be detected by Product FPCA model as it assumes separable eigensurfaces.

\begin{figure}[H]
	\centering 
	\includegraphics[width=0.5\linewidth]{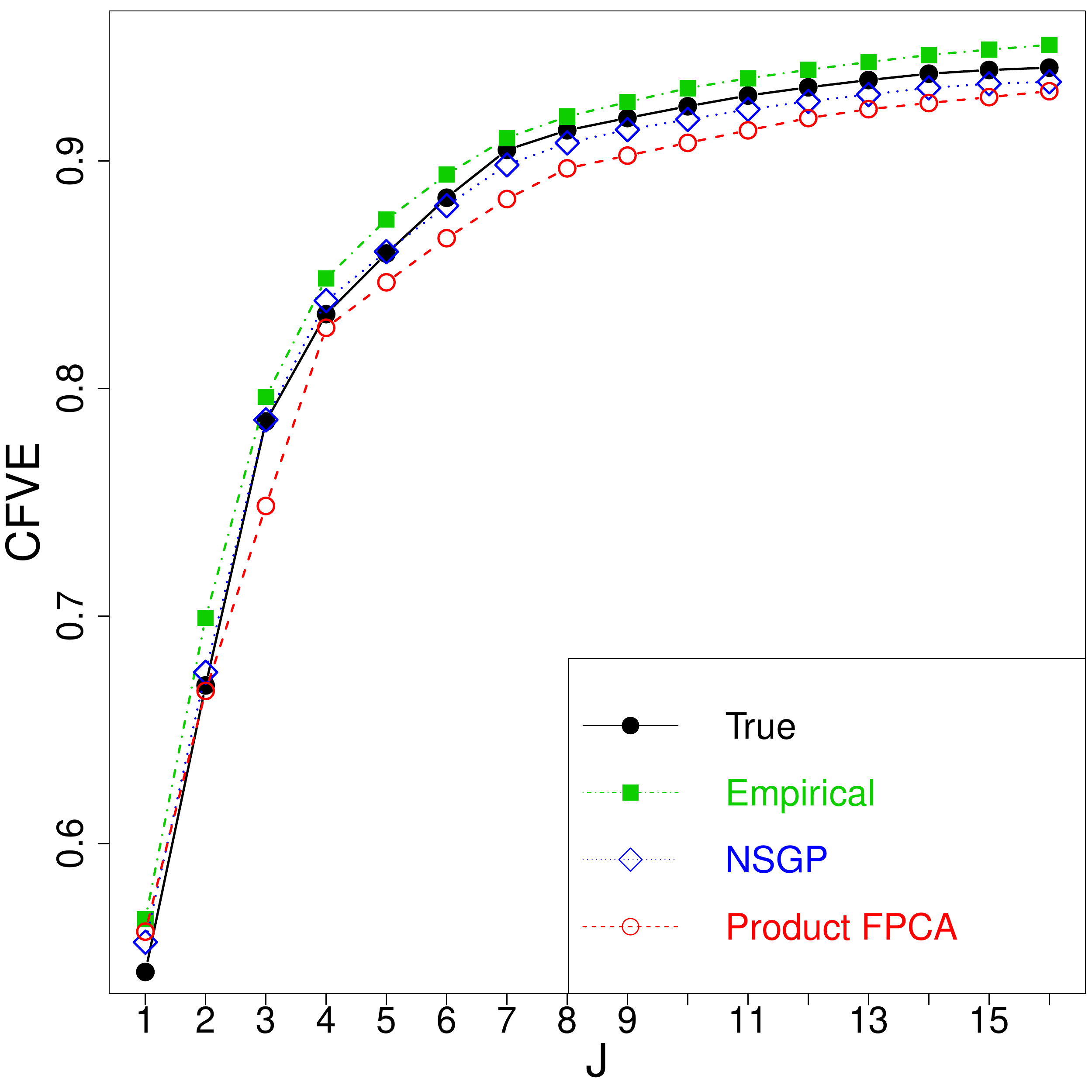} 
	\caption{CFVEs obtained by the true eigensurfaces, the eigensurfaces of the empirical covariance function, and the eigensurfaces obtained by NSGP and Product FPCA models.}
	\label{CFVE_sim3d}
\end{figure}

The CFVEs for the first 16 leading three-dimensional eigensurfaces are illustrated in Figure~\ref{CFVE_sim3d}. As expected, the advantage of the nonseparable model in terms of CFVE is clear not in the first, but in later components. Note that using the empirical covariance function produces slightly larger CFVE than the true model does. This is due to the presence of noisy observations in the sample, which makes nonparametric models explain noise and obtain possibly superior CFVE than the true model does.

\newpage

\section{Application to Relative Humidity Data}\label{sec:AppRelHum}

\begin{figure*}[h!]
	\centering
	\begin{subfigure}[t]{0.32\textwidth}
		\centering
		\includegraphics[width=1\linewidth]{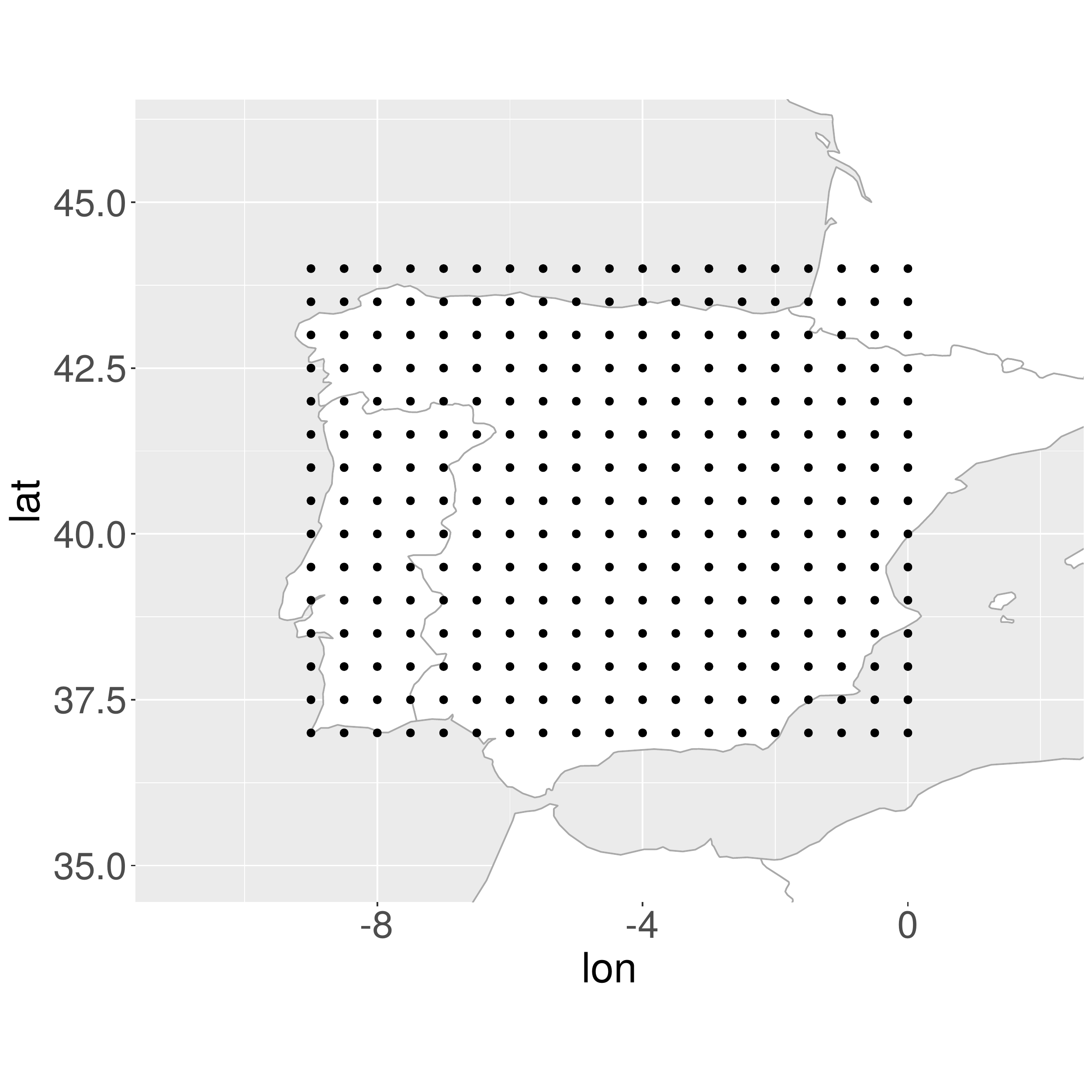}
		\caption{}
		\label{Fig:RelHumsubfig-a}
	\end{subfigure}%
	~ 
	\begin{subfigure}[t]{0.32\textwidth}
		\centering
		\includegraphics[width=0.83\linewidth]{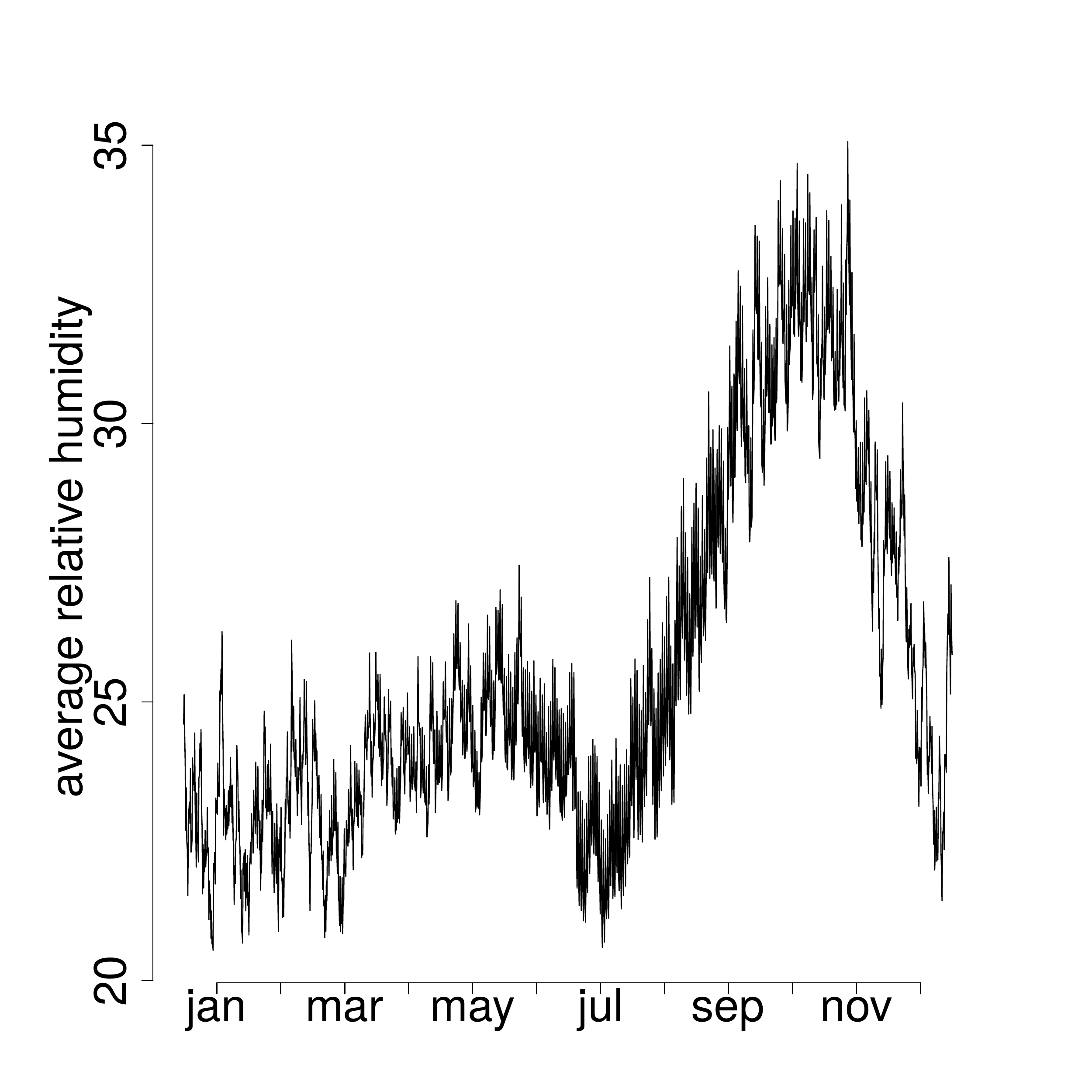}
		\caption{}
		\label{Fig:RelHumsubfig-b}
	\end{subfigure}
	~
	\begin{subfigure}[t]{0.32\textwidth}
		\centering
		\includegraphics[width=0.83\linewidth]{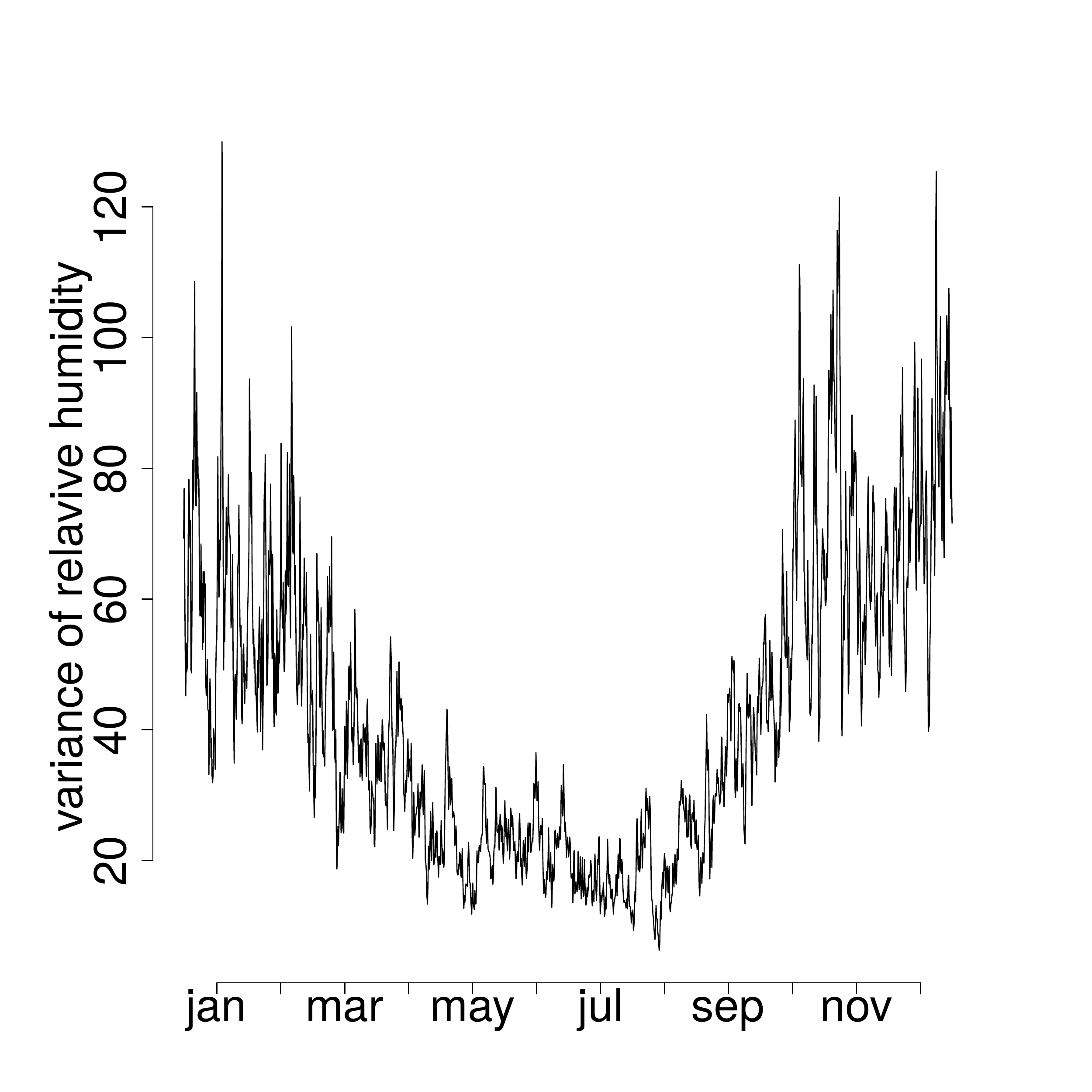}
		\caption{}
		\label{Fig:RelHumsubfig-c}
	\end{subfigure}
	\caption{Relative humidity data: (a) spatial locations; (b) and (c) mean and variance (across stations) over the calendar year.}
	\label{Fig:RelHum}
\end{figure*}

In this application, we use a sample of 6-hourly relative humidity analysis data  \citep{cisl_rda_ds093_0}, which comprises the time period from January 1979 to December 2010. The relative humidity data are of the entire atmosphere (considered as a single layer) and are observed on a two-dimensional grid $0.5^\circ \times 0.5^\circ$, with longitude spanning from $-9^\circ$W to $0^\circ$E and latitude from $37^\circ$N to $44^\circ$N, therefore basically representing a sample from the Iberian Peninsula. Each of the 32 years of the sample is considered as a realization. Figure~\ref{Fig:RelHum} shows the spatial locations and the mean and variance across stations over the calendar years.

Let $s_1, s_2$ and $\tau$ be longitude, latitude and time, respectively, of the relative humidity process $X$. For simplicity of exposition, we estimate the correlation function of the processes $X(s_1, \tau)$ and $X(s_2, \tau)$ considering time-varying parameters. We estimate the correlation structure of both processes using different specifications for correlation function $g$ in \eqref{NScovfun} and cyclic B-splines basis functions with 5 equally spaced knots for modeling the time-varying unconstrained parameters via equation \eqref{bsplines} of the main manuscript.

\begin{figure}[H]
	\centering
	\includegraphics[width=0.48\linewidth]{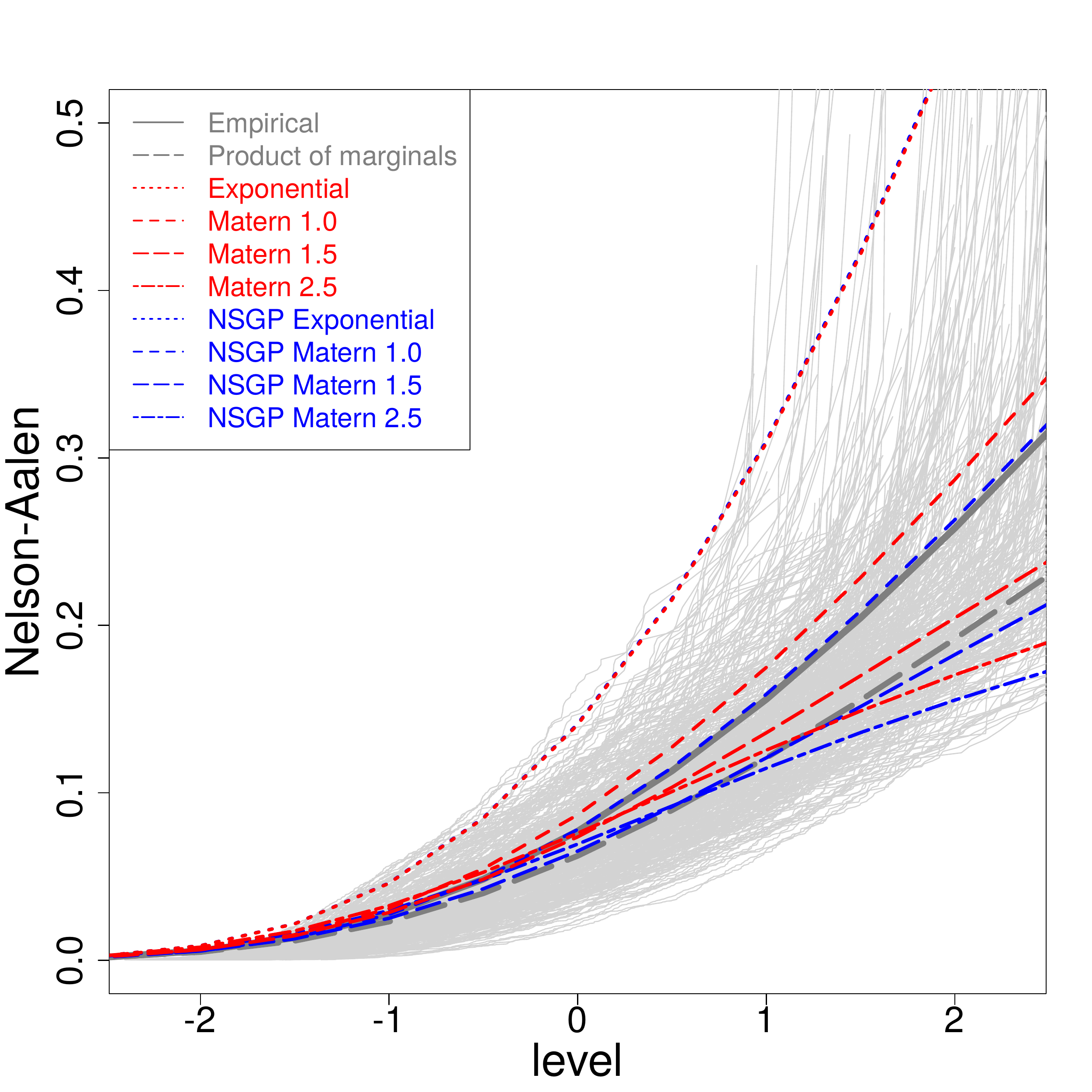}
	\includegraphics[width=0.48\linewidth]{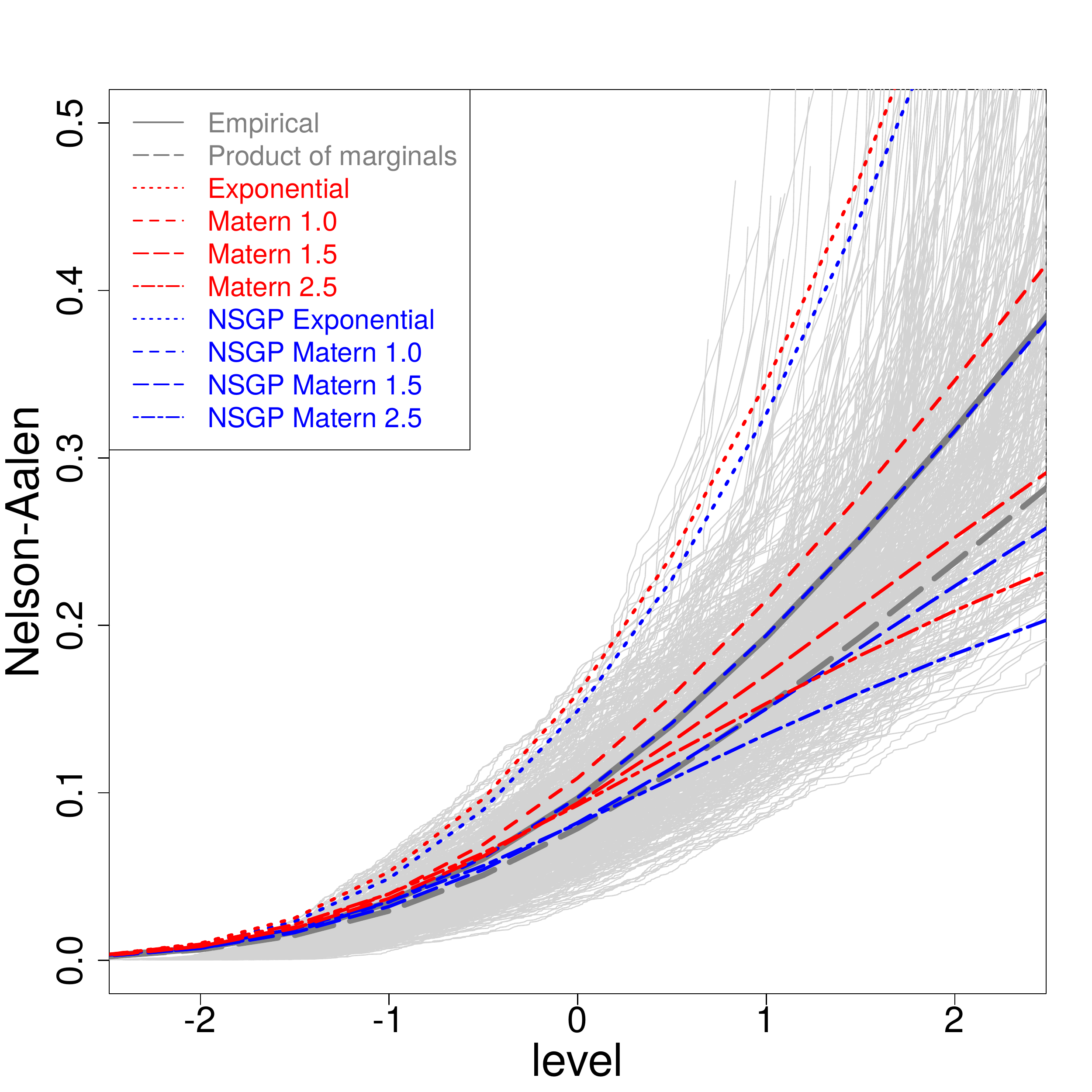}
	\caption{Nelson-Aalen plots for relative humidity data analyzed by the Longitude x Time model (left) and Latitude x Time model (right). The dark grey, continuous line is based on the empirical correlation function. The dark grey, dashed line is on the product of marginal correlation functions estimated separately. The other lines are results based on stationary, separable GP and NSGP models with different specifications for $g$.}
	\label{Fig:RelHum_NelsonAalen_winter}
\end{figure}

Figure~\ref{Fig:RelHum_NelsonAalen_winter} displays Nelson-Aalen plots for both Longitude x Time and Latitude x Time models. While the empirical correlation functions seem to represent well the average behavior of the realizations, the product of marginal correlation functions seems to underestimate the cumulative hazard risk in both models. Note that Gaussian process models using a separable correlation function provide better fitting than the model based on the product of marginal correlation functions. In the latter model, the estimation of the marginal correlation function on space assumes independent realizations over time, an assumption which may be too strong. On the other hand, Gaussian processes with a separable correlation function do not estimate the marginal correlation functions separately. Finally, NSGP model with \Matern ($\nu=1$) produces Nelson-Aalen plots very similar to the model based on the empirical correlation function found in the data.

Using NSGP model with \Matern ($\nu=1$), the next subsections discuss the estimates of the elements of the time-varying anisotropy matrix and show that they are corroborated by the empirical correlation function found in the data.

\subsection{Longitude x Time Model}

\addtolength{\textheight}{.5in}%

\begin{figure}[H]
	\centering
	\includegraphics[width=0.46\linewidth]{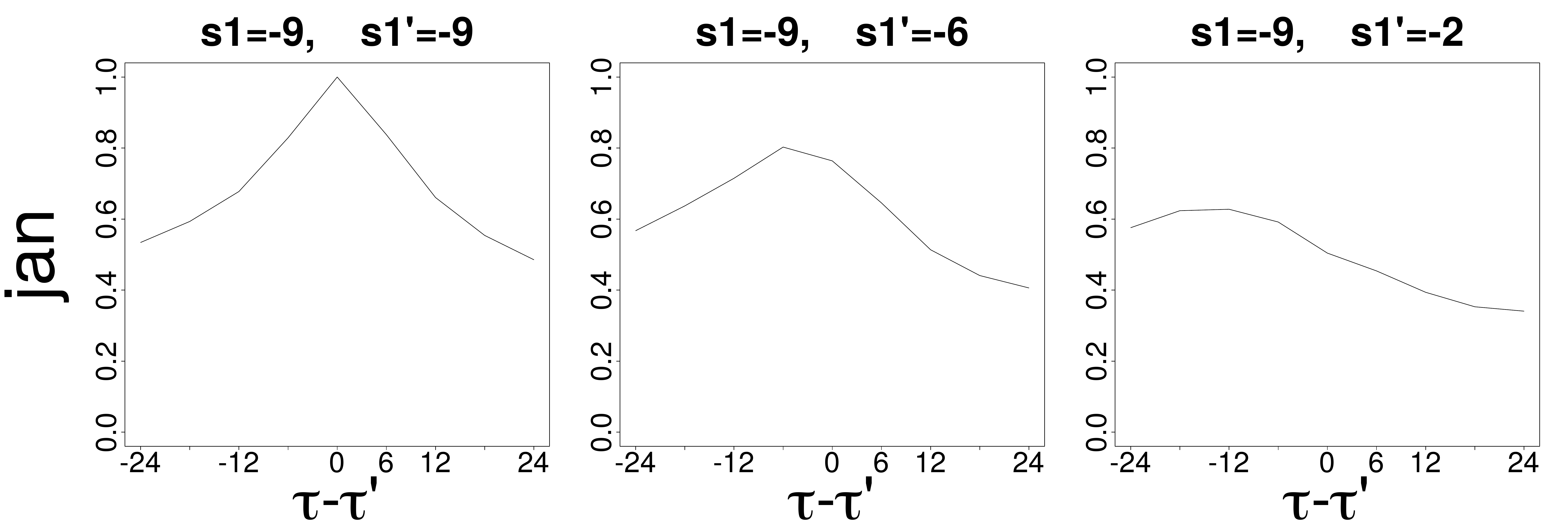} \hspace{1cm}
	\includegraphics[width=0.46\linewidth]{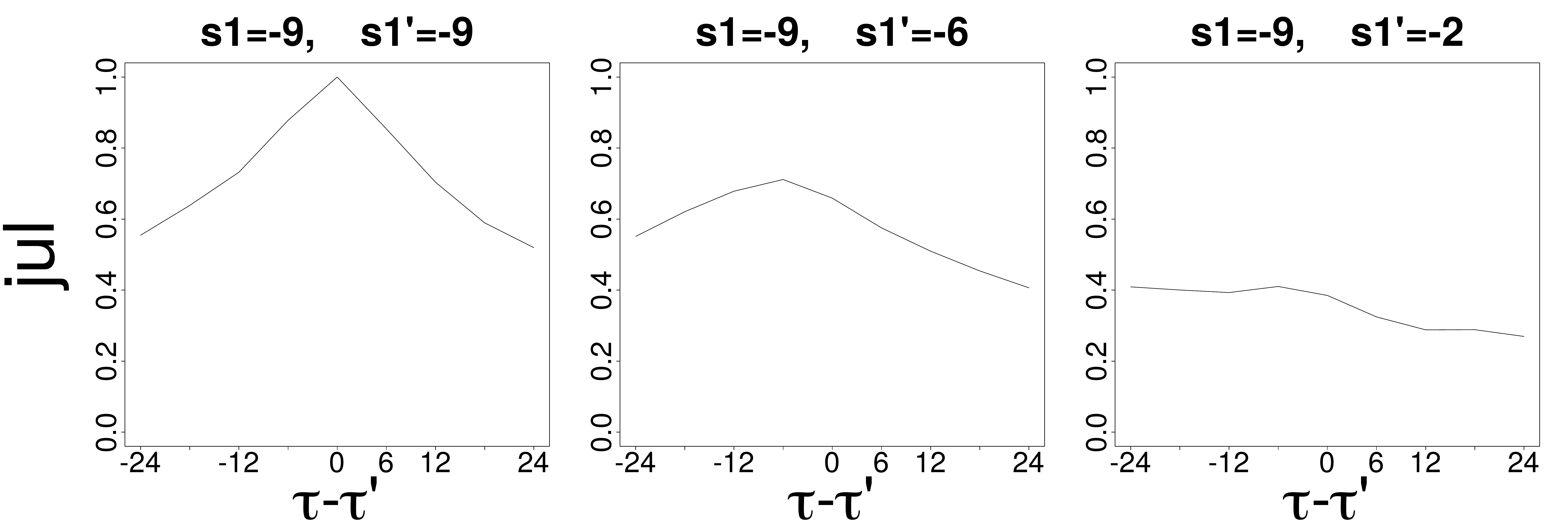}
	\\ \vspace{0.5cm}
	\includegraphics[width=0.46\linewidth]{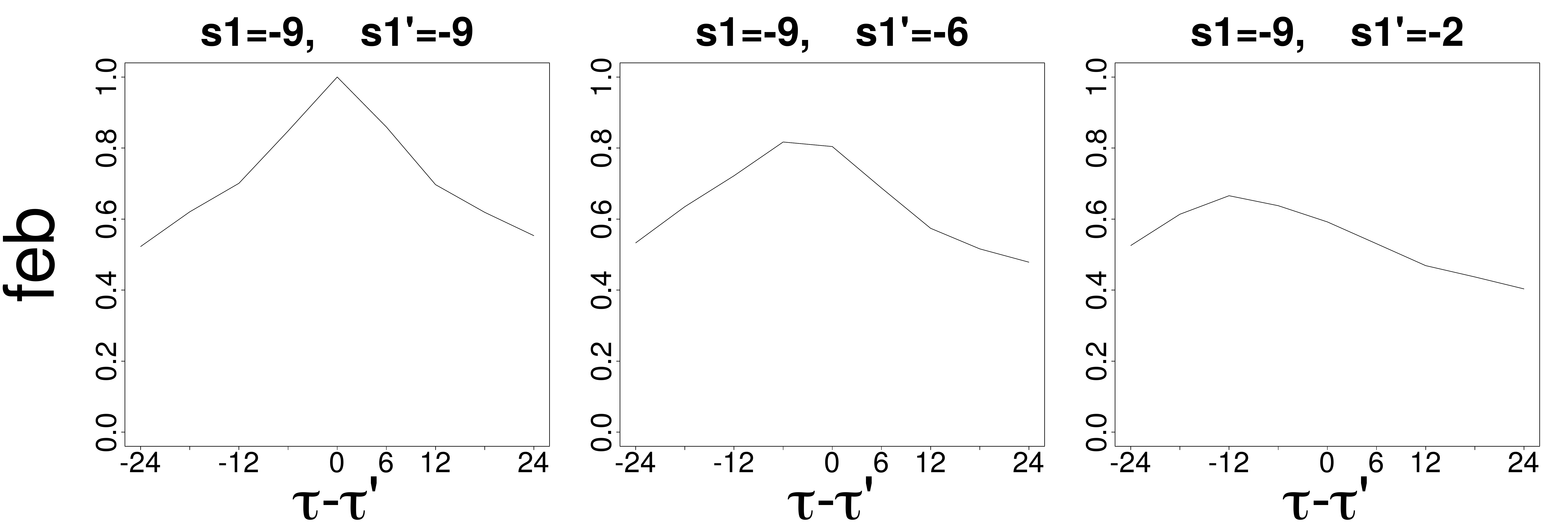} \hspace{1cm}
	\includegraphics[width=0.46\linewidth]{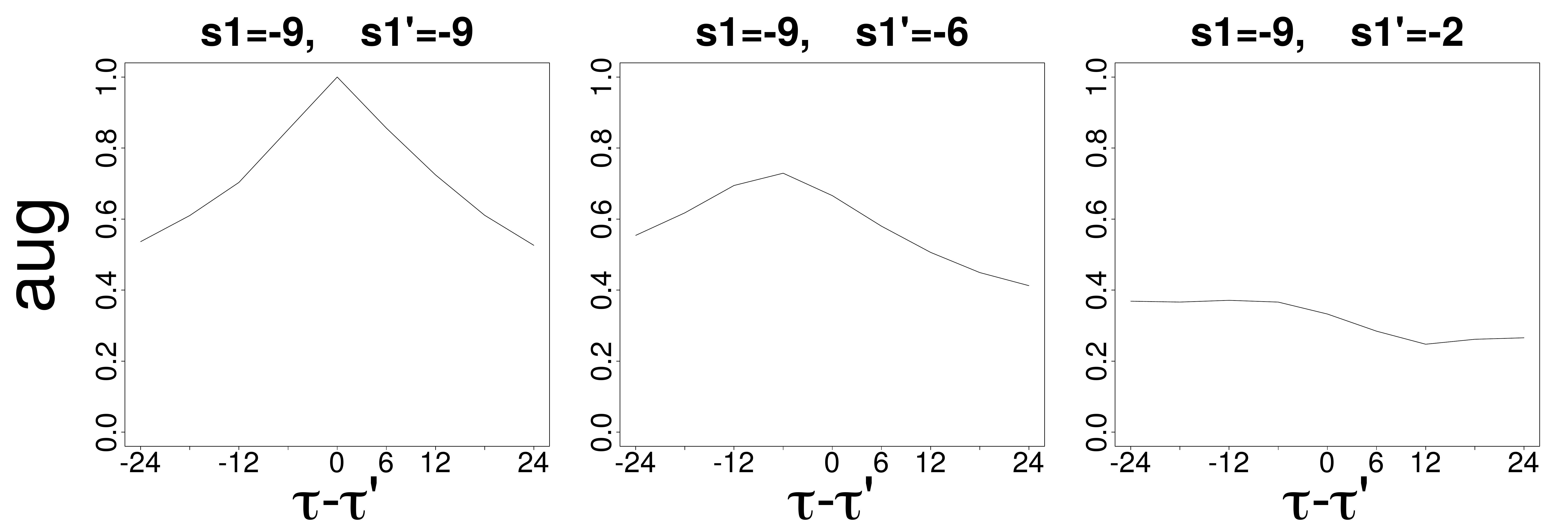}
	\\ \vspace{0.5cm}
	\includegraphics[width=0.46\linewidth]{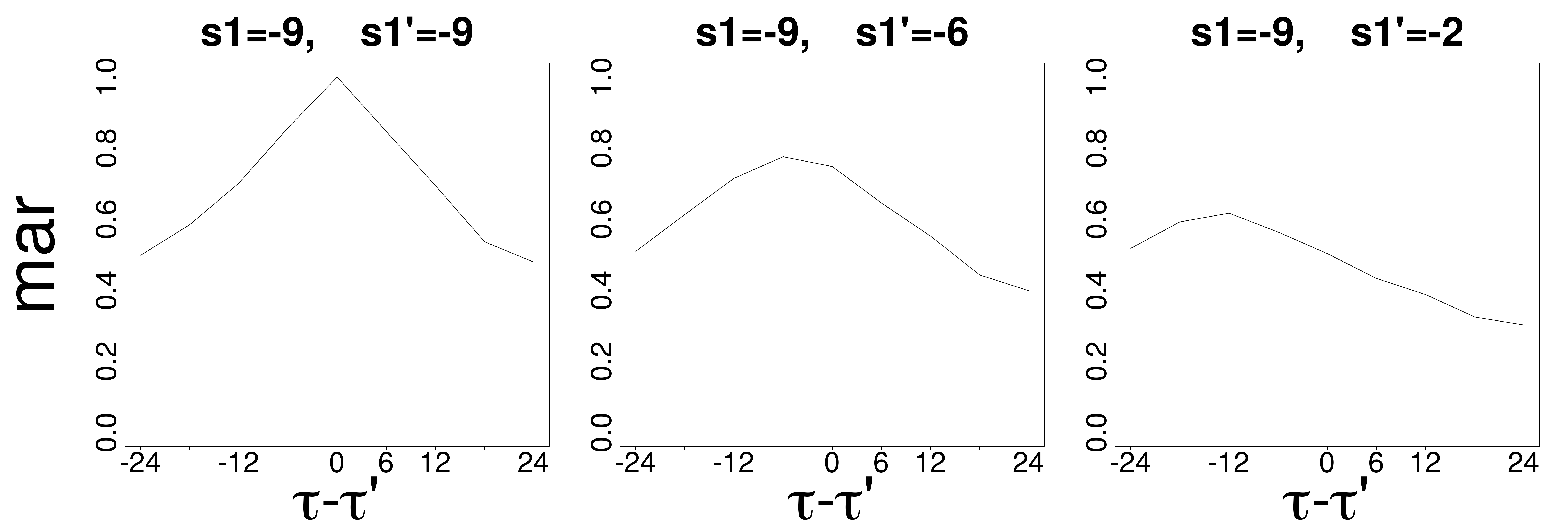} \hspace{1cm}
	\includegraphics[width=0.46\linewidth]{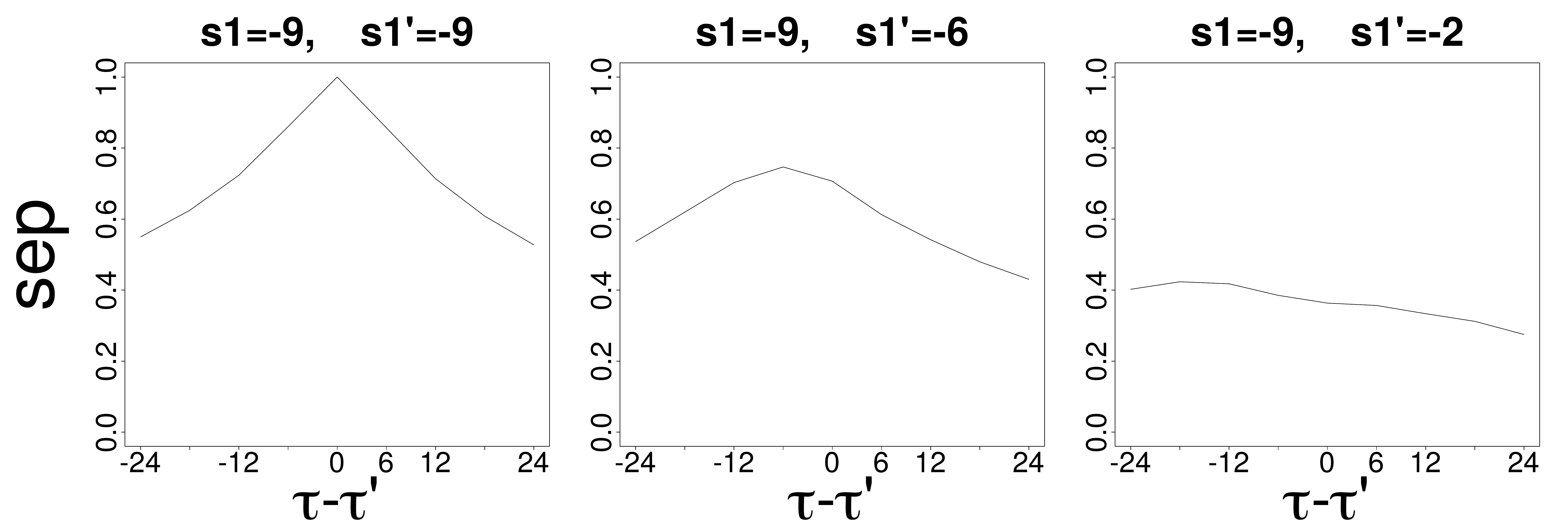}
	\\ \vspace{0.5cm}
	\includegraphics[width=0.46\linewidth]{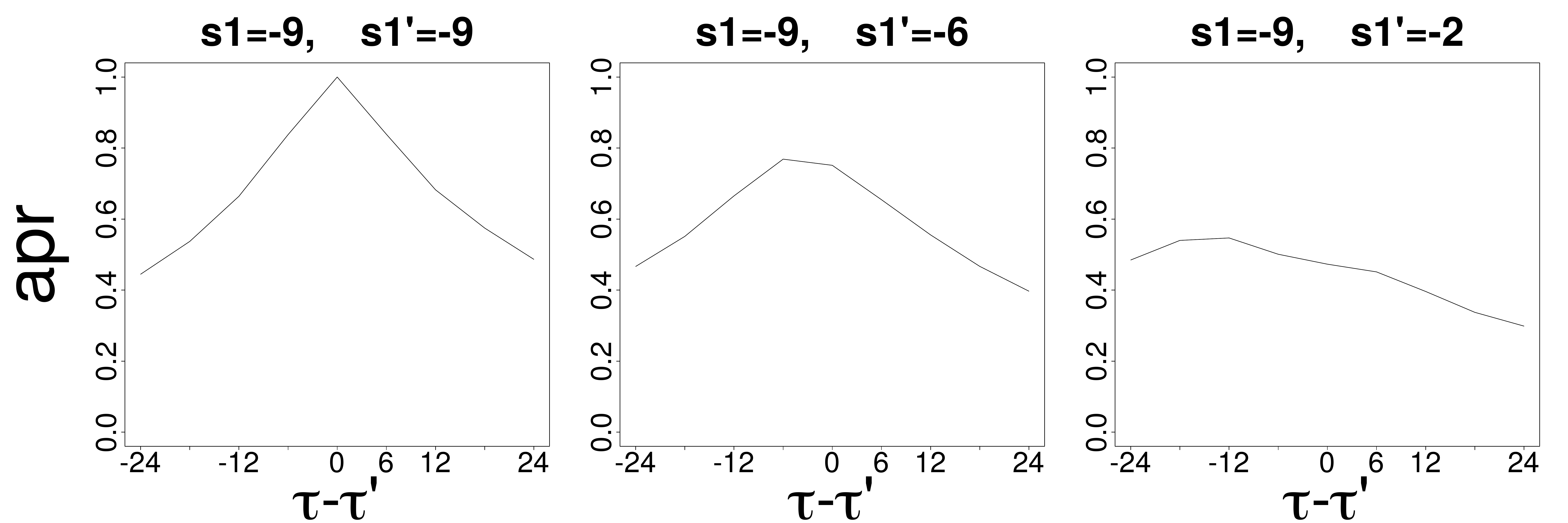} \hspace{1cm}
	\includegraphics[width=0.46\linewidth]{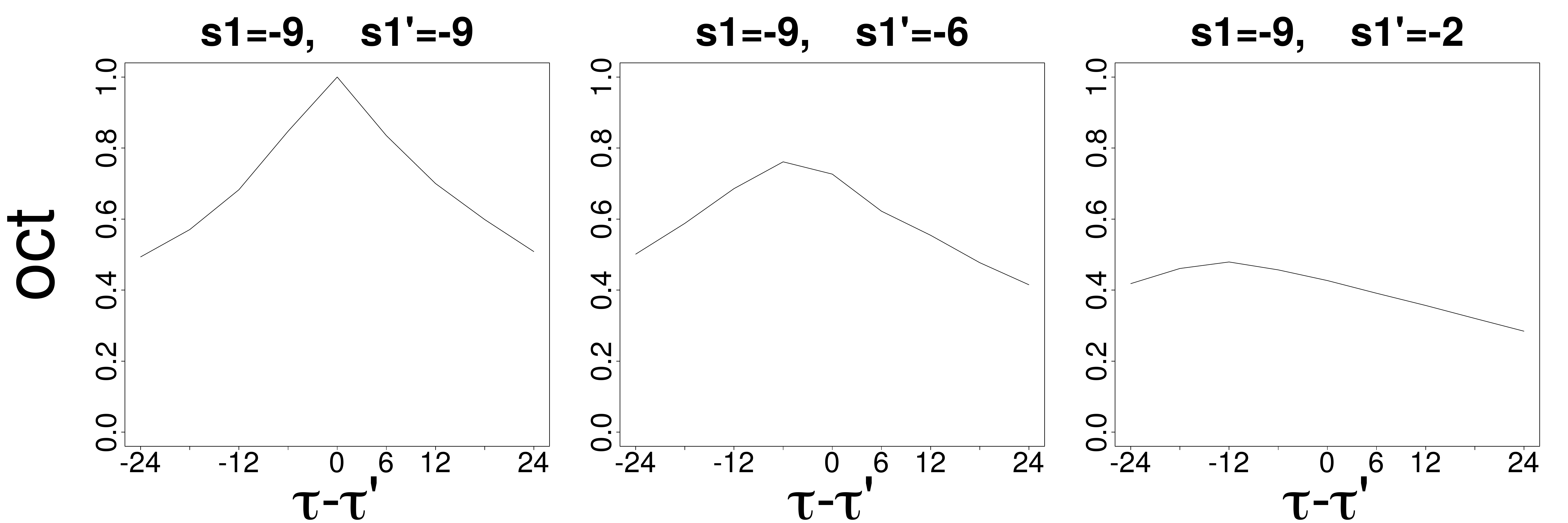}
	\\ \vspace{0.5cm}
	\includegraphics[width=0.46\linewidth]{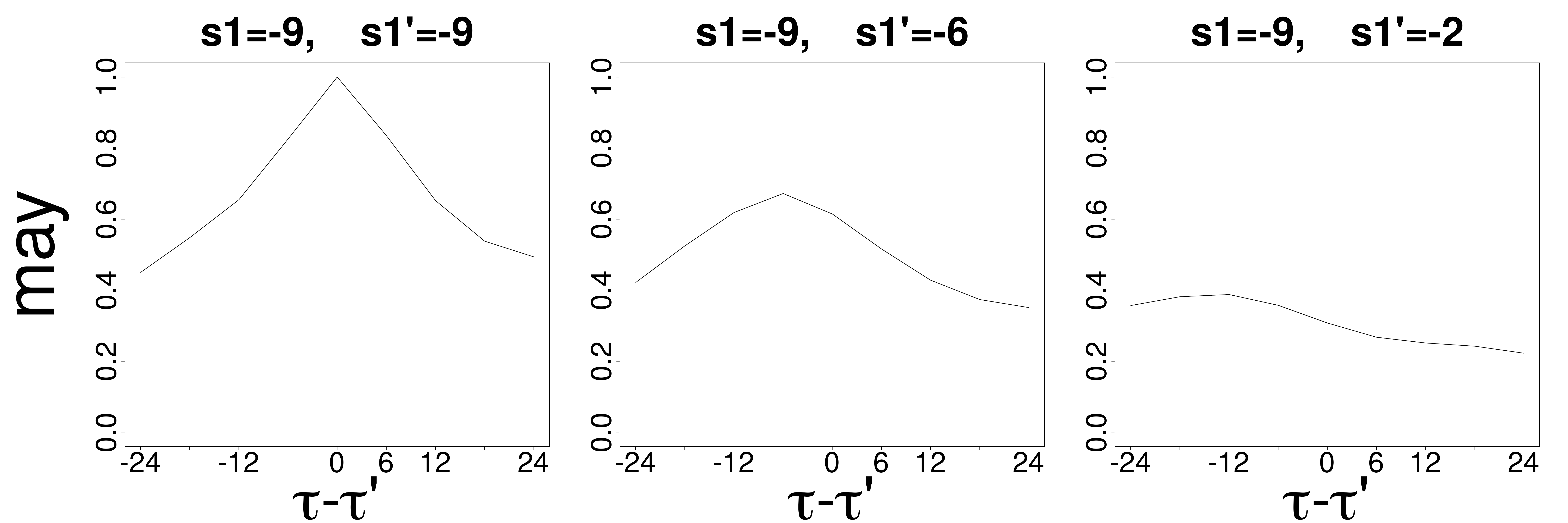} \hspace{1cm}
	\includegraphics[width=0.46\linewidth]{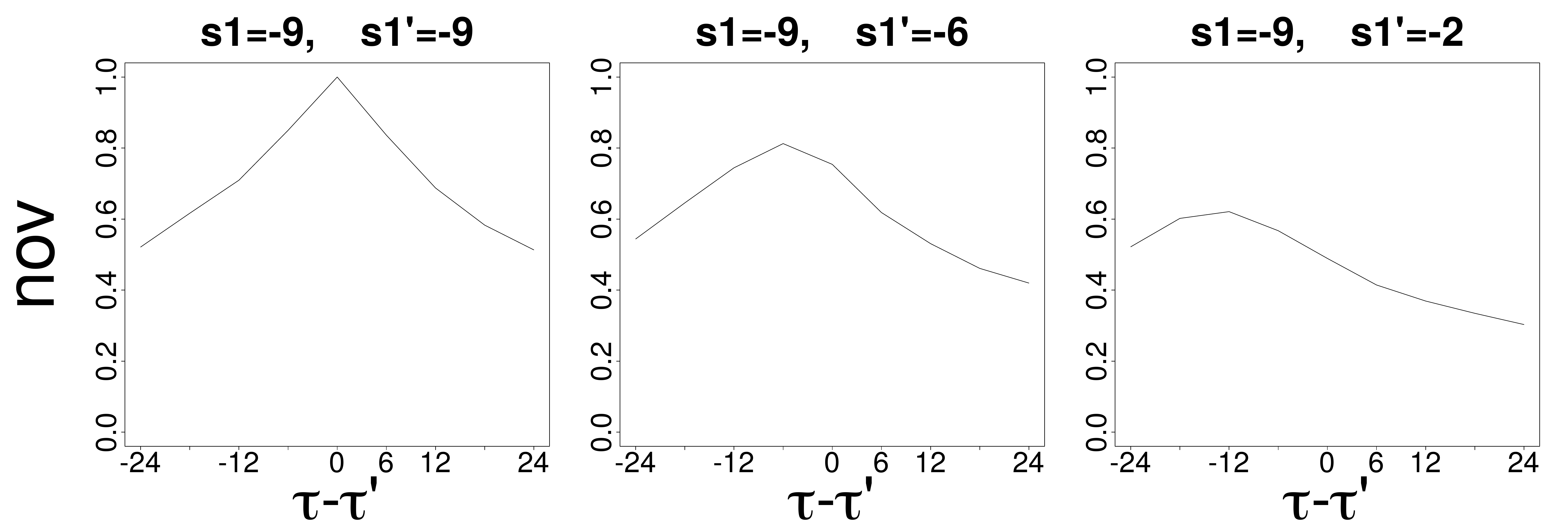}
	\\ \vspace{0.5cm}
	\includegraphics[width=0.46\linewidth]{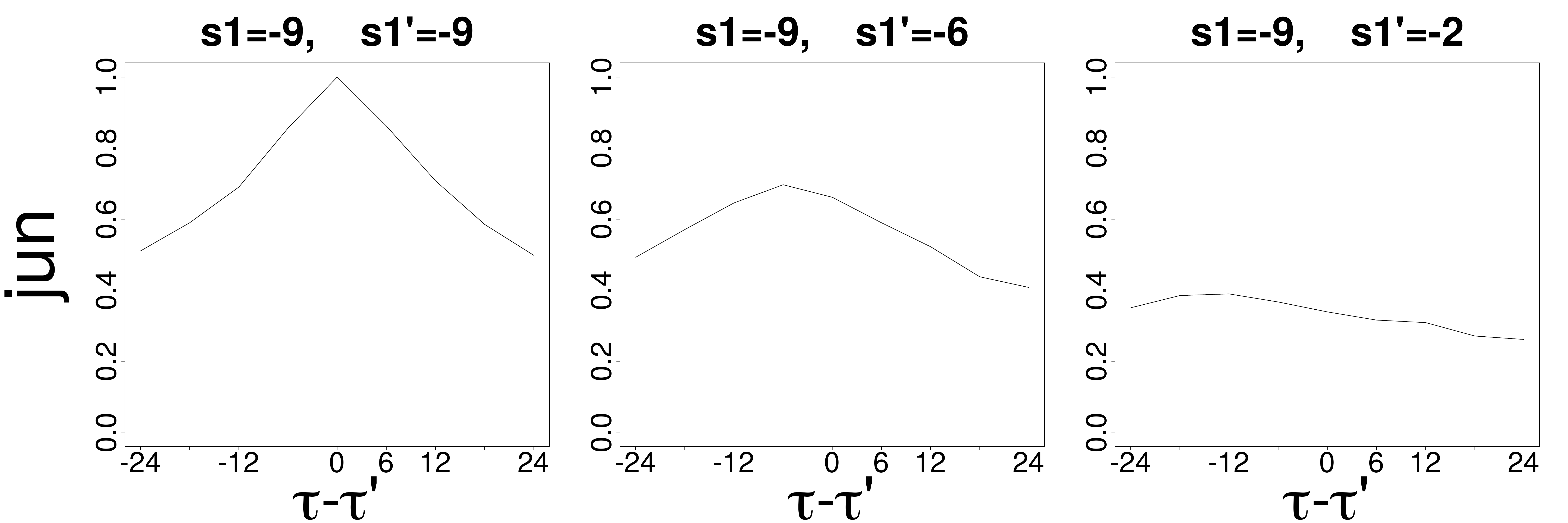} \hspace{1cm}
	\includegraphics[width=0.46\linewidth]{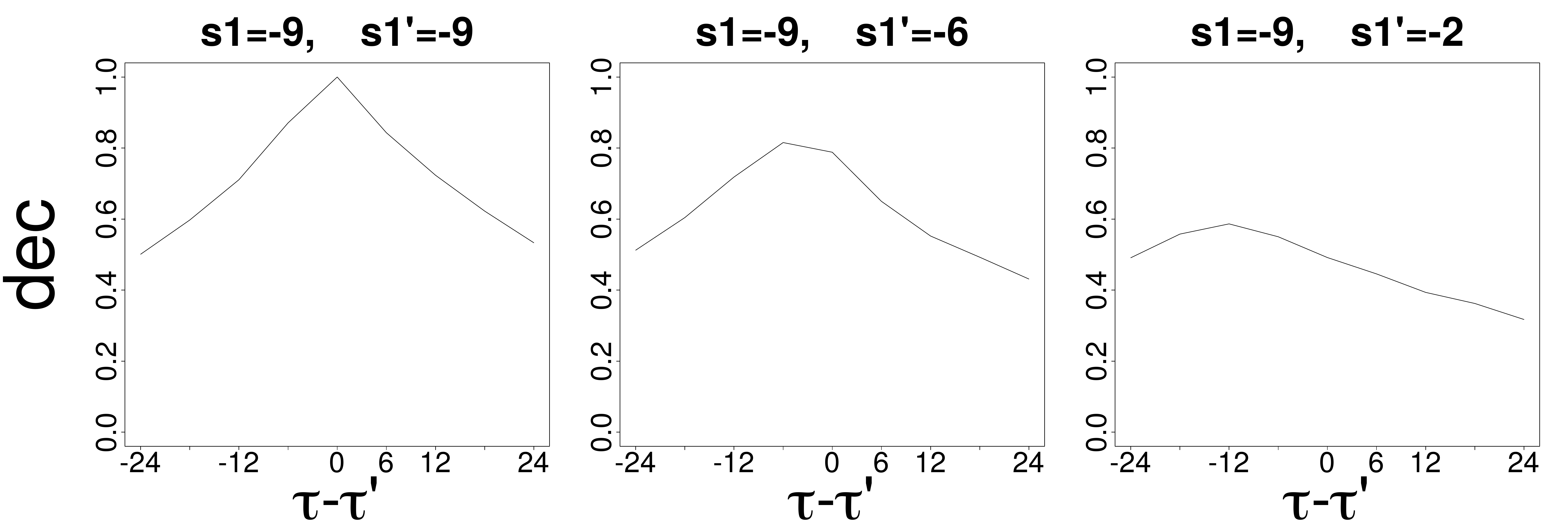}
	\caption{Empirical temporal correlation functions for relative humidity data (Longitude x Time model) from the Iberian Peninsula. Each triple of plots displays the empirical temporal correlation functions $\hat{\Cor} \Big[ X(s_1=-9,\tau), X(s_1',\tau') \Big]$ against $\tau-\tau'$, \ for $s_1'=-9, -6,$ and $-2$  for a particular month. Time distances $\tau-\tau'$ are shown in hours.}
	\label{Plot:RelHumEmpiricalTempCorLon}
\end{figure}

Figure~\ref{Plot:RelHumEmpiricalTempCorLon} displays the estimated monthly empirical temporal correlation functions for spatial locations $s_1=-9$ and $s_1'=-9, -6,$ and $-2$. To obtain plots for each month, we take the average of the empirical temporal correlation functions found in each of the first fourteen 48-hour sub-intervals $[0-48), [48-96), \dots$, of that month. The sub-intervals are centered at their corresponding midpoint $\tau'=24, 72, \dots$, respectively. Each of the 12 triples of plots in Figure~\ref{Plot:RelHumEmpiricalTempCorLon} represents a calendar month and is used to visualize the temporal correlation structure for three different spatial distances in that month. For a fixed spatial distance, the decay of the correlation function over time lag can be seen in each single plot. For a fixed time lag, the decay of the correlation function over spatial distance can be seen by looking at the sets of three plots.

\newpage

\begin{figure}[H]
	\centering
	\includegraphics[width=0.34\linewidth]{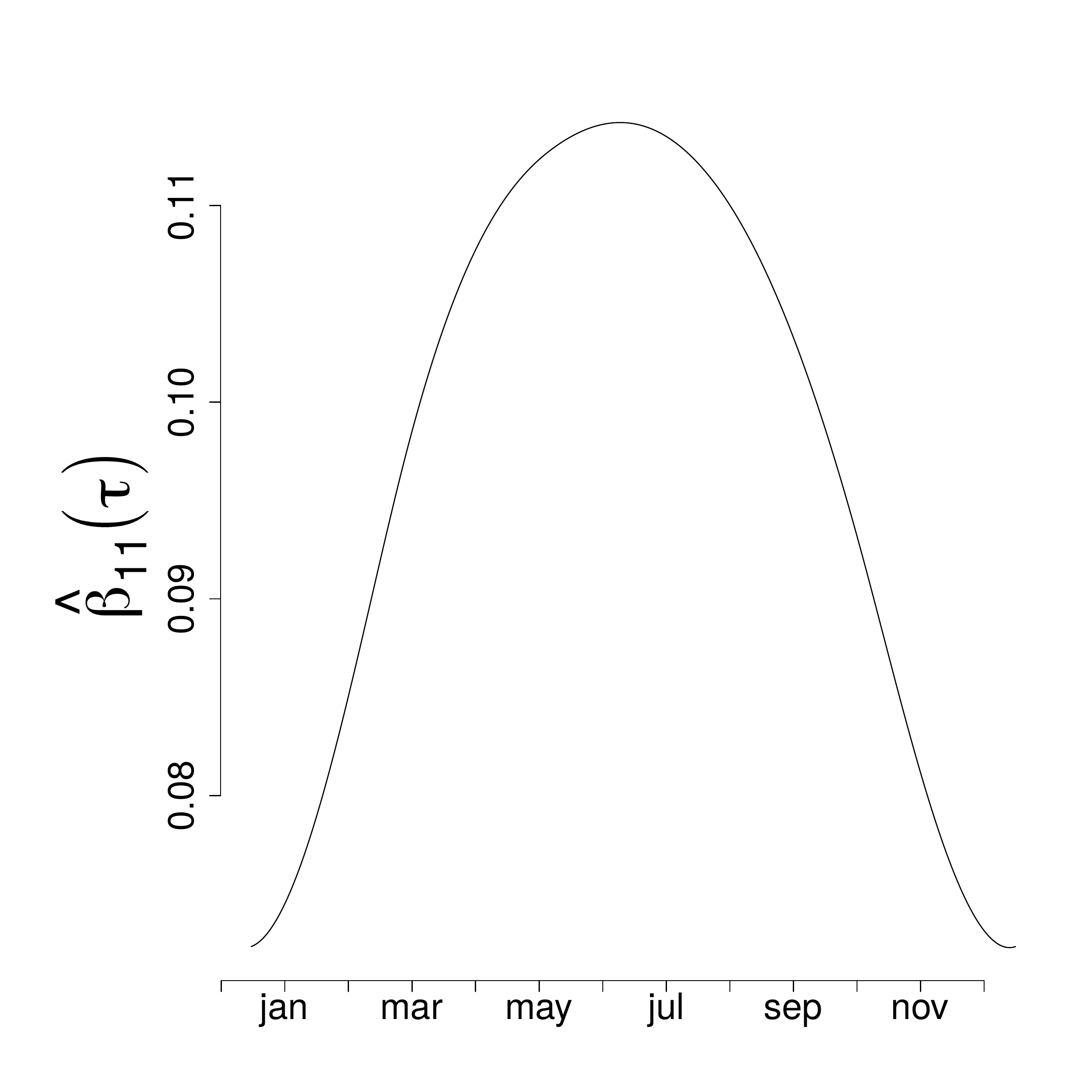}
	\includegraphics[width=0.34\linewidth]{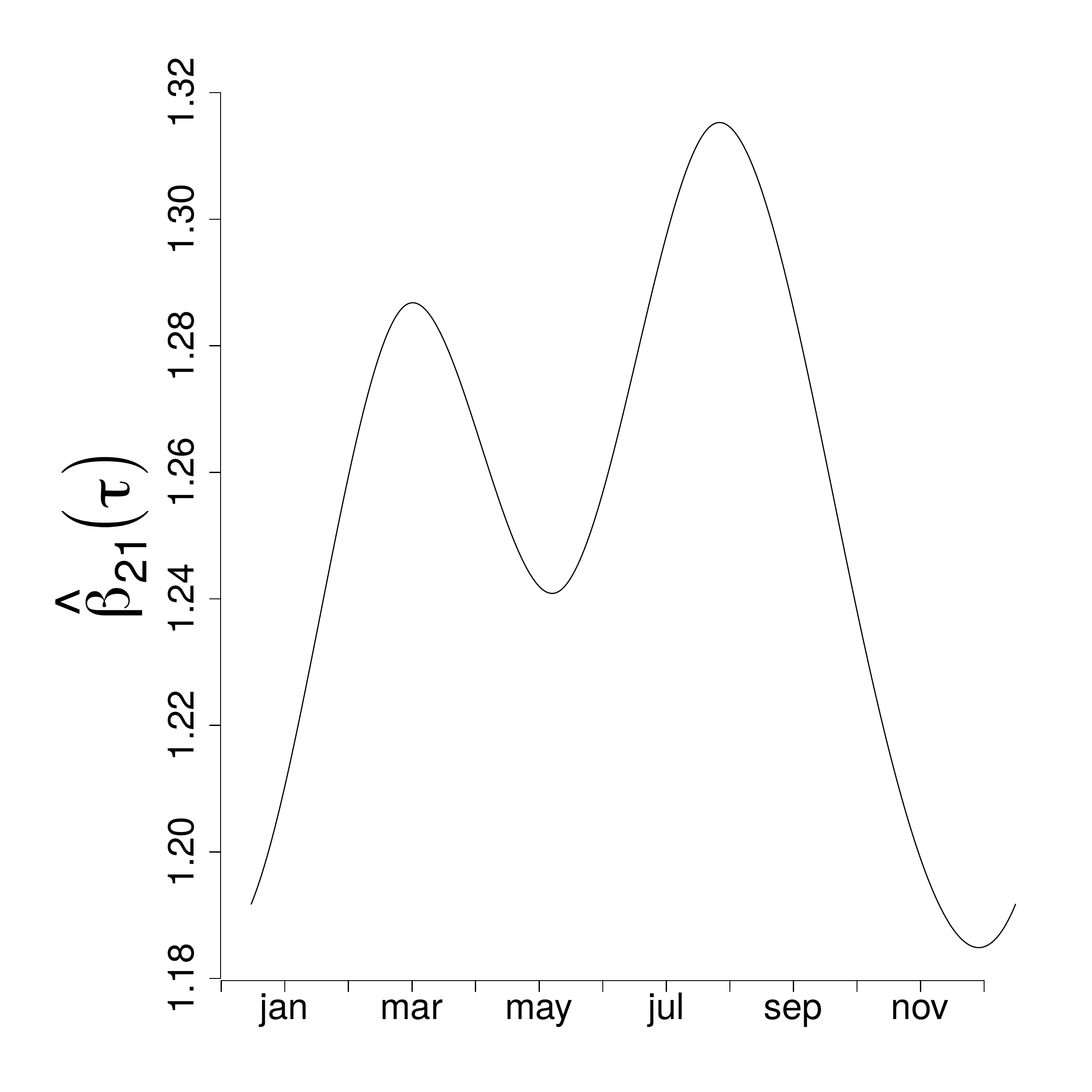}\\
	\includegraphics[width=0.34\linewidth]{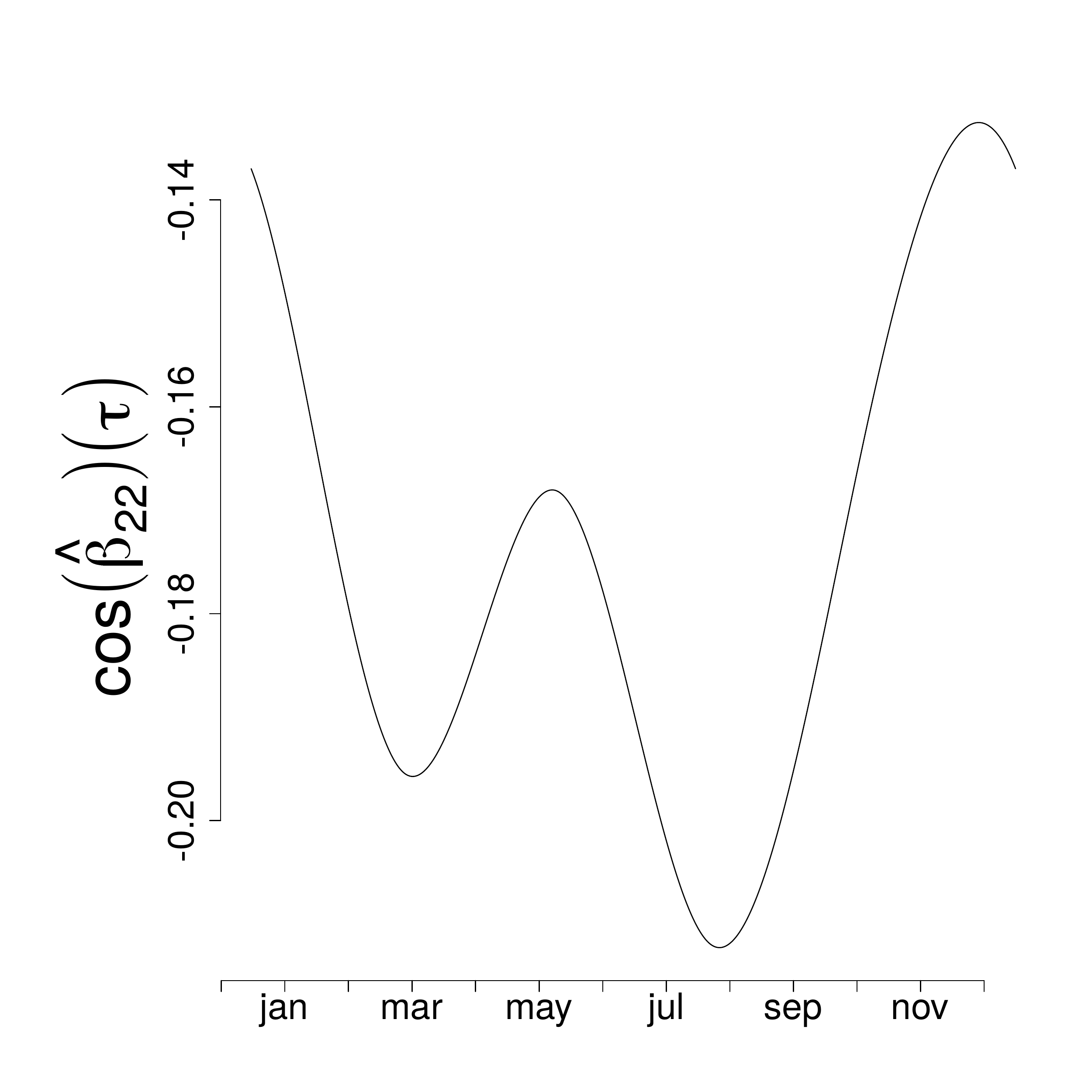}
	\includegraphics[width=0.34\linewidth]{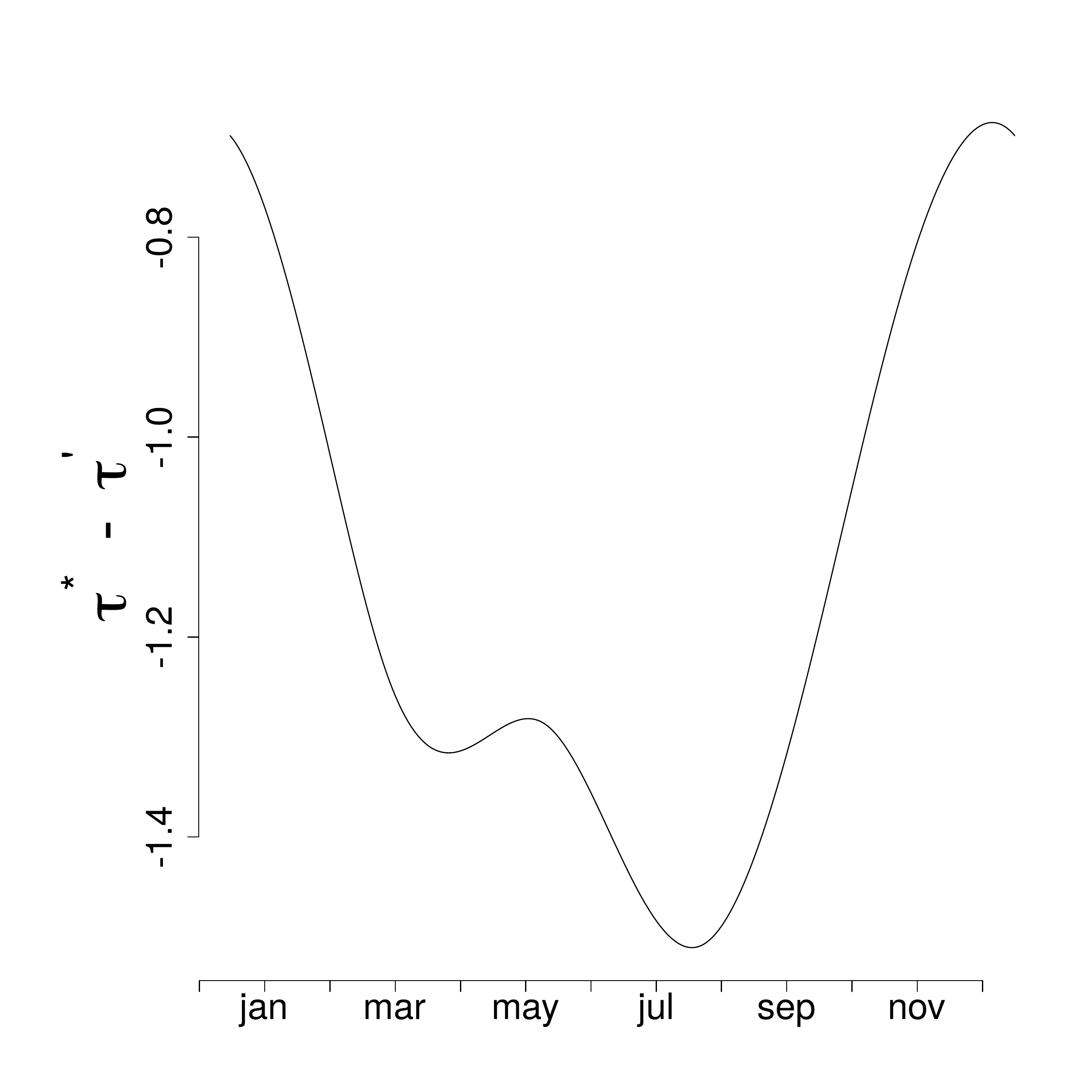} 	
	\caption{Time-varying parameter estimates of the NSGP correlation function of $X(s_1, \tau)$. For simplicity, $\beta_{qp}$ denotes the spherical coordinate $\left[\beta_q \right]_p$. In the last figure, $\tau^*(s_1=-9,s_1'=-2, \tau') - \tau'$ is plotted against time $\tau'$.}
	\label{Fig:RelHumSpainPars_lontime}
\end{figure}

Figure~\ref{Fig:RelHumSpainPars_lontime} displays the estimates of $\left[\beta_1 \right]_1(\tau)$, $\left[\beta_2 \right]_1(\tau)$, $\cos(\left[\beta_2 \right]_2)(\tau)$, and  $\tau^*(s_1=-9,s_1'=-2, \tau') - \tau'$ , which are interpreted as in equation \eqref{taumax2new} of the main manuscript. The estimates illustrate important aspects shown in the empirical temporal correlation functions seen in Figure~\ref{Plot:RelHumEmpiricalTempCorLon}. For example, the larger estimates of $\left[\beta_1 \right]_1$ in summer months indicate that in these months the spatial correlation decays faster, an aspect which seems clear in Figure~\ref{Plot:RelHumEmpiricalTempCorLon}.

In addition, the negative estimates of the degree of nonseparability $\cos(\left[\beta_2 \right]_2)$ explain the negative sign of the time lag $\tau-\tau'$ at which different locations have the largest correlation, which is corroborated by Figure~\ref{Plot:RelHumEmpiricalTempCorLon}. Finally, the plot of $\tau^*(s_1=-9,s_1'=-2, \tau') - \tau'$ shows how this `optimal' time lag evolves over calendar months. Models which assume separable covariance structure inherently assume that this `optimal' time lag is always zero.

\newpage

\begin{figure}[H]
	\centering 
	\includegraphics[width=0.6\linewidth]{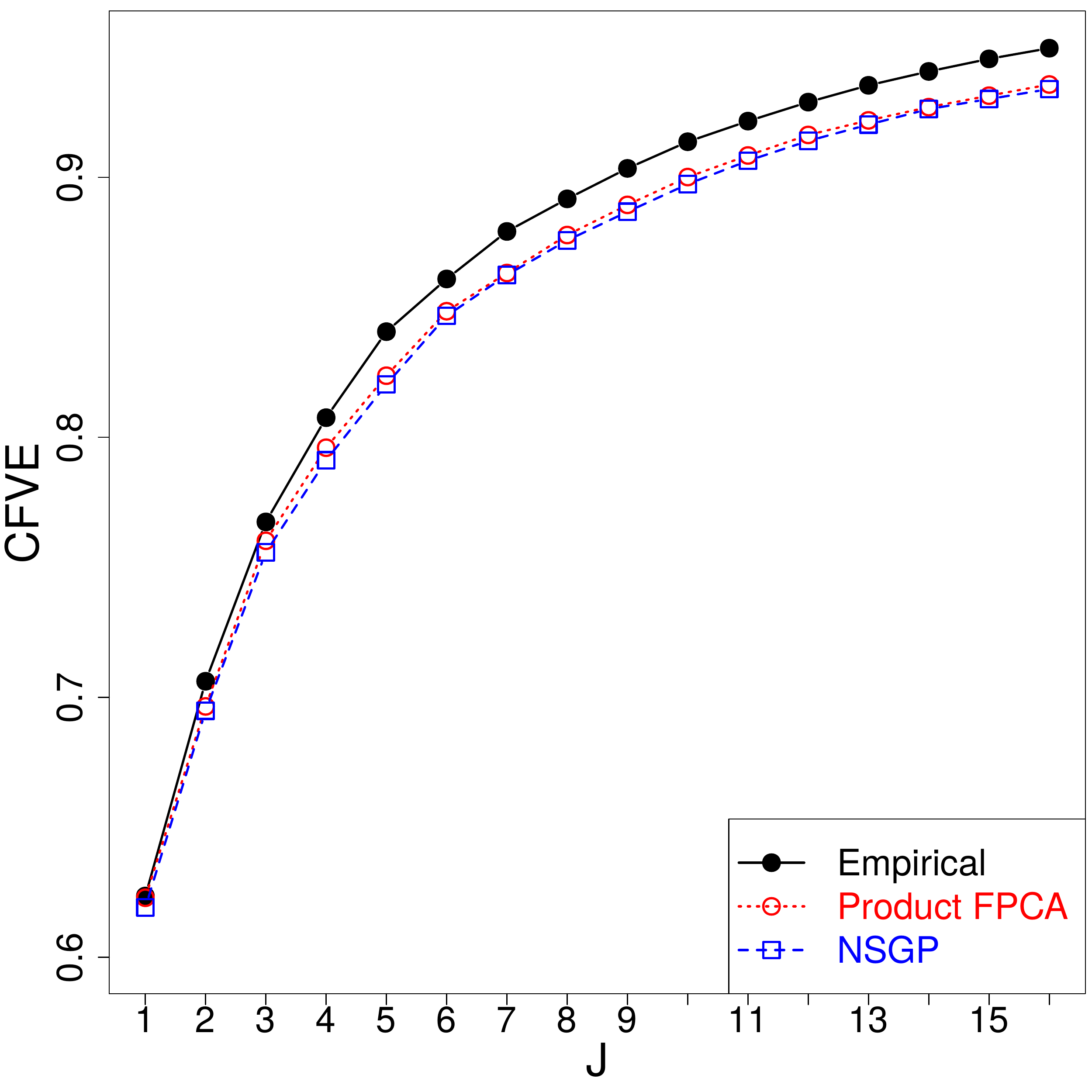} 
	\caption{Cumulative FVEs for Longitude x Time model obtained by using the true eigensurfaces, the eigensurfaces of the empirical covariance function, and the eigensurfaces of the covariance function estimated by NSGP and Product FPCA models.}
	\label{CFVE_RelHumLonTime}
\end{figure}

Figure~\ref{CFVE_RelHumLonTime} shows the CFVEs of the 16 leading eigensurfaces of the empirical correlation function and of NSGP and Product FPCA models. The latter two have very similar performance in terms of CFVE. 

\begin{figure}[H]
	\centering 
	\includegraphics[width=1\linewidth]{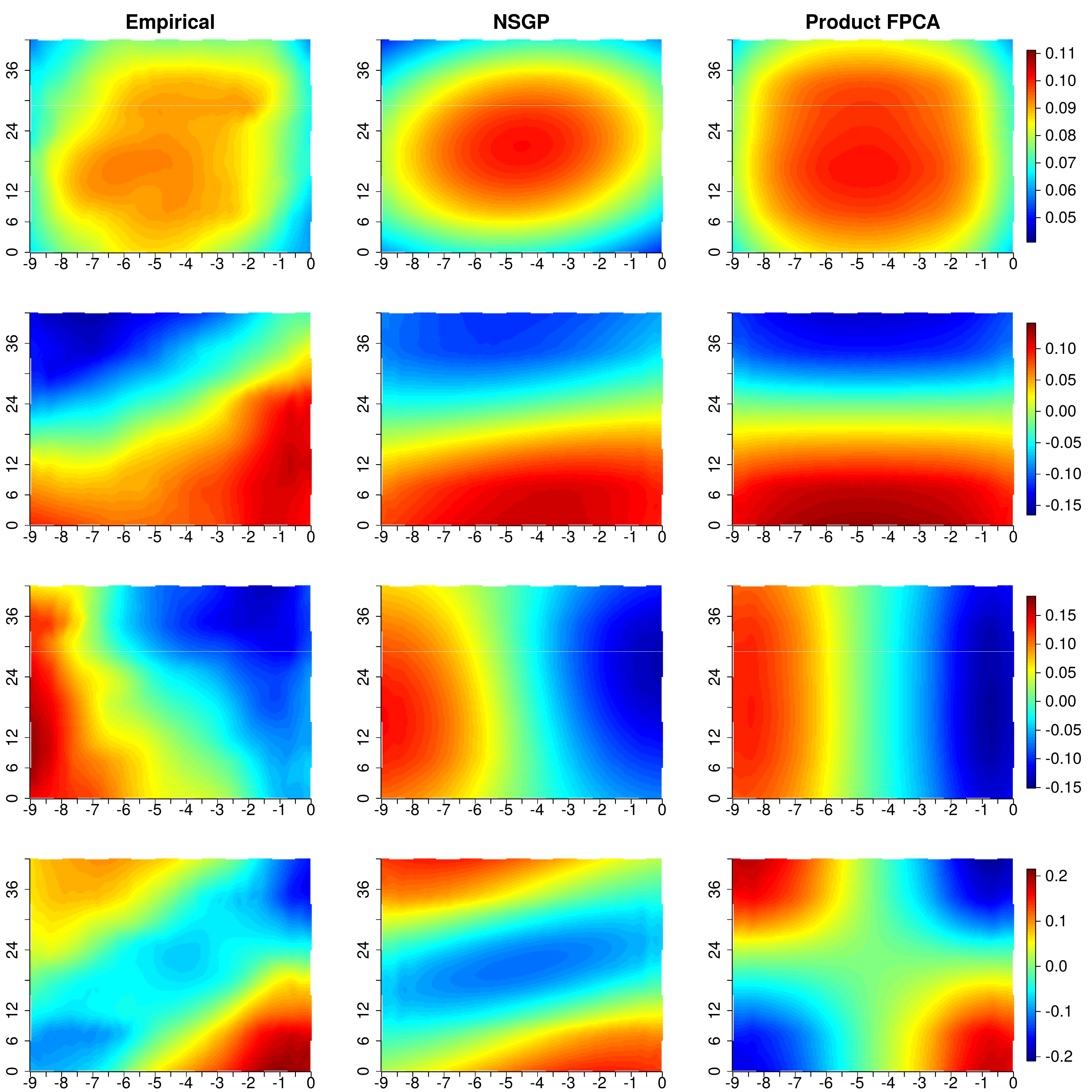} 
	\caption{First four leading eigensurfaces $\phi_j(s_1, \tau)$ of the true model (left column) and the corresponding estimated eigensurfaces $\hat{\phi}_j(s_1, \tau)$ of NSGP model (center) and Product FPCA model (right).}
	\label{Eigensurf_RelHumLonTime}
\end{figure}

Note that, although that the data are observed over the entire calendar year, potential interaction between spatial location and time is only expected to be seen within a short time frame (a few hours or a couple of days). Therefore, we now restrict the analysis for a specific time period only. We look at the first 48 hours of the month of July, where the degree of nonseparability found by NSGP model was fairly large in magnitude as suggested in Figure~\ref{Fig:RelHumSpainPars_lontime}. Figure~\ref{Eigensurf_RelHumLonTime} displays the eigensurfaces found in that time period and indicates that interactions found by NSGP model are corroborated by the empirical correlation function found in the data, while Product FPCA model cannot extract these aspects.

\newpage

\subsection{Latitude x Time Model}

\begin{figure}[H]
	\centering
	\includegraphics[width=0.46\linewidth]{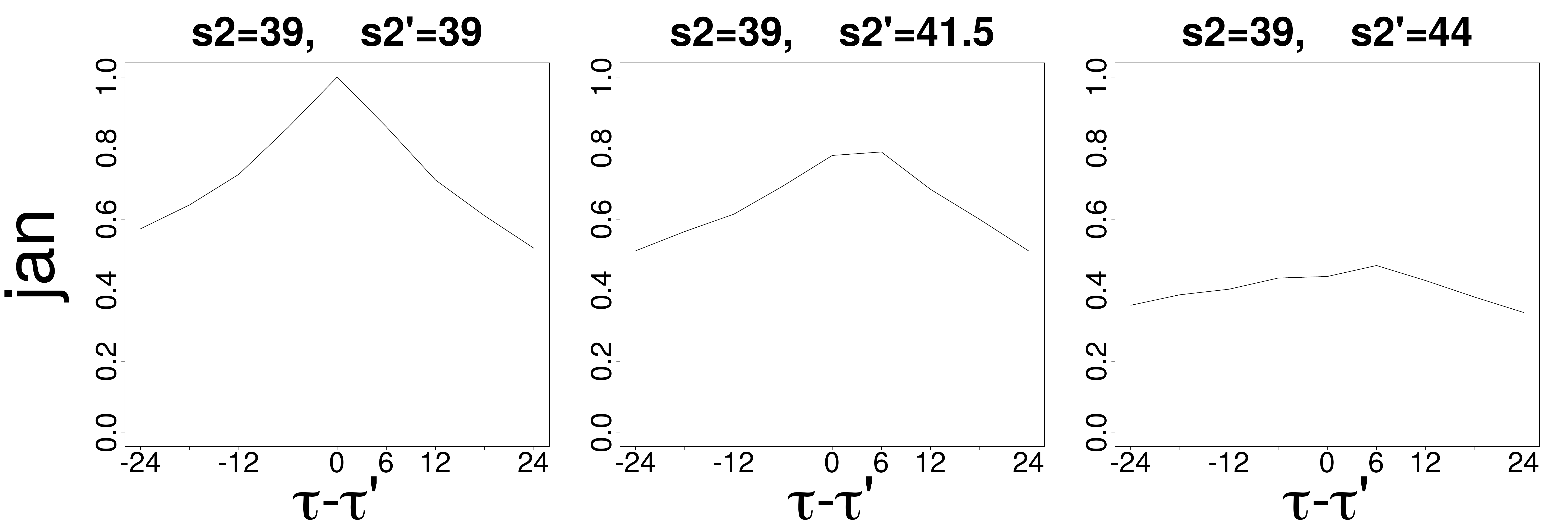} \hspace{1cm}
	\includegraphics[width=0.46\linewidth]{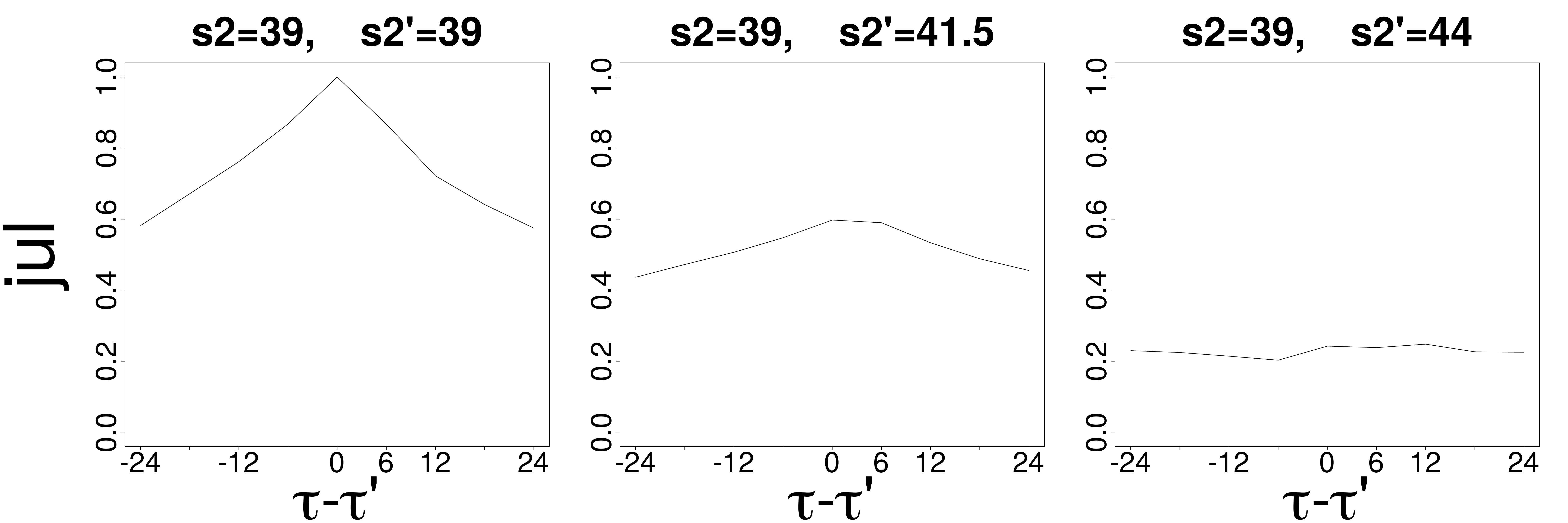}
	\\ \vspace{0.5cm}
	\includegraphics[width=0.46\linewidth]{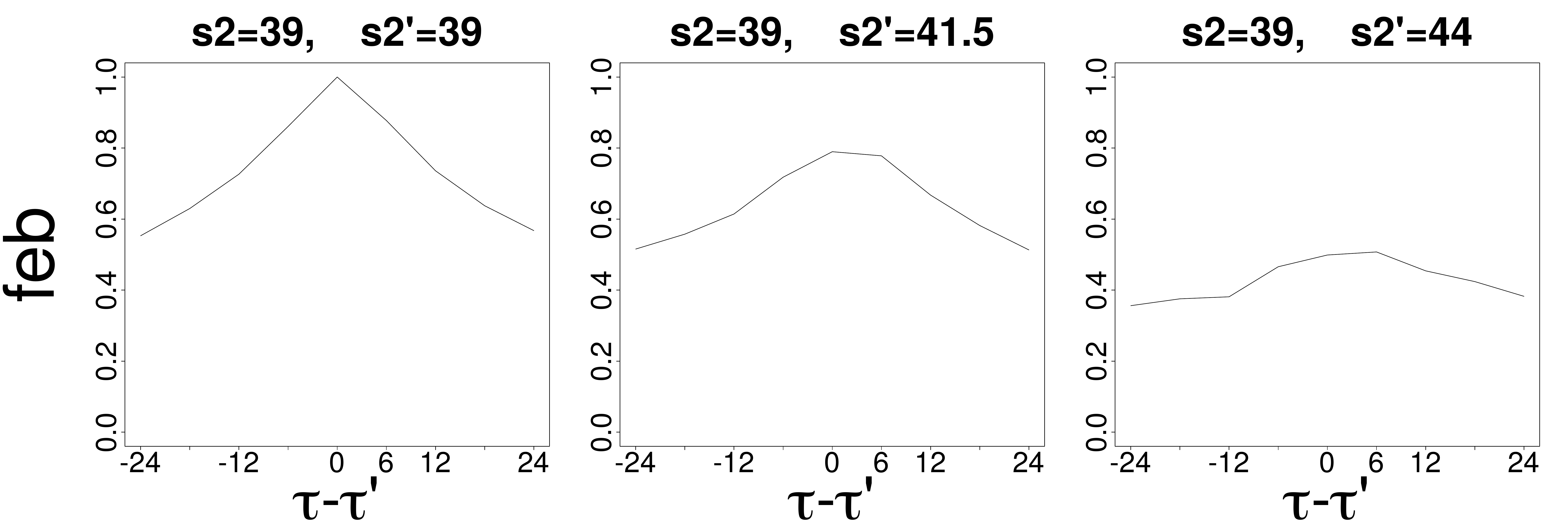} \hspace{1cm}
	\includegraphics[width=0.46\linewidth]{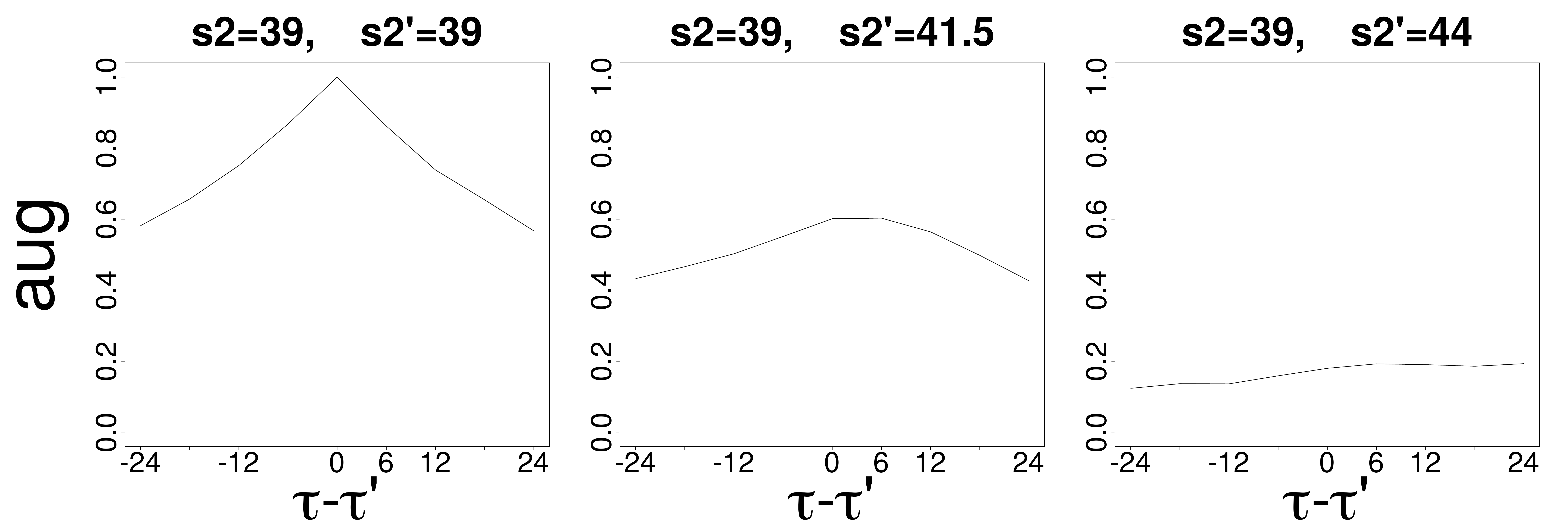}
	\\ \vspace{0.5cm}
	\includegraphics[width=0.46\linewidth]{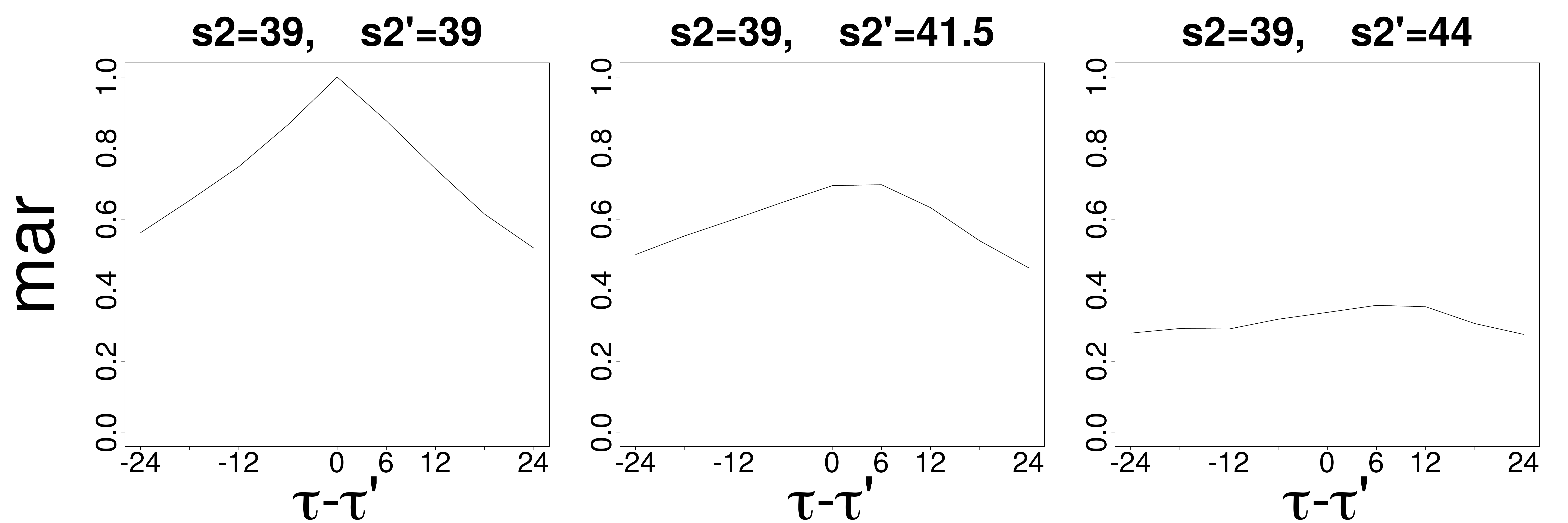} \hspace{1cm}
	\includegraphics[width=0.46\linewidth]{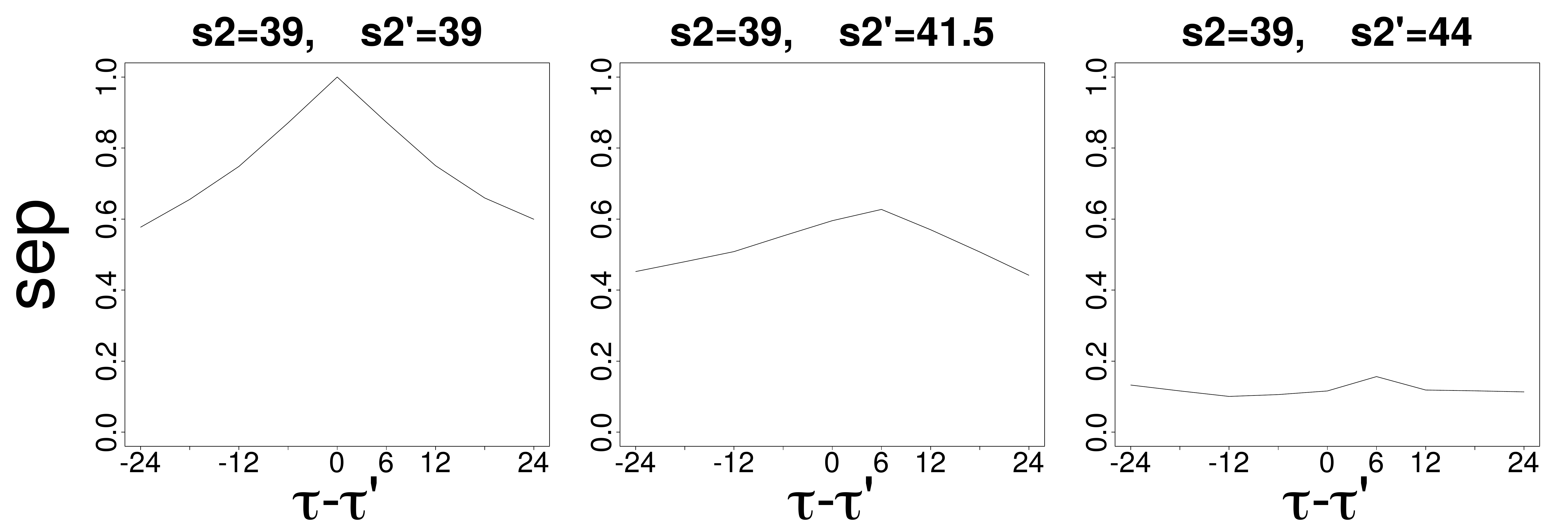}
	\\ \vspace{0.5cm}
	\includegraphics[width=0.46\linewidth]{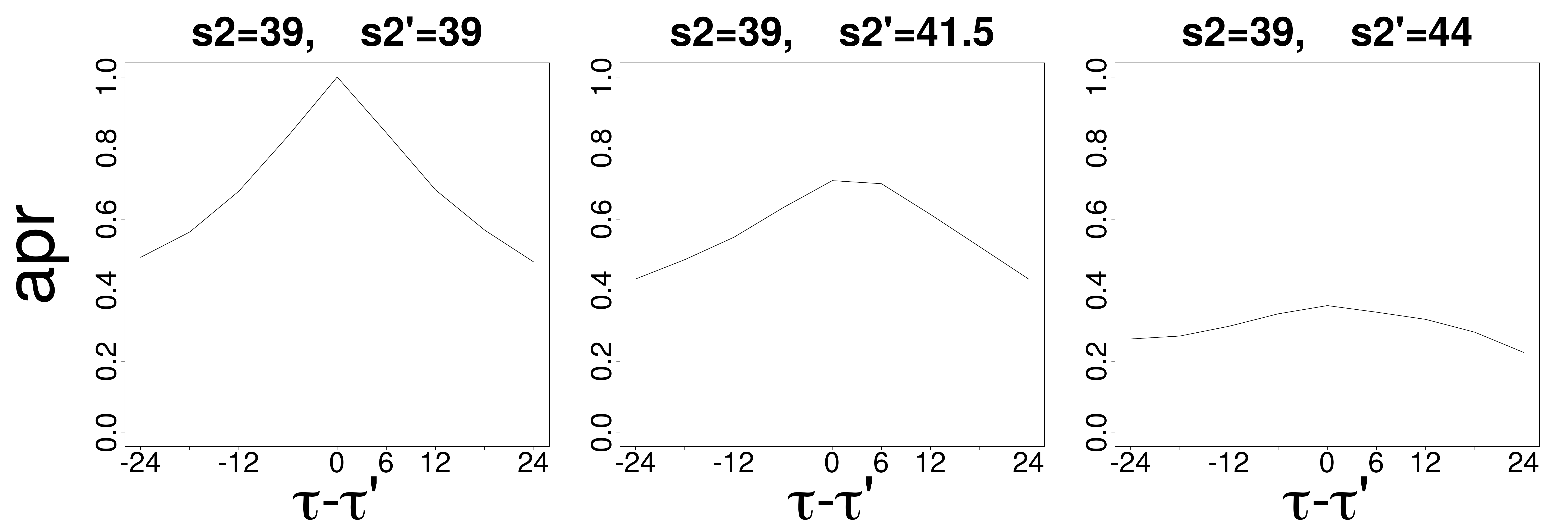} \hspace{1cm}
	\includegraphics[width=0.46\linewidth]{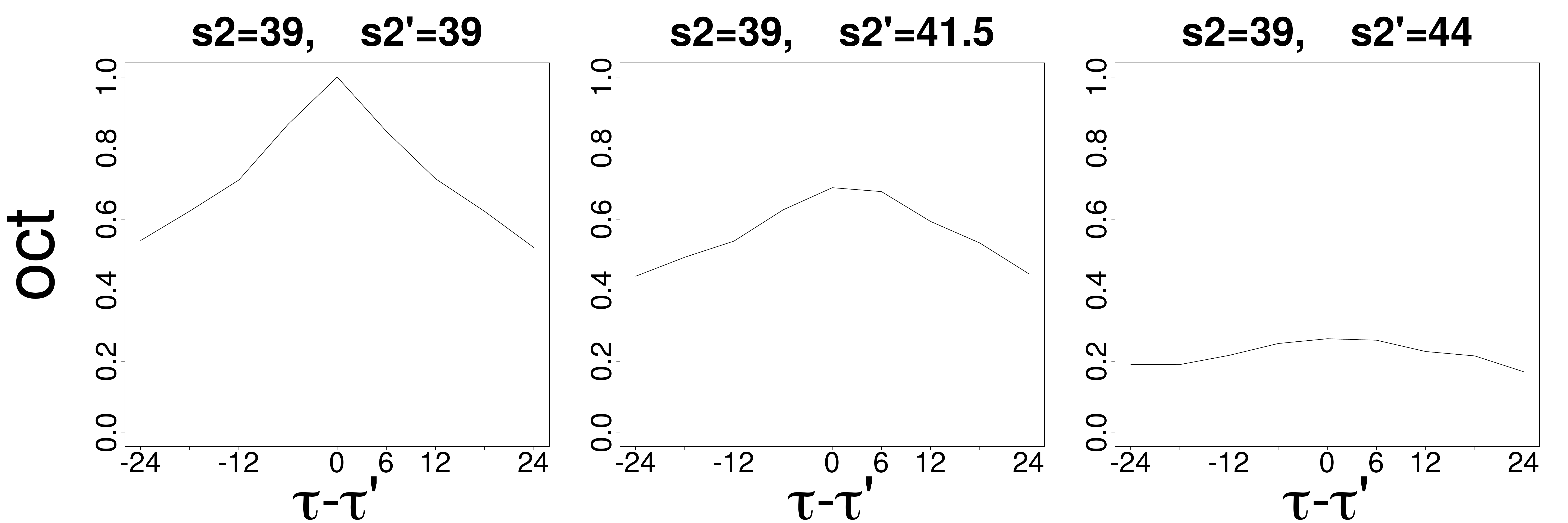}
	\\ \vspace{0.5cm}
	\includegraphics[width=0.46\linewidth]{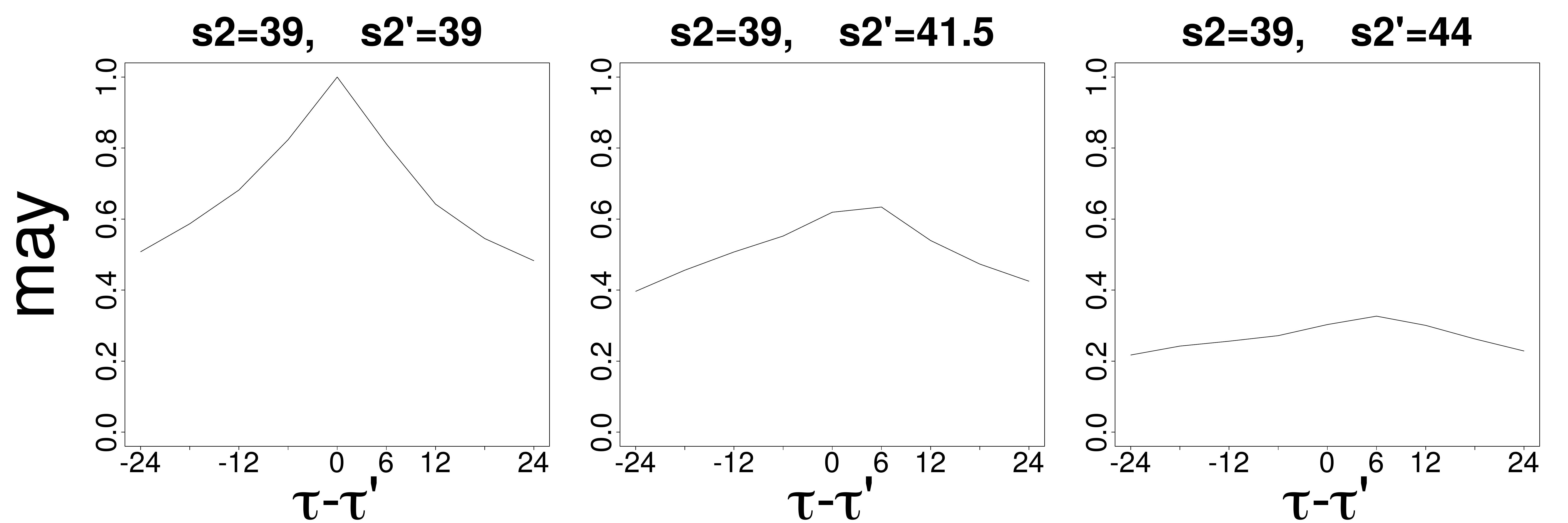} \hspace{1cm}
	\includegraphics[width=0.46\linewidth]{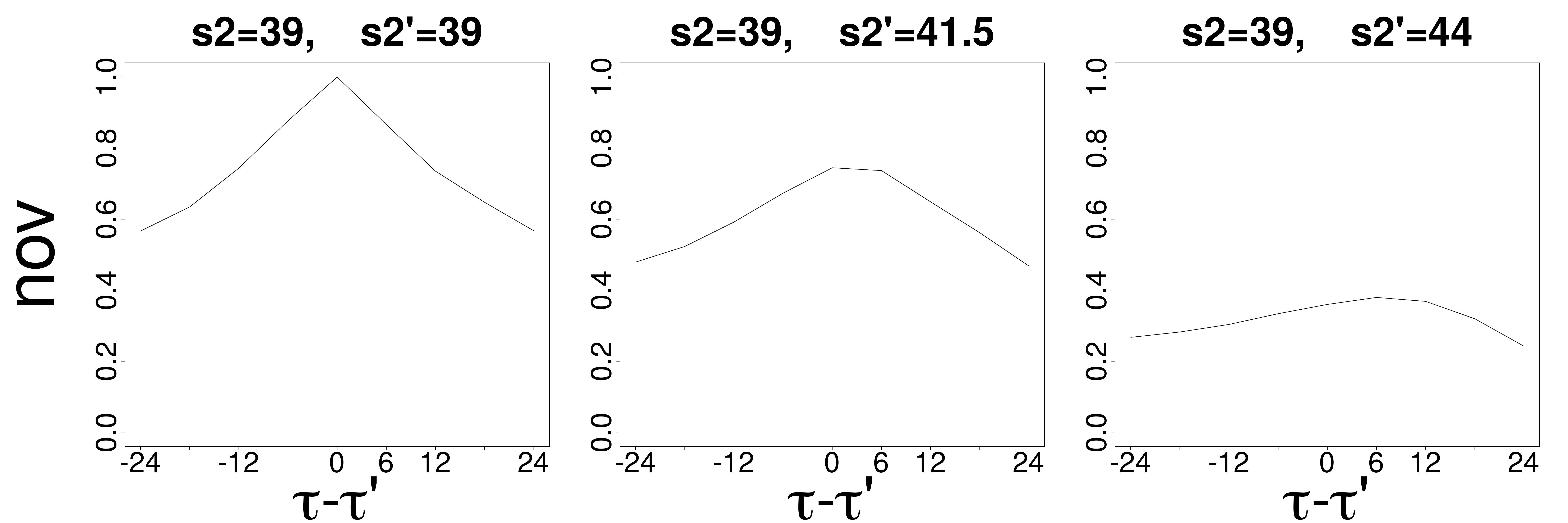}
	\\ \vspace{0.5cm}
	\includegraphics[width=0.46\linewidth]{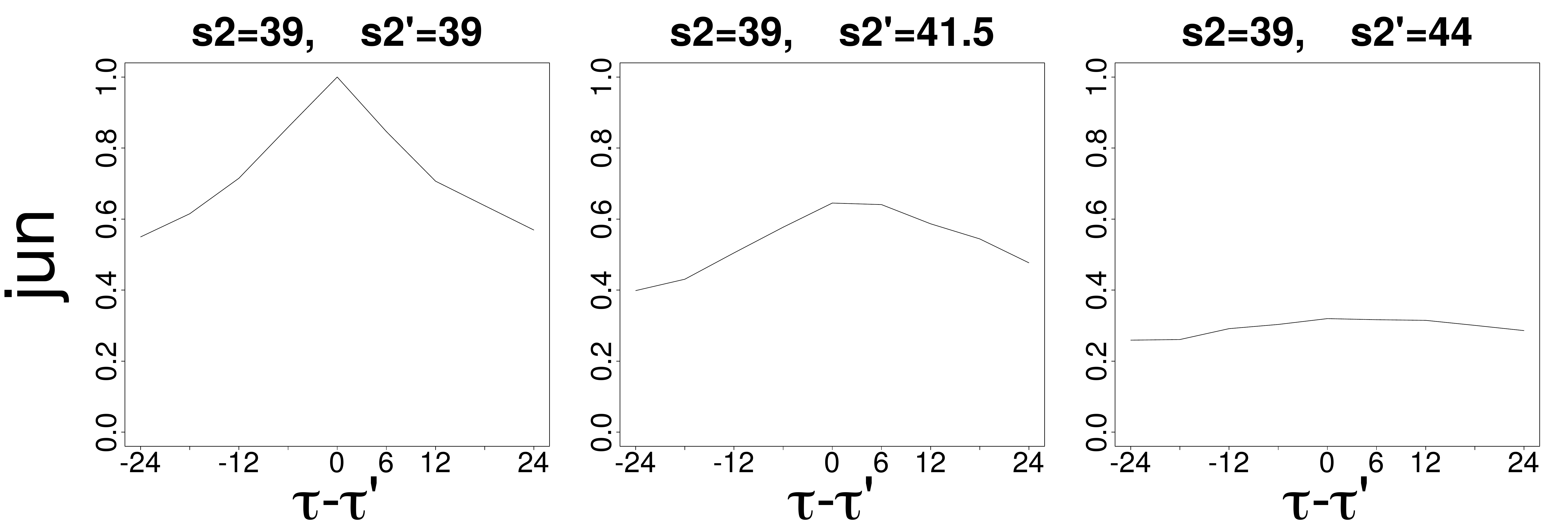} \hspace{1cm}
	\includegraphics[width=0.46\linewidth]{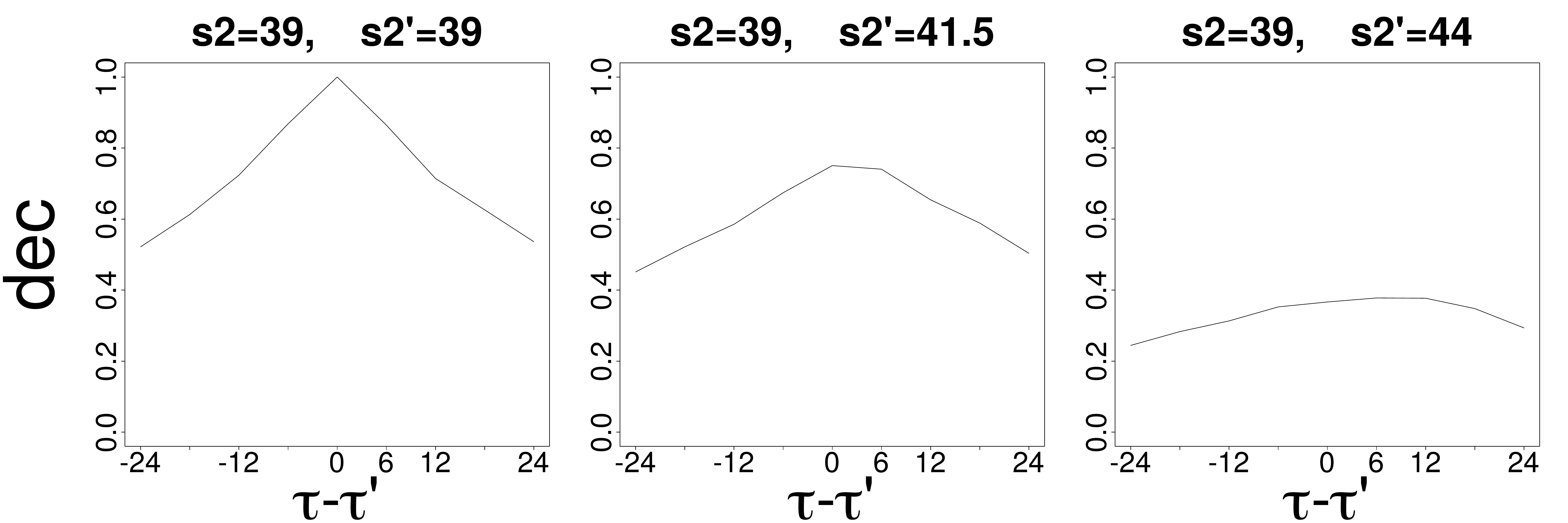}
	\caption{Empirical temporal correlation functions for relative humidity data  (Latitude x Time model) from the Iberian Peninsula. Each triple of plots displays the empirical temporal correlation functions $\hat{\Cor} \Big[ X(s_2=39,\tau), X(s_2',\tau') \Big]$ against $\tau-\tau'$, \ for $s_2'=39, 41.5,$ and $44$ for a particular month. Time distances $\tau-\tau'$ are shown in hours.}
	\label{Plot:RelHumEmpiricalTempCorLat}
\end{figure}

\begin{figure}[H]
	\centering
	\includegraphics[width=0.45\linewidth]{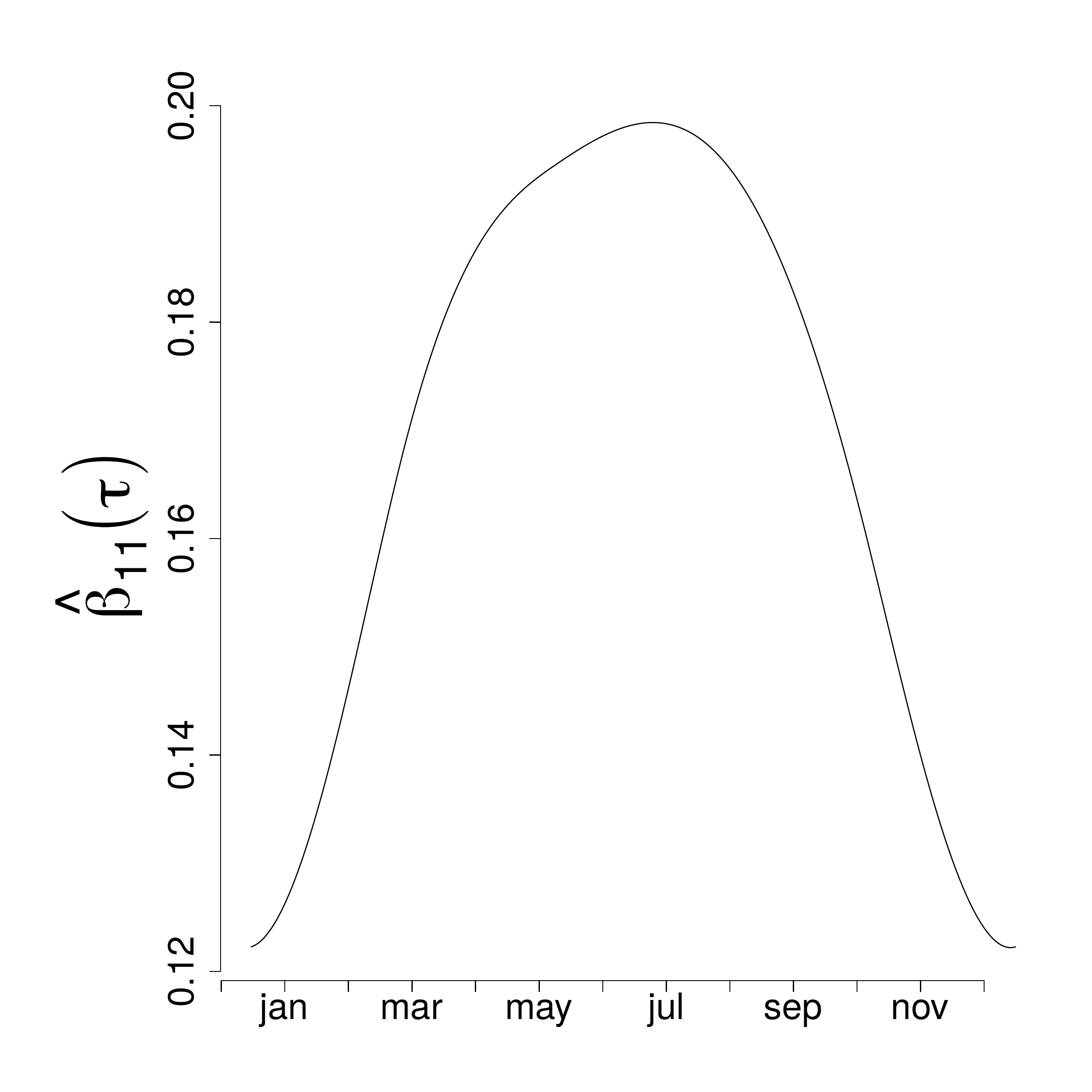}
	\includegraphics[width=0.45\linewidth]{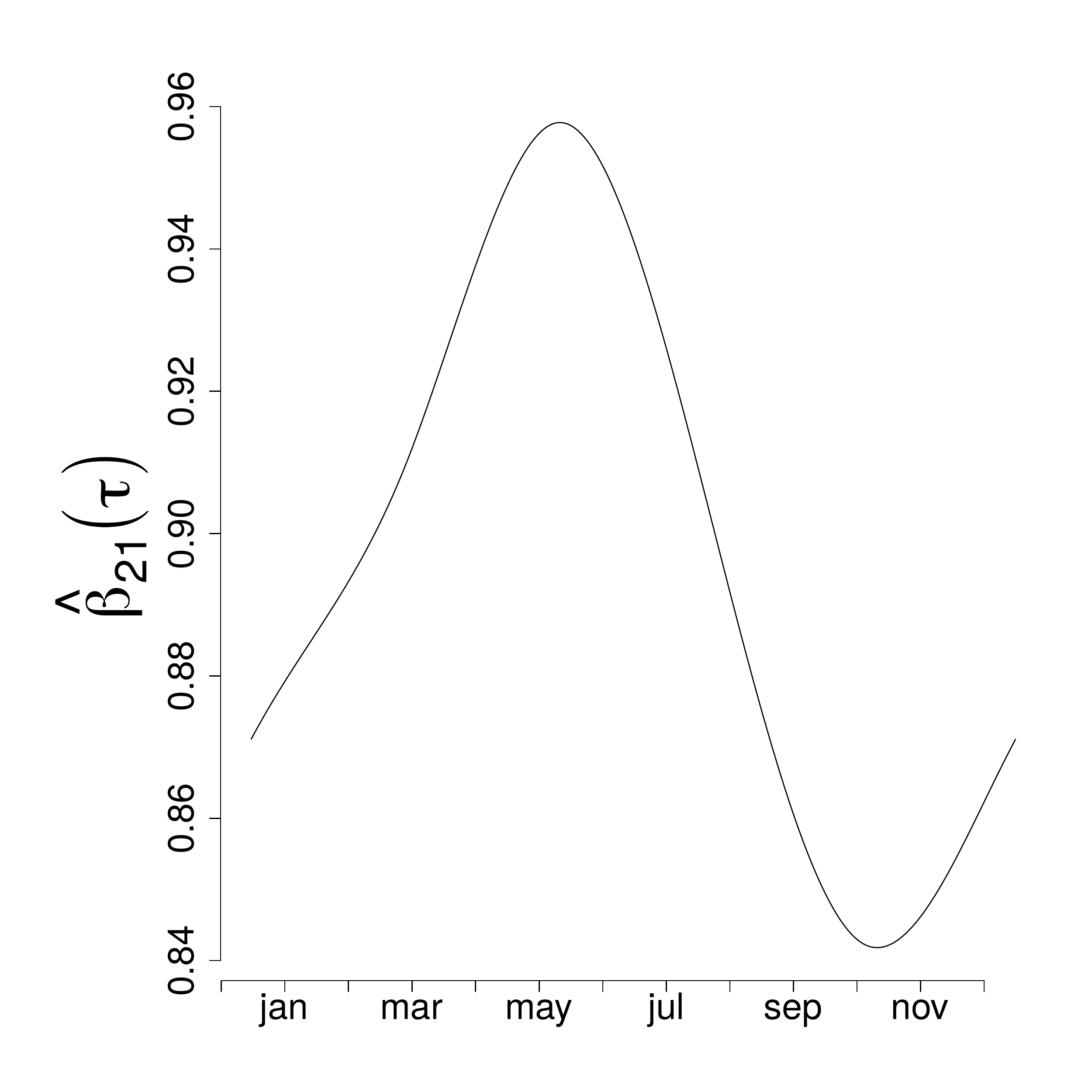}\\
	\includegraphics[width=0.45\linewidth]{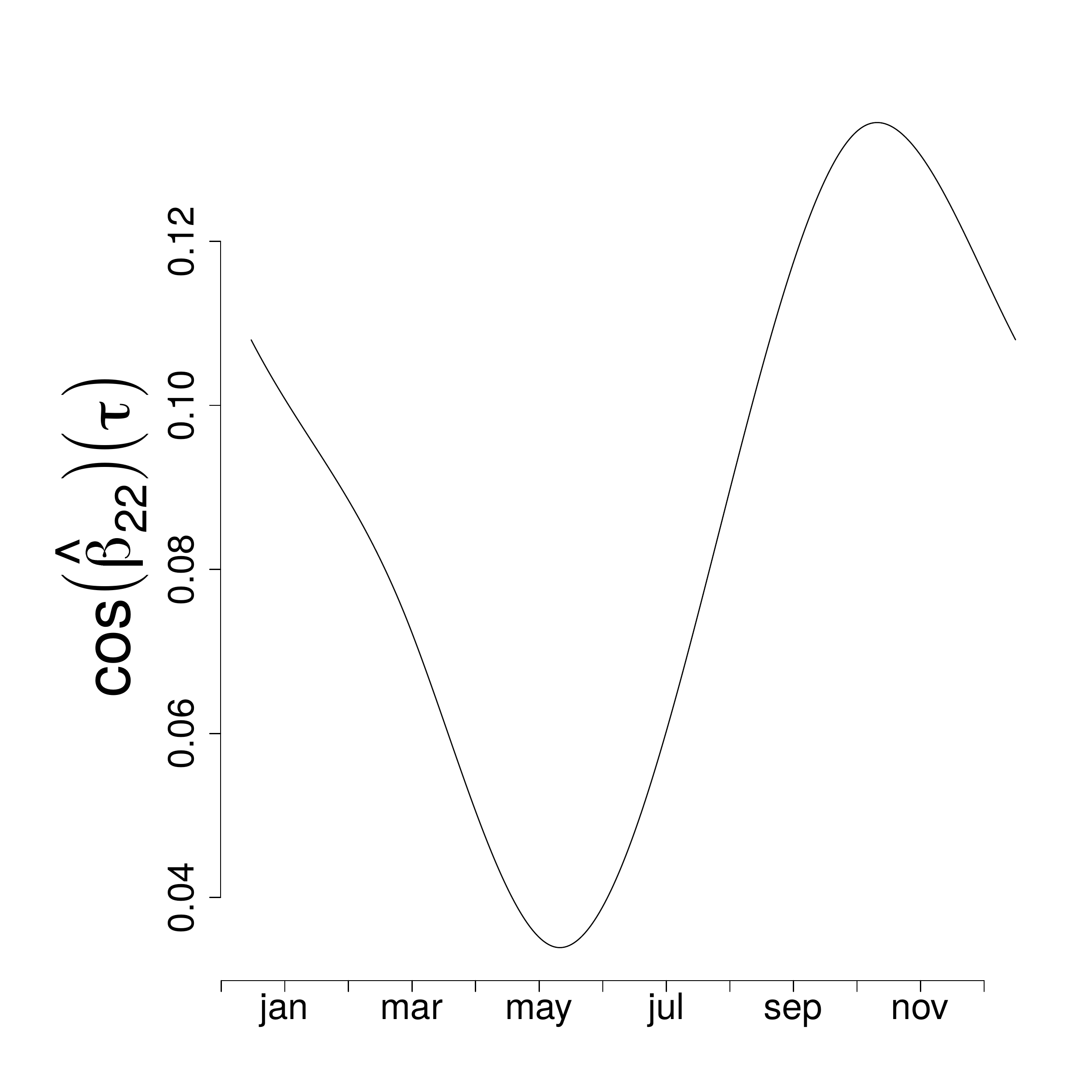}
	\includegraphics[width=0.45\linewidth]{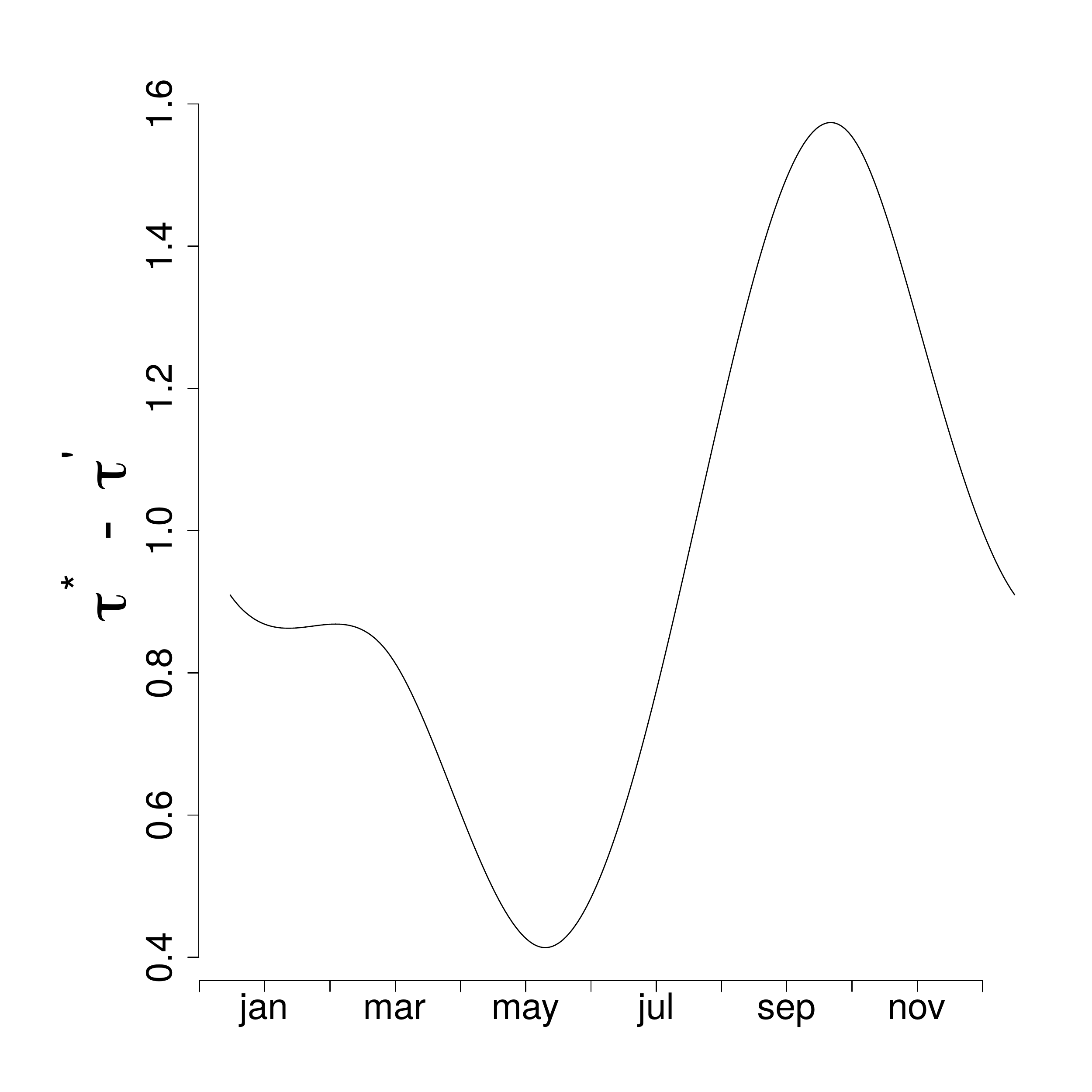} 	
	\caption{Time-varying parameter estimates of the NSGP correlation function of $X(s_2, \tau)$. For simplicity, $\beta_{qp}$ denotes the spherical coordinate $\left[\beta_q \right]_p$. In the last figure, $\tau^*(s_2=39,s_2'=44, \tau') - \tau'$ is plotted against time $\tau'$.}
	\label{Fig:RelHumSpainPars_lattime}
\end{figure}

We now focus on the correlation function of $X(s_2,\tau)$. Figure~\ref{Plot:RelHumEmpiricalTempCorLat} displays the estimated empirical temporal correlation functions \sloppy $\hat{\Cor} \big[ X(s_2=39,\tau), X(s_2',\tau') \big]$, for $s_2'=39, 41.5,$ and $44$, in each month. Figure~\ref{Fig:RelHumSpainPars_lattime} shows the estimates of $\left[\beta_1 \right]_1(\tau)$, $\left[\beta_2 \right]_1(\tau)$, $\cos(\left[\beta_2 \right]_2)(\tau)$, and  $\tau^*(s_2=39,s_2'=44, \tau') - \tau'$. 

Agreeing with Figure~\ref{Plot:RelHumEmpiricalTempCorLat}, the estimates of $\left[\beta_1 \right]_1$ in the summer shows faster spatial correlation decay than in the winter and the estimates of $\left[\beta_2 \right]_1$ change little over time.  The degree of nonseparability $\cos(\left[\beta_2 \right]_2)(\tau)$ is always positive, being lower in April--June. As a consequence, the resulting estimates of $\tau^*(s_2=39,s_2'=44, \tau') - \tau'$ are always positive, being larger for other months. Indeed, the empirical temporal correlation functions in Figure~\ref{Plot:RelHumEmpiricalTempCorLat} show that the `optimal' time lag seems positive in all months, while this is slightly less clear in April--June.

%\newpage

\begin{figure}[H]
	\centering 
	\includegraphics[width=0.6\linewidth]{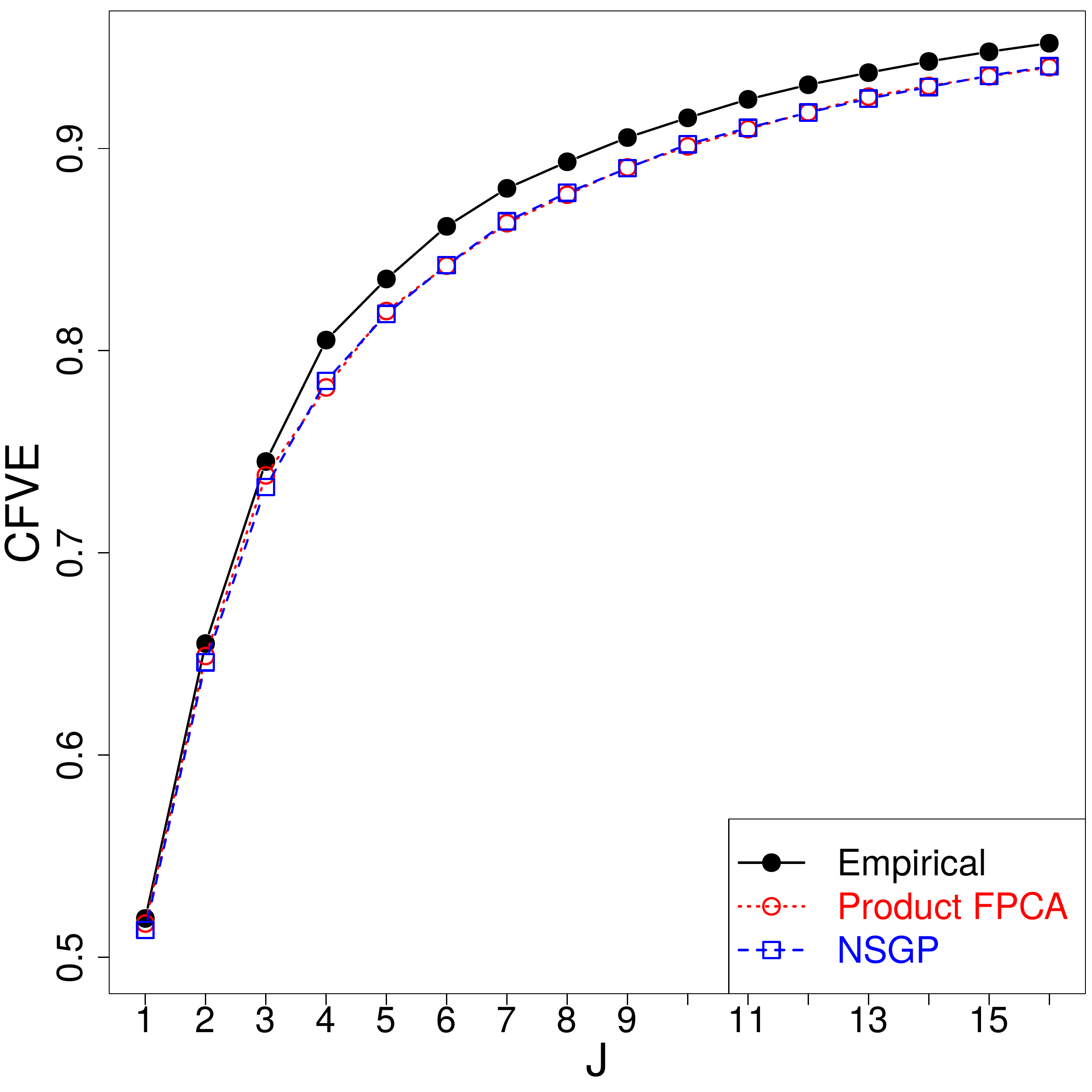} 
	\caption{Cumulative FVEs for Latitude x Time model obtained by using the true eigensurfaces, the eigensurfaces of the empirical covariance function, and the eigensurfaces of the covariance function estimated by NSGP and Product FPCA models.}
	\label{CFVE_RelHumLatTime}
\end{figure}

Figure~\ref{CFVE_RelHumLatTime} shows NSGP and Product FPCA models are similar in terms of CFVE and close to the model based on the empirical correlation function.

\begin{figure}[H]
	\centering 
	\includegraphics[width=1\linewidth]{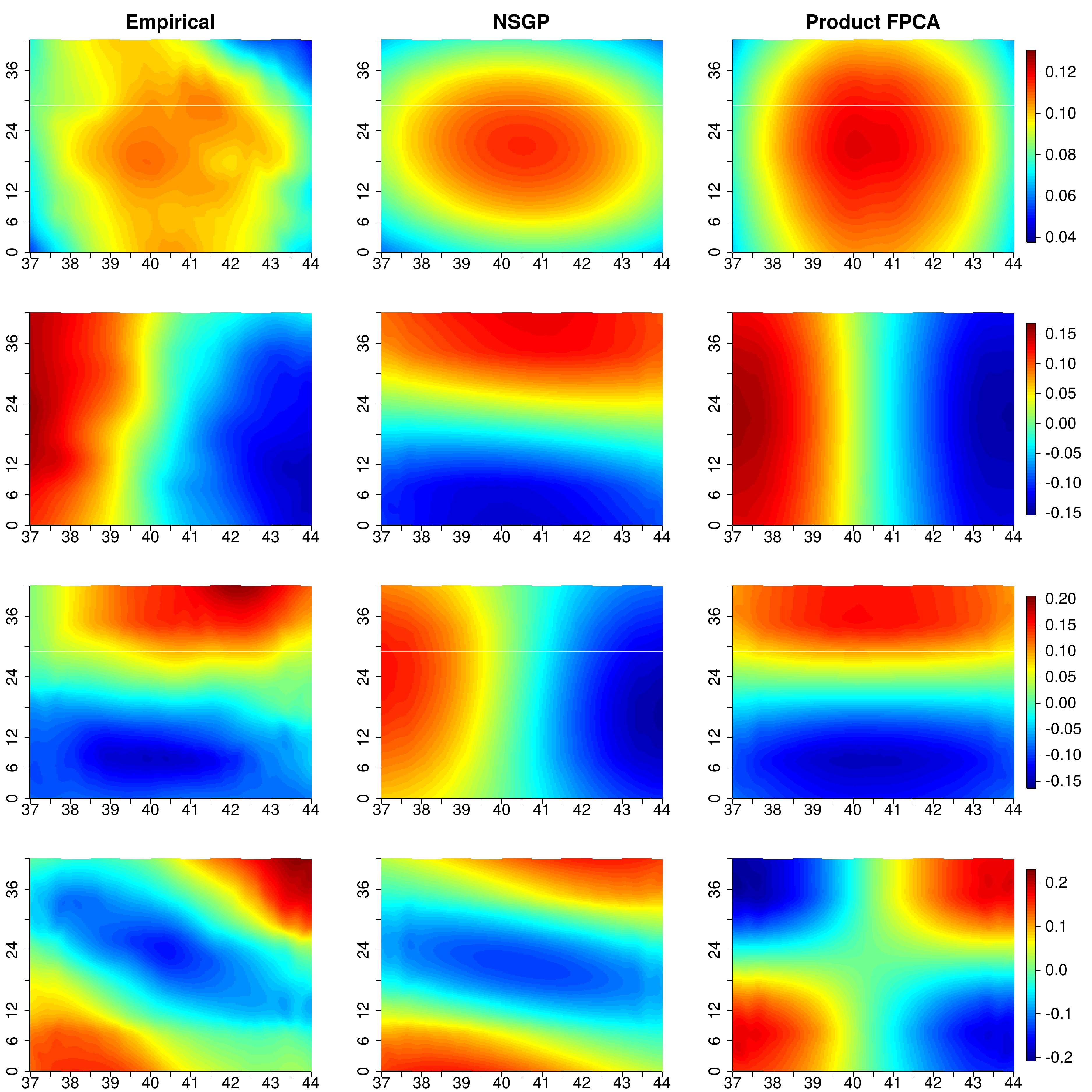} 
	\caption{First four leading eigensurfaces $\phi_j(s_2, \tau)$ of the true model (left column) and the corresponding estimated eigensurfaces $\hat{\phi}_j(s_2, \tau)$ of NSGP model (center) and Product FPCA model (right).}
	\label{Eigensurf_RelHumLatTime}
\end{figure}

In order to look at interactions between latitude and time, we adopt the same strategy used in the Longitude x Time model. We analyze the data corresponding to the first 48 hours of the month of November, a month in which the degree of nonseparability found by NSGP model was fairly large -- see Figure~\ref{Fig:RelHumSpainPars_lattime}. In Figure~\ref{Eigensurf_RelHumLatTime}, we can see the leading eigensurfaces. Especially the fourth one suggests some clear interaction between latitude and time, a feature which seems to be identified by NSGP model and cannot be modeled by Product FPCA model.

\newpage

\section{Nelson-Aalen Estimator}\label{sec:NelsonAalenSuppl}

\addtolength{\textheight}{0.2in}%

In order to measure topological features of random fields, \cite{garside2020topol} propose an estimator for the marginal cumulative event rate which is analogous to the usual Nelson-Aalen estimator. The new estimator is also called Nelson-Aalen and we will explain it briefly, following closely the exposition of \cite{garside2020topol}. 

For a random field defined on a finite discrete space ${\cal S}$, let $x_{\ve s}$  be the field value at location $\ve s \in {\cal S}$ and $x_{(\ve s)}$ the field values at neighbor locations $(\ve s)$. The neighbors can be, for example, those of shared edges (as we use in our analysis) or nearest neighbors. At each location $\ve s$, let $N_{\ve s}(v)$ be the number of individuals born at location $\ve s$ up to level $v$:
\begin{align}
N_{\ve s}(v) =
\begin{cases}
1, \quad & x_{\ve s} \leq v, \ \ x_{(\ve s)} > x_{\ve s} 1_{\ve s}, \\
0, \quad & \text{otherwise},
\end{cases}
\end{align}
where $1_{\ve s}$ is a unit vector of the same length as $x_{(\ve s)}$.
In addition, let $Y_{\ve s}(v)$ be the at-risk indicator at level $v$, given by
\begin{align}
Y_{\ve s}(v) =
\begin{cases}
1, \quad & x_{\ve s} \geq v, \ \ x_{(\ve s)} > v 1_{\ve s}, \\
0, \quad & \text{otherwise}.
\end{cases}
\end{align}
By defining $N(v) = \sum_{\ve s} N_{\ve s}(v)$ and $Y(v) = \sum_{\ve s} Y_{\ve s}(v)$, the proposed Nelson-Aalen estimator for the marginal cumulative event rate is given by
\begin{equation}\label{NelsonAalenIntensity}
\hat{H}(v) = \int_{-\infty}^{v} \frac{dN(u)}{Y(u)}.
\end{equation}
\cite{garside2020topol} point out that $\hat{H}(v)$ is asymptotically Gaussian and consistent for
\begin{equation}\label{NelsonAalenGaussian}
H(v) = \int_{-\infty}^{v} J(u) \frac{\sum_{\ve s} P \big(x_{(\ve s)} > u 1_{\ve s} \ | \ x_{\ve s} = u \big) f_{\ve s}(u)}{ \sum_{\ve s} P \big(x_{\ve s} > u, \ x_{(\ve s)} > u 1_{\ve s}\big)} du,
\end{equation}
where $J(u) = I\{Y(u)>0\}$ and $f_{\ve s}(\cdot)$ is the marginal density of $x_{\ve s}$.

In our implementation, we use \eqref{NelsonAalenIntensity} for obtaining Nelson-Aalen plots for realizations and \eqref{NelsonAalenGaussian} for fitting assessment of covariance functions of $Q$-dimensional function-valued processes.

\bibliographystyle{agsm}
\bibliography{references}